\documentclass[11pt]{article}
\pdfoutput=1 % if your are submitting a pdflatex (i.e. if you have
\usepackage{jcappub} % for details on the use of the package, please
\usepackage[T1]{fontenc} % if needed
\usepackage{amssymb}
\usepackage{epsfig}
\usepackage{rotating}
\usepackage{slashed}
\usepackage{color}
\usepackage[utf8]{inputenc}
\usepackage{extarrows}

\newcommand{\code}[1]{{\tt #1}}

\newcommand{\be}{\begin{equation}}
\newcommand{\ee}{\end{equation}}

\newcommand{\beq}{\begin{equation}}
\newcommand{\eeq}{\end{equation}}
\newcommand{\bea}{\begin{eqnarray}}
\newcommand{\eea}{\end{eqnarray}}

\newcommand{\beqa}{\begin{eqnarray}}
\newcommand{\eeqa}{\end{eqnarray}}
\renewcommand{\Re}{\mathop{\rm Re}}

\newcommand{\N}{N}

\def\lsim{\mathrel{\rlap{\lower4pt\hbox{\hskip1pt$\sim$}}
     \raise1pt\hbox{$<$}}}         %less than or approx. symbol
\def\gsim{\mathrel{\rlap{\lower4pt\hbox{\hskip1pt$\sim$}}
     \raise1pt\hbox{$>$}}}         %grater than or approx. symbol
\def\esim{\mathrel{\rlap{\raise2pt\hbox{$\sim$}}
     \lower1pt\hbox{$-$}}}         %equal to or approx. symbol

\newcommand{\ds}{{\sf DarkSUSY}}
\newcommand{\dst}{{\sf DarkSUSY}}
\newcommand{\dsver}{{\sf 6}}
\newcommand{\dstver}{{\sf 6.1}}

\title{\boldmath \dst\ \dsver: An Advanced Tool to Compute Dark Matter Properties Numerically}

\author[a]{Torsten Bringmann,}
\author[b]{Joakim Edsj\"o,}
\author[c]{Paolo Gondolo,}
\author[d]{Piero Ullio}
\author[b]{and Lars Bergstr\"om}

\affiliation[a]{ Department of Physics, University of Oslo, Box 1048, NO-0371 Oslo, 
   Norway.}
\affiliation[b]{ The Oskar Klein Centre for Cosmoparticle Physics, Department of Physics, 
   Stockholm University, AlbaNova, SE-106 91 Stockholm, Sweden.}
\affiliation[c]{ Department of Physics, University of Utah,115 South 1400 East, Suite 201, 
  Salt Lake City, UT 84112-0830, USA.}
\affiliation[d]{ SISSA and INFN, Sezione di Trieste, via Bonomea 265, 34136 Trieste, Italy.}

\emailAdd{torsten.bringmann@fys.uio.no}
\emailAdd{edsjo@fysik.su.se}
\emailAdd{paolo@physics.utah.edu}
\emailAdd{ullio@sissa.it}
\emailAdd{lbe@fysik.su.se}

\abstract{The nature of dark matter remains
one of the key science questions. Weakly Interacting Massive Particles (WIMPs) are 
among the best motivated particle physics candidates, allowing to explain the measured 
dark matter density by employing standard big-bang thermodynamics. Examples include 
the lightest supersymmetric particle, though many alternative particles have 
been suggested as a solution to the dark matter puzzle. We introduce here a 
radically new version of the widely used \ds\ package, which allows to compute 
the properties of such dark matter particles numerically. With \dst\ \dsver\ one can  
accurately predict a large variety of astrophysical signals from dark 
matter, such as direct detection rates in low-background counting experiments and indirect 
detection signals through antiprotons, antideuterons, gamma rays and positrons from the 
Galactic halo, or high-energy neutrinos from the center of the Earth or of the Sun.
For thermally produced dark matter like WIMPs, high-precision tools are  provided for the 
computation of the relic density in the Universe today,  as well as for the size of the 
smallest dark matter protohalos.  
Furthermore, the code allows to calculate dark matter self-interaction rates, which may
affect the distribution of dark matter at small cosmological scales.
Compared to earlier versions,  \ds\  \dsver\ introduces many significant physics 
 improvements and extensions. The most fundamental new feature of this release, 
 however,  is that the code has been completely re-organized and brought into a highly 
 modular and flexible shape. 
 Switching between different pre-implemented dark matter candidates
  has thus become straight-forward, just as adding new -- WIMP or non-WIMP -- particle models 
  or replacing any given functionality in a fully user-specified way. 
In this article, we describe the physics behind the computer package, along with the 
main structure and philosophy of this major revision of \ds. A detailed manual is provided 
together with the public release at {\tt www.darksusy.org}.
}

\begin{document}

\maketitle
\flushbottom

%%%%%%%%%%%%%%%%%%%%%%%%%%%%%%%%%%%%%%%%%%%%%%%%%%%%%%%%%%%%
\vspace*{1cm}
\section{Introduction}
\label{sec:intro}

The problem of finding the identity of dark matter (DM), one of the most interesting problems 
in our picture of the universe, still lacks a solution despite an impressive improvement of detection 
capabilities.  Indications have, however, grown even stronger 
that an unknown form of matter contributes to the gravitationally attractive matter density of the 
universe, as shown, for example, by analyses of the recent high-precision data from the PLANCK 
satellite \cite{Ade:2015xua}.
Also, the discovery \cite{Clowe:2006eq} of the ``bullet cluster'', two colliding 
galaxy clusters where the mass distribution, as extracted from gravitational lensing, and the baryonic 
mass, as detected by its X-ray emission, are clearly separated, which makes it very difficult to 
modify gravity to circumvent the problem.
We are thus led to believe, as already Zwicky conjectured in the 1930's \cite{Zwicky:1933gu} 
(and where Lundmark reached a 
similar conclusion even a few years earlier \cite{lundmark}), that most of the 
mass of the observable Universe is invisible, or `dark'. A broad consensus has been reached
that the most likely explanation for this excess mass is a so far unknown elementary 
particle \cite{Bertone:2010zza}.

Analyses combining high-redshift supernova luminosity distances,
microwave background fluctuations and baryon acoustic oscillations in the galaxy 
distribution \cite{Komatsu:2010fb,Ade:2015xua}
give tight constraints on the present mass density of matter in
the Universe. This is usually expressed in the ratio  $\Omega_M=\rho_M/\rho_{\rm crit}$, 
normalized to the critical
density $\rho_{\rm crit}=3H_0^2/(8\pi G_N)=h^2\times 1.9\cdot 10^{-29} \ {\rm g\, cm}^{-3}$. 
The value obtained  for the most recent Planck data (Table 4 in \cite{Ade:2015xua}) for cold DM 
 for the (unknown) particle, here called $\chi$, is $\Omega_\chi h^2 = 0.1188\pm 0.0010$, which is 
around 5 times higher  than the value obtained for ordinary baryonic matter, 
$\Omega_Bh^2 = 0.02225\pm 0.00016$.  
Here $h=0.6727\pm 0.0066$ is the derived \cite{Ade:2015xua} present value of the Hubble 
constant in units of
$100$~km~s$^{-1}$~Mpc$^{-1}$. 
In addition, the Planck data is consistent with a flat universe ($\Omega_\mathrm{tot}=1$)
 and a value for the dark energy component, e.g.~a cosmological constant $\Lambda$, of $\Omega_\Lambda = 0.6844\pm 0.0091$.   
Also from the point of view of structure formation, non-baryonic DM
seems to be necessary, and the main part of it should consist
of particles that were non-relativistic at the time when structure
formed (cold dark matter, CDM). This excludes the light standard model neutrinos, 
although there is perhaps a window remaining for sterile neutrinos in the keV range which 
would act like ``warm'' DM (for a review, see \cite{Abazajian:2012ys}).
The Planck collaboration \cite{Ade:2015xua}, 
  limits the contribution of standard model neutrinos to $\Omega_\nu h^2 \lsim 0.005.$

A well-motivated particle physics
candidate which has the required properties has long been the lightest
supersymmetric particle, assumed to be a neutralino
\cite{Goldberg:1983nd,Krauss:1983ik,Ellis:1983ew}.  (For thorough reviews of
supersymmetric DM, see \cite{Jungman:1995df,Bergstrom:2000pn,Bertone:2004pz}.)  
Although supersymmetry
has long been generally accepted as a very promising enlargement of the Standard
Model of particle physics (for instance it would solve the so-called hierarchy problem of 
understanding why the electroweak scale
is protected against Planck-scale corrections), the details of a viable model are largely 
unknown. In fact, simplified templates with only a few free parameters, like the so-called 
constrained minimal supersymmetric model (CMSSM) or mSUGRA, are already very 
strongly constrained by early LHC data \cite{Aaboud:2016wna,Khachatryan:2016nvf}.
All these results seem to push the mass scale of the lightest supersymmetric particle, the 
lightest neutralino,  up to the TeV scale. For a detailed and comprehensive global analysis 
of simple supersymmetric scenarios, we refer to the recent results 
\cite{Athron:2017qdc,Athron:2017yua} of the GAMBIT \cite{Athron:2017ard} collaboration.

Of course, there is no compelling reason why the actual model, if nature
is supersymmetric at all, should be of this simplest kind.  However,
the minimal supersymmetric standard model (MSSM) has served and still serves as a useful 
template with which to test current 
ideas about detection, both in particle physics accelerators and 
in DM experiments,
and contains many features which are expected to be universal for any
weakly interacting massive particle (WIMP) model. Such WIMP DM candidates 
appear almost inevitably in many non-supersymmetric extensions to the standard model
(e.g.~universal extra dimensions \cite{Hooper:2007qk} or little Higgs models \cite{Birkedal:2006fz}) and have 
the additional advantage that the observed DM abundance can be understood from a simple thermal 
production mechanism in the early universe. Although the tendency 
at present is that supersymmetric (SUSY) models are getting slightly out of fashion, we keep  
such models in
\dst, not the least to allow comparison of this well-studied candidate with other DM 
candidates that the user of \dst\ may want to supply as input to the program. 
It should be noted that there is also a large number of DM models that cannot be described
within the WIMP paradigm \cite{Steffen:2008qp,Feng:2010gw,Baer:2014eja}, which will 
become more and more important
alternatives if the classical WIMP searches continue to report the absences of
any (uncontroversial) signals. We stress that many routines in \dst\ can be used, without any 
modification, even for DM candidates that are not WIMPs. 

The main new feature of \dst\ \dsver, as compared to earlier \ds\ \cite{ds4} versions, is a radically 
new modular and flexible structure of the code.  This allows to handle a wide range of 
DM particle models, included in the release or supplied by the user, and quite in general 
makes it much easier to include any user-designed changes or additions. Besides this, 
 we have in \dst\ \dsver\ further developed the analytical and numerical tools for 
 computing the relic density of WIMP-like particles from the early universe, and the 
 ensuing direct and indirect detection rates today. Among the completely newly added 
 features are the possibility to compute the cutoff in the power spectrum of matter density 
 perturbations, related to  the kinetic decoupling of WIMP-like particles in the early 
 universe, the possibility to compute DM self-interaction rates
 as well as a detailed treatment of the impact of radiative 
 corrections on the spectra of DM-induced cosmic-ray spectra relevant for indirect
 detection. Overall, we provide a program package with a rather high level of 
 sophistication, aiming to benefit the scientific community working with problems 
 related to DM.
This paper describes the basic structure and the underlying physical
and astrophysical formulas contained in \dst, as well as examples of its use. Even if it has 
the same name as previous versions, \dst\ \dsver\ constitutes a major revision in that 
it introduces a conceptual change to the very structure of the package. Let us stress that 
these changes in the structure of the code are physics-driven,  greatly enlarging the 
range of potential physics applications. In particular,
the often hard-coded focus on supersymmetric neutralino DM, as implemented in 
previous versions, has now been abandoned. In this sense, one 
may argue that also 
the meaning has changed, and \dst\ should now be interpreted as\footnote{ 
Swedish for (dark) `SUSY and further models'.}
\begin{quote}
\centering {\bf Dark SU}sy {\bf S}amt {\bf Y}tterligare modeller
\end{quote}
To demonstrate this new interpretation, we explicitly include in this release modules for a generic 
WIMP and a generic decaying DM particle, as well as for the scalar singlet model \cite{Cline:2013gha}
(besides, obviously, supersymmetry). Further particle modules will be supplied with future code 
releases, but can at any time also easily be added by the user.

This article is organized as follows. We start, in Section \ref{sec:phys}, by summarizing the main 
new physics features since the last main release of \ds\ (describing version 4.0 of the code
\cite{ds4}). This is closely intertwined with the general structure  and guiding principles of the code,
which we describe in Section \ref{sec:Philo}. In the coming sections we then focus in more detail 
on the various physics aspects behind the implementation, with special emphasis on improvements 
or additions since the previous publication \cite{ds4}. We first describe 
the thermal production and decoupling of WIMPs (Section \ref{sec:thermal}), the direct 
detection of DM particles  (Section \ref{sec:direct}) and models for the DM distribution in the Milky 
Way or other halos (Section \ref{sec:halo}). As discussed in Section \ref{sec:si}, the structure of such
DM halos is expected to be affected by strong DM self-interactions, thus providing potential DM
observables.
We then move on to discuss the particle yield from DM annihilation or decay 
(Section \ref{sec:yields}) that are needed for the indirect detection of DM signals using gamma rays and 
neutrinos from the Galactic halo (Section \ref{sec:gamma}), charged cosmic rays (Section \ref{sec:cr}) or
neutrinos from the interior of the Sun or Earth (Section \ref{sec:se}). 
We summarize, and provide our 
conclusions in Section \ref{sec:conc}. In four Appendices, we describe the specific particle physics models
that are currently implemented, namely the generic WIMP case (Appendix \ref{app:genwimp}), a 
generic decaying DM scenario (Appendix \ref{app:gendecay}), the MSSM (Appendix \ref{app:mssm}), the 
scalar singlet model (Appendix \ref{app:singlet}) and a module for self-interacting DM confined to a 
dark sector (Appendix \ref{app:vdsidm}). Finally, in Appendix \ref{app:technical}, we 
provide a quick-start guide and, for users already familiar with \ds\ 5 or 
earlier, list the main technical  differences of this release compared to earlier code versions.

The version of the package described in this paper is \dst\ \dstver.
To download the latest version of \dst, and for a more technical
manual, please visit the official \dst\ website at \url{
www.darksusy.org}. This webpage also contains an updated list of contributed 
particle modules and other extensions of general interest that have been externally developed by 
users of the code.

%%%%%%%%%%%%%%%%%%%%%%%%%%%%%%%%%%%%%%%%%%%%%%%%%%%%%%%%%%%%%%%%%%%%%%
\section{Physics highlights}
\label{sec:phys}

Let us start with a quick reading guide, mostly for users already familiar with earlier code versions, 
where we list some of the most important new or 
considerably improved \emph{physics} features since the last \ds\ publication \cite{ds4}, 
which described version 4.0 of the code
(see also Appendix \ref{app:changes} for a list of the most important \emph{technical} 
changes).

\begin{itemize}

\item As described in more detail in Section \ref{sec:Philo}, \ds\ can now directly be used 
with particle models different from supersymmetry. An example of a fully implemented,
UV-complete new model is Scalar Singlet DM (Appendix \ref{app:singlet}). Another
example is self-interacting DM in a secluded dark sector (Appendix \ref{app:vdsidm}). 

\item The \emph{relic density routines} have been made more general (allowing for
different particle models, including situations where the dark sector and photon temperatures differ), while 
retaining their ability to treat co-annihilations, thresholds and resonances with high 
numerical accuracy; partial parallelisation of the code has led to a %considerable 
speed-up 
that allows to compute the relic density even for `critical' models (Section \ref{sec:RelDens}).

\item A newly added feature is the calculation of the \emph{kinetic decoupling} decoupling
of WIMPs, along with the size of the smallest protohalos (Section \ref{sec:minihalo}).

\item The routines for direct detection have been significantly revised, and generalised 
to include both spin-dependent and spin-independent scattering, in the corresponding limit, 
as well as descriptions using nonrelativistic effective operators
(Section \ref{sec:direct}).

\item Another completely newly added feature is the possibility to calculate DM self-interaction
rates  (Section \ref{sec:si}).

\item Extensive runs of {\sf Pythia} \cite{Sjostrand:2006za} have been tabulated to provide the \emph{yields
of stable particles} relevant for DM indirect detection, improving both the statistics of
previous tables and extending them to new channels (e.g.~light quarks) and final states
(anti-deuterons)
 (Section \ref{sec:yields}).

\item The routines for the numerical \emph{propagation of cosmic rays} have been completely 
re-written; they are now both much more flexible and stable than in previous versions  (Section \ref{sec:cr}). 
This allows in particular to efficiently scan over propagation parameters, as well as to treat cuspy
DM density profiles.

\item New capture rate calculations for WIMP capture in the Sun. We now include much more detailed models of the solar composition from Ref.~\cite{Serenelli:2009yc}. We include up to 289 different isotopes for spin-independent capture and 112 elements for spin-dependent capture. We also perform the form factor integration numerically, which allows us to use better form factors than the usual Gaussian ones.

\item Astrophysics and particle physics have now been completely disentangled, allowing to link the master
library to \emph{any particle physics} module. This release, in particular, ships with complete modules
for a generic WIMP  (App.~\ref{app:genwimp}), a generic decaying DM scenario (App.~\ref{app:gendecay}), 
the MSSM (App.~\ref{app:mssm}), the scalar singlet model (App.~\ref{app:singlet}) and  a range of simplified
DM models with velocity-dependent self-interactions (App.~\ref{app:vdsidm}).

\item For the \code{mssm} module, the \emph{internal bremsstrahlung}  of
both $U(1)$, $SU(2)$  and $SU(3)$ gauge bosons has been fully implemented, 
resulting in highly accurate predictions of all cosmic-ray spectra relevant for 
indirect DM detection (Section \ref{sec:chi_radcorr}).

\item For the relic density calculations in the \code{mssm} module, a new effective 
framework to take into account the
effect of \emph{QCD corrections} to neutralino-neutralino annihilation has been 
implemented; an interface to  {\sf DM@NLO}  \cite{dmnlo} for the full NLO result is in 
preparation (Section \ref{sec:chi_loop}).

\item The \code{mssm} module has an updated {\it SLHA reader}, which is more versatile  
and flexible than in previous versions of the code. For example, \ds\ can now handle more supersymmetric models.

\item The core library contains a new set of functionalities to compute limits 
\cite{Bringmann:2011ut,Clark:2015sha} on DM
models from the formation of {\it ultracompact minihalos} 
\cite{Berezinsky:2003vn,Ricotti:2009bs,Scott:2009tu}.\footnote{
We are very grateful to Pat Scott for providing this part of the code.
} 
\end{itemize}

\noindent
Apart from these additions included in the released version \dsver\ of the code, there currently
exist several publicly available packages to be used with \ds. The full list will be constantly 
updated
at \url{www.darksusy.org}, but let us point out in particular dedicated code to
compute the Sommerfeld effect in the MSSM \cite{Hryczuk:2011tq} as well as detailed likelihoods for 
neutrino signals from the IceCube experiment \cite{IC22Methods,IC79_SUSY}.

After this overview of the basic structure and guiding principles of the code, we will next turn to the physics that is 
implemented in \ds\ \dsver. In the remaining Sections of the main text of this article, 
we will describe in some detail those parts of \ds\ that are independent of any 
\emph{specific} particle-physics framework beyond the standard model (BSM), as 
implemented in the main library \code{ds\_core}, and refer to 
the Appendix for any functionality contained in the particle modules. Throughout,
most of our focus will be on new aspects compared to earlier code versions. 
Before we start, and as a quick reading guide to users familiar with earlier code versions, 
let us briefly list some of the most important new or 
considerably improved \emph{physics} features since the last \ds\ publication \cite{ds4}, 
which described version 4.0 of the code
(see also Appendix \ref{app:changes} for a list of the most important \emph{technical} 
changes).

%%%%%%%%%%%%%%%%%%%%%%%%%%%%%%%%%%%%%%%%%%%%%%%%%%%%%%%%%%%%%%%%%%%%%%
\section{Guiding principles}
\label{sec:Philo}

\dst\ \dsver\ has a new structure compared to earlier versions of the code. The most 
striking difference is that we have split the particle physics model dependent parts from 
the rest of the code. This means in practice that we have separated \ds\ into one set of 
routines, \code{ds\_core}, which contains no reference to any specific particle model, as 
well as distinct sets of routines for each implemented model of particle physics. For 
supersymmetry, 
 for example, all routines that require model-dependent information now reside in the 
 \code{mssm} module. The advantage with this setup is that \code{ds\_core} and the 
 particle physics modules may be put in separate FORTRAN libraries, which implies
 that \ds\ can now have several particle physics modules side by side. The user 
 then simply decides at the linking stage, i.e.~when making the main program, which 
 particle physics module to include. 
 
 For this to work, the main library needs each particle physics module to supply
 so-called \emph{interface} functions (or subroutines), with 
 pre-defined signatures and functionalities.  Note that a 
 particle physics module does not have to provide all of these predefined functions: 
 which of them are required is ultimately determined only when the user links the main
 program to these libraries. Assume for example that the main program wants to calculate 
 the gamma-ray flux from DM (a functionality provided by \code{ds\_core}). This is only
 possible if the particle module provides an interface function for the DM contribution to the
 local cosmic ray source function; if it does not, the main program will not compile and a  
 warning is issued that points to the missing interface function. If, on the other hand,
 interface functions required by direct detection routines would be missing in this example,
 this would not create any problems at either runtime or the compile stage.

%%%%%%%%%%%%%%%%%%%%%%%%%%
\begin{figure}[t!]
\centering
\includegraphics[width=\textwidth]{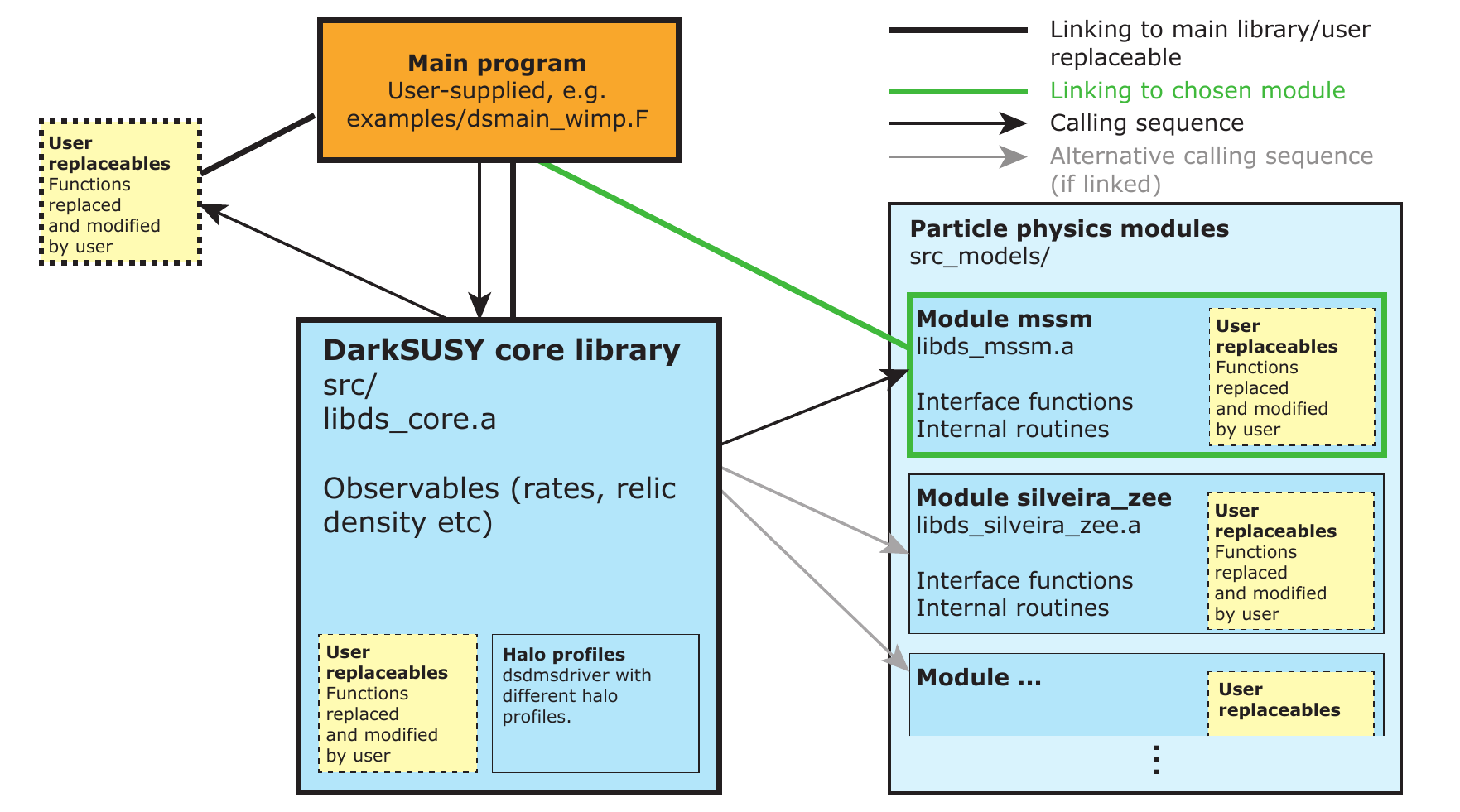}
\vspace{-0.5cm}
\caption{ 
Conceptual illustration of how to use \ds\ \dsver. The main program links
to both  the main library, \code{ds\_core}, and to \emph{one} of the available particle 
physics modules.  
User-replaceable functions are optional and may be linked to directly 
from the main program, or indirectly by including them in the various libraries. 
See \code{examples/dsmain\_wimp.F} for an example of a main program that demonstrates
typical usage of \ds\ for different particle physics modules.}
\label{fig:concept}
\end{figure}
%%%%%%%%%%%%%%%%%%%%%%%%%%		

 In addition, we have added the concept of 
 \emph{replaceable functions}, which allows users to replace essentially any function in 
 \ds\ with a user-supplied version. \ds\ ships with dedicated tools to assist you setting up 
 both replaceable functions and new particle physics modules 
 (see also Appendix \ref{app:technical}).
 Fig.~\ref{fig:concept} illustrates these concepts by showing how a typical program 
 would use \ds; below, we describe each of them in more detail. 

%%%%%%%%%%%%%%%%%%%%%%%%%%
\begin{table}[t!]
%\centering
\begin{tabular}[l]{l|l}
subdirectory name & description\\
\hline
\code{src/ini} & Initialization routines\\
\code{src/aux...} & General utility functions of the core library\\
    & (incl.~contributed code like \code{quadpack} \cite{1983qspa.book.....P}, 
    \code{diag} \cite{Hahn:2006hr} and  \code{CERNLIB} \cite{CERNLIB}) \\[1ex]

\code{src/dmd\_aux} & general, auxiliary functions needed by halo routines\\
\code{src/dmd\_mod} &  main DM halo driver routine and DM profile definitions\\
\code{src/dmd\_astro} & Astrophysical source functions\\
\code{src/dmd\_vel} &  DM velocity distributions \\[1ex]

\code{src/si} &  DM self-interactions \\[1ex]

\code{src/an\_yield} & Simulated yield tables, e.g.~from $\bar bb$ final states\\[1ex]

\code{src/cr\_aux} & general, auxiliary functions needed by cosmic ray routines\\
\code{src/cr\_axi} & (Anti-)nucleon propagation routines for a generic 
                               \\&axially symmetric diffusive halo\\
\code{src/cr\_gamma} & Gamma-ray routines\\
\code{src/cr\_nu} & Routines for neutrinos from the halo\\
\code{src/cr\_ps} & Positron propagation routines\\[1ex]

\code{src/se\_aux} & Auxiliary routines needed by capture rate routines\\
\code{src/se\_yield}& Simulated neutrino yield tables from the center of the Sun or Earth\\
%\code{src/se\_ic}& Neutrino flux likelihood calculations, using IceCube data\\
\code{src/se\_mod} & Density and composition models for the Sun and the
Earth\\
\code{src/se\_nu} & Capture rate for dark matter in the Sun and in the Earth,
\\& as well as neutrino fluxes from the center of Sun and Earth\\[1ex]

\code{src/dd} & Direct detection routines\\[1ex]

\code{src/kd} & Kinetic decoupling\\
\code{src/rd} & Relic density%\\[1ex]
\end{tabular}
\caption{Organization of the main library  \code{ds\_core}: all functions and subroutines 
reside in the \code{src/} folder of the \ds\ installation, with the names of the subdirectories 
indicating the subject area.}
\label{tab:dsmain}
\end{table}
%%%%%%%%%%%%%%%%%%%%%%%%%%		

\subsection{The \ds\ core library}
As introduced above, the library \code{ds\_core} is in some sense the heart of the 
new \ds\ \dsver\,,
offering all the functionality that a user typically would be interested in without having 
explicitly to refer to specific characteristics of a given particle physics model (after
initialization of such a model).
 The main library thus contains routines for, e.g., 
cosmic ray propagation, solar models, capture rates for the Sun/Earth, a 
Boltzmann solver for the relic density calculation, yield tables from annihilation/decay etc. 
None of the routines in the main library contains any information about the particle 
physics module. Instead any information needed is obtained by calling an interface 
function that resides in the  particle physics module which is linked to (see below). 

The source code for all functions and subroutines in the main library can be found in the 
\code{src/} directory of the \ds\ installation folder, with subdirectory names indicating  
subject areas as summarized in Table \ref{tab:dsmain}.

\subsection{Particle physics modules}
The particle physics modules are collected in \code{src\_models/} and contain the parts of the 
code that depend on the respective 
particle physics model. Examples are cross section calculations, yield calculations etc. 
The routines in the particle physics module have access to all routines in 
\code{ds\_core}, whereas the reverse is in general not true (with the exception of a very 
limited set of interface functions that each particle module provides). 

\ds\ 5 and earlier was primarily used for supersymmetric, and specifically neutralino DM, 
and those 
parts of the code now reside in the MSSM module \code{mssm}. However, many people 
used \ds\ even before for e.g.\ a generic WIMP setup, which was doable for parts of the 
code, but not all of it. We now provide a generic WIMP module \code{generic\_wimp} that 
can be used for these kinds of calculations in a much more general way. In a similar spirit,
we also provide a module \code{generic\_decayingDM} for phenomenological studies
of decaying DM scenarios. As an example for an actual particle physics model
other than supersymmetry,
\ds\ \dsver\ furthermore includes a module \code{silveira\_zee} which implements the
DM candidate originally proposed by Silveira and Zee \cite{Silveira:1985rk} and which 
now is often
referred to as Scalar Singlet DM \cite{McDonald:1993ex,Burgess:2000yq,Cline:2013gha}. 
We furthermore provide a set of simplified renormalizable models with light mediators that are fully confined 
to a dark sector, in the \code{vdSIDM} module; the characteristic feature of these models are 
Sommerfeld-enhanced DM annihilation rates, strong self-interactions and late kinetic decoupling.
We also include an empty model, \code{empty}, 
which is of course not doing any real calculations, but contains (empty versions of) all
interface functions that the core library is aware of -- which is very useful for debugging and
testing purposes. Designing new particle modules is 
a straight-forward exercise, see Appendix \ref{sec:newmodel}, and it is generally a good idea 
to start with the most similar module that is already available.

The source code for the particle modules can be found in the directory
\code{src\_models/}  of the \ds\ installation folder, each of the subdirectories 
(e.g.~\code{src\_models/mssm/}, \code{src\_models/generic\_wimp/}) typically 
reflecting a (sub)subdirectory structure analogous to what is shown in Table \ref{tab:dsmain}
for the core library. A given model, defined by its model parameters, is then in general initialized 
with a call to a routine like \code{dsgivemodel\_decayingDM} for decaying DM, 
or \code{dsgivemodel\_25} for a pMSSM model with 25 parameters (see the Appendix for more 
examples), followed by a call to \code{dsmodelsetup}. 
Many dark matter models, furthermore, constitute only relatively simple extensions to the 
standard model, inheriting most of its structure. For convenience,
we therefore also provide various auxiliary routines, in \code{src\_models/common/sm}, that each particle module 
automatically has access to, and which return basic standard model quantities like, e.g., the masses
of standard model particles and their running (thus, additional BSM effects have to be implemented in the
respective particle module).

\subsection{Interface functions}
Interface functions (or subroutines) are the limited set of functions that are provided by 
the particle physics module and which routines in \code{ds\_core} might need to call; if they 
are not provided, a main 
program that makes use of those routines in  \code{ds\_core} will not compile. 
For this inter-library communication to work seamlessly, the behaviour of each interface 
function must be uniquely defined by its usage in the core library, and is explicitly specified in 
the function header. Furthermore, interface functions must have a well defined signature; the number and 
types of all their arguments, in other words is fixed.

Examples of interface functions include \code{dsddsigma} that 
returns the equivalent DM nucleus cross section (relevant for direct detection), \code{dscrsource} that 
returns the source term for DM-induced cosmic rays (relevant for indirect detection), and
\code{dsanxw} that returns the invariant annihilation rate (in the case of WIMPs).
All interface functions  contain the keyword `interface' in the function or subroutine header.
A complete list of all interface functions known to the core library, about a dozen for this version
\dsver\ of the code, can be found by looking in the  \code{empty} module.

\subsection{Replaceable functions}
\label{sec:replace}
The concept of replaceable functions introduces the possibility to replace a \ds\ routine with one 
of your own. The way it works is that the user-provided function will be linked 
when you make your main program instead of the \ds\ default one. If, for example, you want to 
replace the yields from a typical final state of DM annihilation or decay (like $\bar bb$)
with a new function of your own -- e.g.~because you are interested in comparing the tabulated
{\sf Pythia} \cite{Sjostrand:2006za} yields with those provided by PPPC \cite{Cirelli:2010xx} --  
you create your own version of the routine \code{dsanyield\_sim} and let \ds\ use this one 
instead.  Note that both routines in \code{ds\_core} and in any of the particle physics modules 
can be  replaced in this way,  including interface functions.\footnote{%
In fact, the particle modules themselves can use this concept to replace pre-defined routines in \code{ds\_core}.
We have set up all makefiles for the example programs such that, whenever a routines with identical
names exists in both \code{src} and \code{src\_models}, the latter version is chosen by the linker.
}

To help you with this setup, 
we provide tools that can create (or delete) a replaceable function from any \ds\ function, 
and update the makefiles to use this user-supplied function instead. We also provide a
simple way of managing large `libraries' of user-supplied functions, via a list imported 
by the makefiles, where the user can determine on the fly which user-replaceable 
functions should be included and which ones should not.

\subsection{Halo models}

Several routines in the core library need information about which DM halo should be adopted for the 
calculations. With this \ds\ version, we introduce a new and flexible scheme that avoids 
pre-defined hardcoded functions to describe the DM density profiles, and allows to 
consistently define different DM targets at the same time. For convenience, we still provide
several pre-defined options for such halo parameterizations, and the user can choose 
between, e.g., the Einasto \cite{1965TrAlm...5...87E} and the Navarrow-Frenk-White profile 
\cite{Navarro:1995iw}, or read in any tabulated axi-symmetric (or spherically symmetric) profile. 
On a {\it technical} level, the halo models are handled by the \code{dsdmsdriver} routine which 
contains a database of which  halo profiles the user has set up, and consistently passes this
information to all parts in the code where it is needed. For more details, see Section \ref{sec:halo}.

%%%%%%%%%%%%%%%%%%%%%%%%%%%%%%%%%%%%%%%%%%%%%%%%%%%%%%%%%%%%%%%%%%%%%%
\section{Thermal production and decoupling}
\label{sec:thermal}

The evolution of the DM phase-space density $f(\mathbf{p})$ 
is described by the Boltzmann equation, which in an expanding 
Friedmann-Robertson-Walker universe reads \cite{Kolb:1990vq,Bringmann:2006mu}
\be
  \label{diff_boltzmann}
  E\left(\partial_t-H\mathbf{p}\cdot\nabla_\mathbf{p}\right)f=C[f]\,.
\ee
Here, $H=\dot a/a$ is the Hubble parameter, and all interactions between DM and
standard model particles are contained in the collision term $C[f]$. For WIMPs, for 
example, the right-hand side is sufficiently large at high 
temperatures that these DM candidates have been in full  equilibrium with the thermal
bath in the  very early universe.
In general, one can distinguish two types of interactions: \emph{i)} processes that 
change the number of the involved DM particles, or other new particles close in mass, 
and  \emph{ii)} proper scattering events that are particle number-conserving.
\ds\ provides advanced routines to handle both, and to compute the corresponding 
observables with high precision; the DM relic density for the first class of processes 
(Section \ref{sec:RelDens}) and the mass of the smallest protohalos for the second 
(Section \ref{sec:minihalo}).

%%%%%%%%%%%%%%%%%%%%%%%%%%%%%%%%%%%%%%%%%%%%%%%%%%%%%%%%%%%%%%%%%%%%
\subsection{Chemical freeze-out and relic density}
\label{sec:RelDens}

To compute the DM relic density, one is typically not only interested in the interactions
of the DM particle itself, but also in the possibility that other new particles may be 
thermally produced as well. 
To take into account the 
effect of such {\it coannihilations}, one needs to consider an equation like 
(\ref{diff_boltzmann}) for each particle that is involved, and the collision terms will  in 
general also involve interactions between those new particles. In most scenarios the 
stability of DM is guaranteed by a discrete parity, like $R$-parity in supersymmetry, 
implying that all other particles will eventually decay into the DM particle.
In \ds\ we therefore currently only keep track of  the total number density of  
particles $n\equiv \sum_{i=1}^N n_{i}$, where $n_1$ denotes the number density of the lightest (i.e.~DM) particle. The evolution equation for this 
quantity is obtained by summing over all individual Boltzmann equations 
of the form (\ref{diff_boltzmann}), and then integrating over the DM momenta  \cite{Edsjo:1997bg}:
\begin{equation} \label{eq:Boltzmann2}
   \frac{dn}{dt} =
   -3Hn - \langle \sigma_{\rm{eff}} v \rangle
   \left( n^2 - n_{\rm{eq}}^2 \right)\,,
\end{equation}
where $n_\mathrm{eq}$ denotes the total number density in thermal equilibrium 
with the heat bath and $\langle...\rangle$ denotes a velocity average. 
Note that this assumes  kinetic decoupling (see below) to happen well after chemical
decoupling. 
Motivated by the 
WIMP case, \ds\ provides this velocity average %at the moment only 
under the assumption 
that the DM particles follow a Maxwell-Boltzmann distribution. In that case, the thermal
average of the effective cross section weighted by the relative velocity of the incoming 
particles is given by
\begin{equation} \label{eq:sigmaveffdef}
   \langle \sigma_{\rm{eff}} v \rangle \equiv \sum_{ij} \langle
   \sigma_{ij}v_{ij} \rangle \frac{n_{i}^{\rm{eq}}}{n^{\rm{eq}}}
   \frac{n_{j}^{\rm{eq}}}{n^{\rm{eq}}}
   =\frac{\int_0^\infty
   dp_{\rm{eff}} p_{\rm{eff}}^2 W_{\rm{eff}} K_1 \left(
   \frac{\sqrt{s}}{T_{\tilde\gamma}} \right) } { m_1^4 T_{\tilde\gamma} \left[ \sum_i \frac{g_i}{g_1}
   \frac{m_i^2}{m_1^2} K_2 \left(\frac{m_i}{T_{\tilde\gamma}}\right) \right]^2},
\end{equation}
where $K_1$ ($K_2$) is the modified Bessel function of the second kind of order 1 (2), 
$T_{\tilde\gamma}$ is the heat bath temperature (which can, but does not have to coincide with 
the  photon temperature $T$), and $s$ is one of the Mandelstam variables. The effective
invariant rate introduced above is defined as 
\begin{equation} \label{eq:weff}
   W_{\rm{eff}} \equiv  \sum_{ij}\frac{p_{ij}}{p_\mathrm{eff}}
   \frac{g_ig_j}{g_1^2} W_{ij}\,,
\end{equation}
where the $g_i$ denote internal degrees of freedom and the relative momenta are given by
\begin{equation}
    p_{ij} =
   \frac{\left[s-(m_i+m_j)^2\right]^{1/2}
   \left[s-(m_i-m_j)^2\right]^{1/2}}{2\sqrt{s}},
\end{equation}
with $p_\mathrm{eff}\equiv p_{11}$. The relation between the invariant rate $W_{ij}$ and the more 
familiar cross section $\sigma_{ij}$ for the annihilation of two particles $i$ and $j$ is\footnote{
Further useful relations between $W_{ij}$ and $\sigma_{ij}$ include
%\begin{equation} \label{eq:Wijcross2}
   $W_{ij} = 4 p_{ij} \sqrt{s} \sigma_{ij} = 4 E_{i} E_{j} \sigma_{ij} v_{\text{\rm M{\o}l},ij}$,
%\end{equation}
where $v_{\text{\rm M{\o}l},ij}=[|{\bf v}_i-{\bf v}_j|^2-|{\bf v}_i\times{\bf v}_j|^2]^{1/2}$ is the M{\o}ller 
velocity. For the special case of self-annihilation or particle-antiparticle annihilation of particles 
of mass $m$, this becomes
\be
\label{Weffsimp}
W = W_{\rm eff}=2 (s-2m^2) \sigma v\,, 
\ee
where
$ v = {\sqrt{s(s-4m^2)}}/{(s-2m^2)} = {2v_{\rm cm}}/{(1+v_{\rm cm}^2)}$
and
$v_{\rm cm} = {(1 - \frac{4m^2}{s})^{1/2} }$.
} 
\begin{equation} \label{eq:Wijcross}
   W_{ij} = 4 \, p_i \cdot p_j \, \sigma_{ij} \, v_{ij} =  \frac{4m_i m_j v_{ij}}{(1-v_{ij}^2)^{1/2}} \sigma_{ij},
\end{equation}
where $v_{ij}$ is the invariant relative velocity, i.e., the magnitude of the velocity of one particle in the rest frame of the other,
\begin{align}
v_{ij} = \frac{\left[s-(m_i+m_j)^2\right]^{1/2}
   \left[s-(m_i-m_j)^2\right]^{1/2}}{s-m_i^2-m_j^2} 
   =\frac{p_{ij} \sqrt{s}}{p_i \cdot p_j}.
\end{align}
Note that the sum in Eq.~(\ref{eq:sigmaveffdef}) only involves processes that
change the total number $n$ of non-SM particles. For more details on the standard
way of calculating the relic density, as implemented in \ds, see 
Refs.~\cite{Gondolo:1990dk,Edsjo:1997bg}. 

For complex particle models involving many coannihilations, like the MSSM, the 
calculation of the invariant rate $W_{\rm eff}$ can be rather time-consuming. Given
that it is independent of temperature, \ds\ therefore tabulates it once and
for all for every set of model parameters (with a particularly dense sampling of potentially 
critical regions like thresholds and resonances).
The thermal average is then performed by means of an adaptive gaussian integration, 
using a spline routine for interpolation between the tabulated points.
In order to integrate the density evolution equation (\ref{eq:Boltzmann2}), we follow
Ref.~\cite{Edsjo:1997bg} and rephrase it as an evolution equation for the abundance, $Y=n/s$ 
(with $s$ being the entropy density); instead of time, we use the dimensionless quantity 
$x=m_\chi/T$ as our integration variable. The numerical integration is still subtle, 
because the differential equation is stiff, and we developed a dedicated  implicit 
trapezoidal method with adaptive stepsize to guarantee a high precision result. For more 
details on the numerical implementation of these routines, see Ref.~\cite{ds4}. 
As a default for the energy and entropy degrees of freedom in the early universe,
finally, we implemented the results by Drees et al \cite{Drees:2015exa}.

Despite those similarities at the detailed implementation level, the relic 
density routines are now considerable more general than in previous versions of the 
code. This allows to link them to a given particle physics model in a simple and 
straightforward way. The effective invariant rate $W_{\rm eff}$ in Eq.~(\ref{eq:weff}), 
in particular, must be supplied by the particle module 
in the form of the interface function 
\code{dsanwx}. In order to calculate the thermal average in Eq.~(\ref{eq:sigmaveffdef}),
the main library also needs to know the masses and internal degrees of freedom of 
the coannihilating particles
and, in order to obtain an as precise result as possible, the location of possible  
resonances or thresholds in $W_\mathrm{eff}(p_\mathrm{eff})$. This information is 
provided by the only other interface function
in the relic density part, \code{dsrdparticles}.
Another new feature since \ds\ 6.1 is that the relic density routines now can handle heat bath 
temperatures different from the photon temperature, parameterized in terms of the ratio
\be
\label{eq:xidef}
\xi(T)\equiv T_{\tilde\gamma}/T\,.
\ee
This ratio is returned by the replaceable function \code{dsrdxi} which simply returns a constant
factor of unity in its default version -- as supplied by the \code{core} library -- but can take
any user-supplied form more appropriate for the particle model in question.\footnote{%
\label{foot:xi}%
For example, if DM is confined to a dark sector that has been in thermal contact with the 
visible sector until some decoupling temperature $T_\mathrm{dc}$, entropy is afterwards typically
conserved separately in the two sectors. This implies that the temperature ratio changes
with the ratios of the degrees of freedom (see, e.g., Ref.~\cite{Feng:2008mu, Bringmann:2013vra}):
\be
 \xi(T) = 
 \frac{\left[{g_*^\mathrm{SM}}(T)/{g_*^\mathrm{DS}(T)}\right]^\frac13}
 {\left[{g_*^\mathrm{SM}}(T_\mathrm{dc})/{g_*^\mathrm{DS}(T_\mathrm{dc})}\right]^\frac13}\,,
 \ee
where $g_*^{\mathrm{SM}, \mathrm{DS}}$ are the entropy degrees of freedom in the visible
and dark sector, respectively.
In Appendix \ref{app:vdsidm}, we will present a module already implemented in \ds\ where 
we make use of this relation.
}

The DM relic density (of both components, in case the DM particle is not its own anti-particle) 
in terms of $\Omega h^2$ is provided by a call to the function 
\code{dsrdomega},  which manages the calls to  the \ds\ relic density routines. 
In their full version, these routines return the relic density with a
precision beyond per-cent accuracy -- assuming that $W_\mathrm{eff}$ is provided with 
corresponding precision -- but there are also options for considerably faster 
calculations (for example by taking into account only coannihilations up to mass 
differences of $f_{\rm co}=1.4$). This can be very useful in situations where an exceptional 
error of a few per cent is an acceptable trade-off in view of significantly improved 
performance with respect to computation time (e.g.~in large global scans). For 
comparison, the DM density today is now observationally determined 
%as $\Omega_\chi h^2 = 0.1198 \pm 0.0015$, 
with a precision of about 1\% \cite{Ade:2015xua}.

%%%%%%%%%%%%%%%%%%%%%%%%%%%%%%%%%%%%%%%%%%%%%%%%%%%%%%%%%%%%%%%%%%%%%%
\subsection{Kinetic decoupling and the smallest protohalos}
\label{sec:minihalo}

Even after the DM particles have chemically decoupled from the thermal bath, they are generally 
still in local equilibrium through frequent scattering processes (typically with SM particles). 
As long as this is the case, the DM 
temperature $T_\chi\equiv\frac{g_\chi}{3m_\chi n_\chi}\int\frac{d^3p}{(2\pi)^3}\mathbf{p}^2f(\mathbf{p})$
is the same as the temperature of the scattering partners. After kinetic
decoupling, the DM `temperature' will simply decrease due to the expansion of the universe and, 
for non-relativistic DM, scale like $T_\chi\propto a^{-2}$. It is thus natural
 to define the moment of decoupling as the transition between these two regimes \cite{Bringmann:2009vf}.
Allowing for the scattering partners again to have a different temperature ($T_{\tilde\gamma}$) than the photons 
($T$), this implies
\be
\label{TchiT}
\,T_\chi(T)=\left\{
\begin{array}{cc}
T_{\tilde\gamma}(T) & \mathrm{for}~T\gtrsim T_\mathrm{kd} \\
T_{\tilde\gamma} (T_\mathrm{kd})\left(\frac{a(T_\mathrm{kd})}{a(T)}\right)^2& \mathrm{for}~T\lesssim T_\mathrm{kd}
\end{array}
\right. 
\ee
The evolution of $T_\chi(T)$ is obtained by solving the second moment of the full Boltzmann equation
\ref{diff_boltzmann}. Introducing  $x\equiv m_\chi/T$ and $y\equiv{m_\chi T_\chi}{s^{-2/3}}$, where
$s$ is the total entropy density, this is to leading order in $\mathbf{p}^2/m_\chi^2$ given by 
\cite{Bringmann:2016ilk}\footnote{
We note that this expression assumes a  covariantly conserved DM number density. While typically
satisfied, this is not the case, e.g., in the presence of strongly Sommerfeld-enhanced DM annihilation 
or because kinetic decoupling happens very early. In such situations one needs instead to solve a 
coupled system of differential equations for $T_\chi$ and $n_\chi$ \cite{Aarssen:2012fx, Binder:2017rgn}. 
This will be available in a future version of \ds.
}
\be
\label{dydx}
\frac{d\log y}{d\log x}=\left(1-\frac{1}{3}\frac{d \log g_{*S}}{d\log x}\right)
\frac{\gamma(T_{\tilde\gamma})}{H(T)}\left(\frac{y_\mathrm{eq}}{y}-1\right)\,.
\ee
Here, $g_{*S}$ is the number of effective entropy degrees of freedom, $y_\mathrm{eq}$ is the 
value of $y$ in thermal equilibrium and $\gamma$ denotes the momentum transfer rate, 
\be
\label{fT}
 \gamma (T_{\tilde\gamma})=\frac{1}{48\pi^3g_\chi T_{\tilde\gamma}m_\chi^3}
 \sum_i
 \int d\omega\,k^4
 \left(1\mp g_i^\pm\right)g_i^\pm(\omega)
  \mathop{\hspace{-12ex}\left|\mathcal{M}\right|^2_{t=0}}_{\hspace{4ex}s=m_\chi^2+2m_\chi\omega+m_\mathrm{\tilde\gamma}^2}\,,
\ee
where the sum runs over all DM scattering partners (counting, e.g., electrons and positrons separately), 
$k\equiv\left| \mathbf{k}\right|$ is their momentum and $\omega$ their energy. The scattering amplitude
squared in this expression, $|\mathcal{M}|^2$, is understood to be {\it summed} over all internal (spin or 
color) degrees of freedom, including initial ones. If it is not Taylor expandable around $t=0$, one has to 
make the replacement \cite{KasaharaPHD,Gondolo:2012vh} 
\be
\label{maverage}
 \mathop{\hspace{-12ex}\left|\mathcal{M}\right|^2_{t=0}}_{\hspace{4ex}s=m_\chi^2+2m_\chi\omega+m_\mathrm{\tilde\gamma}^2}
\!\!\! \longrightarrow
\left<\left|\mathcal{M}\right|^2\right>_t 
\!\!\equiv \frac{1}{8k^4}\int_{-4k^2_{CM}}^0
\!\!\!\! dt(-t)\left|\mathcal{M}\right|^2
\ee
in the above expression.

As long as DM is in local thermal equilibrium, the frequent scattering processes will wash out 
any small-scale perturbations in the DM fluid as soon as they enter the horizon. For temperatures
close to $T_\mathrm{kd}$, first a remaining viscous coupling and then the free-streaming of the 
DM particles generates an exponential cutoff in the power spectrum of density fluctuations 
\cite{Green:2005fa}, which translates to the smallest mass a protohalo can have once structure
formation enters the non-linear regime. Acoustic oscillations of the DM fluid also generate an effective 
exponential cutoff in the power spectrum \cite{Loeb:2005pm,Bertschinger:2006nq}. 
Which of the two mechanisms dominates, 
i.e.~leads to a larger cutoff mass $M_\mathrm{cut}$, depends on the combination of decoupling
temperature $T_\mathrm{kd}$ and DM mass $m_\chi$ \cite{Bringmann:2009vf}.

In \ds, the kinetic decoupling temperature is obtained with a call to the routine \code{dskdtkd}, which
solves the Boltzmann equation and determines the asymptotic solution of $T_\chi/T$ to return
$T_\mathrm{kd}$ as defined in Eq.~(\ref{TchiT}). A call to \code{dskdmcut} then takes $T_\mathrm{kd}$
and the DM mass as input, and returns $M_\mathrm{cut}$ (as given in Ref.~\cite{Bringmann:2009vf}). 
These functions reside in the core library.
There are three interface functions that a particle physics module must provide if the main program 
uses this functionality: \code{dskdm2} returns the full scattering matrix element squared, evaluated 
at $t=0$ or averaged over $t$, while \code{dskdm2simp} returns only the leading contribution
for small $\omega$ (expressed as a simple power-law in $\omega$, in which case there exists an
analytic solution for $T_\mathrm{kd}$ \cite{Bringmann:2006mu}). Lastly, the particle physics
module must provide a routine \code{dskdparticles} which, in analogy to the routine  \code{dsrdparticles}
for the case of chemical decoupling, sets the location of resonances in $\left|\mathcal{M}\right|^2$. This allows the corresponding routine
in the core library to perform the energy integration in Eq.~(\ref{fT}) to a much higher 
precision. Non-standard heat bath temperatures $T_{\tilde\gamma}\neq T$ are automatically handled by supplying a user-defined 
function \code{dsrdxi}, in the same way as described in the previous Section.

%%%%%%%%%%%%%%%%%%%%%%%%%%%%%%
\section{Direct detection}
\label{sec:direct}

The rate for nuclear recoil events in direct detection experiments (often expressed in {\it dru}, or differential rate unit,
i.e.~counts/kg/day/keV) can be written as
\begin{equation}
\frac{dR}{dE_R} = \sum_T c_T \frac{\rho_\chi^0}{m_T  m_\chi} \int_{v>v_{\rm min}} \frac{d\sigma_{\chi T}}{dE_R}  \frac{f({\bf v},t)}{v} d^3 v\,.
\label{eq:dRdE} 
\end{equation}
The sum here runs over nuclear species in the detector, $c_T$ being the detector mass fraction in nuclear species 
$i$. $m_T$ is the nuclear (target) mass, and $\mu_{\chi T} = m_\chi m_T /(m_\chi + m_T)$ is the reduced DM--nucleus 
mass. 
Furthermore, $\rho_\chi^0$ is the local DM density, ${\bf v}$ the DM velocity relative to the detector, $v=|{\bf v}|$, and 
$f({\bf v},t)$ is the (3D) DM velocity distribution.  In order to impart a recoil energy $E_R$ to the nucleus, the DM
particle needs a minimal speed of $v_{\rm min}=\sqrt{M_TE_R/2\mu^2_{\chi T}}$. 

Finally, ${d\sigma_{\chi T}}/{dE_R}$ describes the unpolarized differential scattering cross section for DM 
scattering off a target nucleus of mass $m_T$, differential with respect to the nucleus recoil energy $E_R$ in the 
initial rest frame of the target nucleus. This is of the form
\begin{align}
\frac{d\sigma_{\chi T}}{dE_R} = \frac{m_T}{2\pi  v^2} \, \overline{ {\cal S} }.
%\frac{d\sigma_{\chi T}}{dE_R} = \frac{m_T}{2\pi \hbar^4 v^2} \, \overline{ {\cal S} }.
%\qquad
%\frac{{\rm GeV}}{c^2 \hbar^4 c^2}  \left(\frac{\hbar^3 c^3}{{\rm GeV}^2} \right)^2=
%\frac{\hbar^2c^2}{{\rm GeV}^3} = \frac{1}{\rm GeV} \left( \frac{\hbar c}{\rm GeV} \right)^2
\label{eq:dsigmadER}
\end{align}
Here $v$ is the  WIMP-nucleus relative velocity, and $\overline{ {\cal S} }$ is a Lorentz-invariant 
scattering factor (the overline indicates a sum over final polarizations and an average over initial 
polarizations). If the WIMP and the nucleus are treated as elementary particles, the scattering factor is 
given by ${\cal S} = \left| {\cal M}/(4k\cdot p) \right|^2$, where $k$ and $p$ are the initial four-momenta of 
the WIMP and nucleus and ${\cal M}$ is the invariant amplitude in the normalization of the Review of 
Particle Physics \cite{Olive:2016xmw}. For a nucleus of finite size, with nonrelativistic single-nucleon 
interactions governed by WIMP-nucleon contact operators ${\cal O}_i$, one can separate the scattering 
factor ${\cal S}$ into the contributions of single operators and their interference. One can write
\begin{align}
\overline{ {\cal S} } = \sum_{ij} \, \overline{ {\cal S} }_{ij} 
\label{eq:defScal1}
\end{align}
with
\begin{align}
\overline{ {\cal S} }_{ij} = \Re \, \sum_{\N\N'} \, G_i^{\N*} \, G_j^{\N'} \, P_{ij}^{\N\N'} .
\label{eq:defScal2}
\end{align}
The sum in Eq.~(\ref{eq:defScal1}) is over the operator indices $i$ and $j$. A ``square'' term 
${\cal S}_{ii}$ is the contribution from a single operator ${\cal O}_i$, and a ``cross'' term ${\cal S}_{ij}$ 
with $i\ne j$ is half of the contribution from the interference of operators ${\cal O}_i$ and ${\cal O}_j$ (the 
other half is the term ${\cal S}_{ji}={\cal S}_{ij}$). The sum in Eq.~(\ref{eq:defScal2}) is over proton and 
neutrons ($\N={\rm p},{\rm n}$), 
the $G_i^{\N}$'s are effective WIMP-nucleon coupling constants generalizing the Fermi coupling 
constant $G_F$, given in the same units ($\hbar^3c^3$GeV$^{-2}$), and the $P_{ij}^{\N\N'}$'s are 
dimensionless factors containing the nuclear structure functions for specific nucleon currents (see below 
for details).

In place of the differential cross section $d\sigma/dE_R$, \ds\ provides an {\it equivalent cross section} 
$\tilde \sigma_{\chi T}$, and partial cross sections $\tilde \sigma_{ij}$, defined by the relation
\begin{align}
\tilde \sigma_{\chi T} \equiv \frac{2\mu_{\chi T}^2v^2}{m_T} \, \frac{d\sigma_{\chi T}}{dE_R}  %,
%\end{align}
%or equivalently
%\begin{align}
%\sigma_{\chi T} 
%= \frac{\mu_{\chi T}^2}{\pi}  \, \, \overline{ {\cal S} }.
\equiv \sum_{ij}\tilde \sigma_{ij}\,.
\label{eq:sigmachiT}
\end{align}
Here $\mu_{\chi T} = m_\chi m_T/(m_\chi + m_T)$ is the  WIMP-nucleus reduced mass, and the 
prefactor is equal to the maximum recoil energy for elastic scattering,
\begin{align}
E_{\rm max} =  \frac{2\mu_{\chi T}^2v^2}{m_T} .
\end{align}
Notice that $\tilde\sigma_{\chi T}$ is in general not equal to the total WIMP-nucleus cross section 
$\sigma_{\chi T}$. The latter is 
obtained by integration of $d\sigma_{\chi T}/dE_R$ over the recoil energy $E_R$ and the dependence of 
the nuclear structure functions on $E_R$ may be important. In the limit of vanishing $E_R$ and $v$, however,
the two quantities coincide.
Moreover, the definition of $\tilde\sigma_{\chi T}$ 
applies to both elastic and inelastic scattering, and is merely a convention. 
%Similarly, we define partial cross sections
%\begin{align}
%\tilde \sigma_{ij} \equiv \frac{\mu_{\chi T}^2}{\pi} \, \overline{ {\cal S} }_{ij}
%\end{align}
%for the contributions of single operators ($\tilde\sigma_{ii}$) and their interference ($\tilde\sigma_{ij}$, $i\ne j$).

%
%There are various definitions of the coupling constants $G_a$ and their respective factors $P_{ab}$ in Eq.~(\ref{eq:defScal}). \ds\ distinguishes these definitions by means of the integer variable \code{ddscheme}. One  DD scheme has been implemented (\code{ddscheme=1}) and another is in progress (\code{ddscheme=2}). DD scheme 1 is the usual spin-independent/spin-dependent formalism, while DD scheme 2 is the effective operator approach of Fitzpatrick et al. \cite{Fitzpatrick:2012ix}. In addition, the value \code{ddscheme=0} is used to indicate that the WIMP-nucleus cross section (\ref{eq:sigmachiT}) is coded directly without the use of Eq.~(\ref{eq:defScal}).

In the usual spin-independent/spin-dependent treatment, the equivalent cross section $\tilde \sigma_{\chi T}$ 
contains only two terms, in this case without interference terms, the spin-independent term $\sigma_{\rm SI} $ and the spin-
dependent term $\sigma_{\rm SD}$,
\begin{align}
\tilde \sigma_{\chi T} = \tilde\sigma_{\rm SI} + \tilde\sigma_{\rm SD}
\mathop{\longrightarrow}_{E_R=0} \sigma_{\rm SI} + \sigma_{\rm SD} \, .
\end{align}
In the convention of \cite{Fitzpatrick:2012ix}, these two terms arise from the operators ${\cal O}_1$ 
and ${\cal O}_4$, thus we will use the indices 1 and 4 for 
$\tilde \sigma_{\rm SI}  = \tilde \sigma_{11} $ and 
$\tilde \sigma_{\rm SD}  = \tilde \sigma_{44} $. The coupling constants $G_1^{\rm p}$, 
$G_1^{\rm n}$, $G_4^{\rm p}$, $G_4^{\rm n}$ are hence equal to the previous \ds\ constants 
$G_s^{\rm p}$, $G_s^{\rm n}$, $G_a^{\rm p}$, $G_a^{\rm n}$, respectively. They are related to the 
coupling constants $f_{\rm p}$, $f_{\rm n}$, $a_{\rm p}$ and $a_{\rm n}$ in \cite{Savage:2008er} by the 
following equations.
\begin{align}
& G_{1}^{\rm p} = G_s^{\rm p} = 2 f_{\rm p},  && G_{1}^{\rm n} = G_s^{\rm n} = 2 f_{\rm n} ,
\label{eq:gpn1}\\
& G_{4}^{\rm p} = G_a^{\rm p} = 2 \sqrt{2} G_F a_{\rm p} , && G_{4}^{\rm n} = G_a^{\rm n} = 2 \sqrt{2} G_F a_{\rm n} . \label{eq:gpn2}
\end{align}
The quantities $P_{ij}^{NN'}$  in Eq.~(\ref{eq:defScal2}) corresponding to the constants $G_1^{\rm p}$, 
$G_1^{\rm n}$, $G_4^{\rm p}$, $G_4^{\rm n}$ are
\begin{align}
\label{eq:PabSI}
& P_{11}^{\rm pp} = Z^2 \, |F(q)|^2, 
\\
& P_{11}^{\rm pn} = P_{11}^{\rm np} = Z \, (A-Z) \, |F(q)|^2, 
\\
& P_{11}^{\rm nn} = (A-Z)^2 \, |F(q)|^2 ,
\\
& P_{44}^{\rm pp} = \frac{4s_\chi(s_\chi+1)}{3} \frac{4\pi}{2J+1} \, S_{\rm pp}(q), 
\\
& P_{44}^{\rm pn} = P_{44}^{\rm np} = \frac{4s_\chi(s_\chi+1)}{3} \frac{4\pi}{2J+1} \, \frac{S_{\rm pn}(q)}{2}, 
\\
& P_{44}^{\rm nn} = \frac{4s_\chi(s_\chi+1)}{3} \frac{4\pi}{2J+1} \, S_{\rm nn}(q).
\label{eq:PabSD}
\end{align}
Here $s_\chi$ is the spin of the WIMP, $J$ is the spin of the nucleus, $A$ and $Z$ are the mass number 
and atomic number of the nucleus, $q=\sqrt{2m_T E_R}$ is the magnitude of the momentum transfer, 
$F(q)$ is the spin-independent nuclear form factor (e.g., the Helm form factor) normalized to $F(0)=1$, 
and the $S_{\rm pp}(q)$, $S_{\rm pn}(q)$, and $S_{\rm nn}(q)$ are the nucleus spin structure functions 
as defined in \cite{Engel:1991wq}. The factor $4s_\chi(s_\chi+1)/3$ has been introduced in 
$P_{44}^{\N\N'}$ because the original spin structure functions $S_{\N\N'}(q)$ were computed for 
$s_\chi=\tfrac{1}{2}$.

In the nonrelativistic effective operator approach of \cite{Fitzpatrick:2012ix}, there are several operators 
${\cal O}_i$ with corresponding coupling constants $G_i^{\rm p}$ and $G_i^{\rm n}$. As physical 
quantities, the constants $G_i^{\N}$ are equal to the constants $c_i^{\N}$ in \cite{Anand:2014kea},
%\begin{align}
$G_i^{\N} = c_i^{\N}$ ,
%\end{align}
but the $c_i^{\N}$ in the code provided by \cite{Anand:2014kea} is in units of $\sqrt{2}G_F$ (there is a 
typo in their Eq.~(20)). So numerically
\begin{align}
G_i^{\N} = (c_i^{\N} \text{ in \cite{Anand:2014kea}'s code}) \times \sqrt{2} \, G_F .
\end{align}
The quantities $P_{ij}^{\N\N'}$ in Eq.~(\ref{eq:defScal2}) corresponding to the $G_i^N$'s just defined 
coincide with the quantities $F^{(N,N')}_{i,j}$ in Eq.~(77) of \cite{Fitzpatrick:2012ix} and are given in 
terms of the WIMP and nuclear response functions of \cite{Anand:2014kea} by 
\begin{align}
P_{ij}^{\N\N'} = \quad
& R^{[ij]}_{M}(q,v^{\perp}) \, F^{\N\N'}_{M}(q)
+ R^{[ij]}_{\Sigma'}(q,v^{\perp}) \, F^{\N\N'}_{\Sigma'}(q)
+ R^{[ij]}_{\Sigma''}(q,v^{\perp}) \, F^{\N\N'}_{\Sigma''}(q)
+ \nonumber \\
& \frac{q^2}{m_N^2} \bigg(
   R^{[ij]}_{\Delta}(q,v^{\perp}) \, F^{\N\N'}_{\Delta}(q)
+ R^{[ij]}_{{\tilde\Phi}'}(q,v^{\perp}) \, F^{\N\N'}_{{\tilde\Phi}'}(q)
+ R^{[ij]}_{\Phi''}(q,v^{\perp}) \, F^{\N\N'}_{\Phi''}(q)
+ \nonumber \\
& \phantom{\frac{q^2}{m_N^2} \bigg( \mkern160mu}
R^{[ij]}_{\Delta\Sigma'}(q,v^{\perp}) \, F^{\N\N'}_{\Delta\Sigma'}(q)
+ R^{[ij]}_{\Phi'' M}(q,v^{\perp}) \, F^{\N\N'}_{\Phi'' M}(q)
\bigg) .
\end{align}

Alternative formulation %(here the $R'$ may have an extra factor $q^2/m_N^2$ with respect to the $R$'s):
(where the factor $q^2/m_N^2$ in front of  the $R$s in the equation above is included in the $R'$s):
\begin{align}
\tilde\sigma_{\chi T} 
& = \frac{\mu_{\chi T}^2}{\pi}  \, \, \sum_{\alpha ijNN'} \Re \, G_i^{\N*} \, G_j^{\N'} \, R^{\prime[ij]}_{\alpha}(q,v^{\perp}) \, F^{\N\N'}_{\alpha}(q) 
\\ &
= \frac{\mu_{\chi T}^2}{\mu_{\chi N}^2}  \, \, \sum_{\alpha NN'} \sigma_{\alpha}^{NN'}(q,v^{\perp}) \, F^{\N\N'}_{\alpha}(q) 
\end{align}
with %($8\times3=24$)
\begin{align}
\sigma_{\alpha}^{NN'}(q,v^{\perp}) = \frac{\mu_{\chi N}^2}{\pi} \, \, \Re \, G_i^{\N*} \, G_j^{\N'} \, R^{\prime[ij]}_{\alpha}(q,v^{\perp}) .
\end{align}

Here the $F_{\alpha}^{\N\N'}(q)$ (with 
$\alpha=M,\Sigma',\Sigma'',\Delta,\tilde\Phi',\Phi'',\Delta\Sigma',\Phi''M$) are the nuclear response 
functions called $F^{(N,N')}_{X}$ and $F^{(N,N')}_{X,Y}$ in Eqs.~(73) and (74) of 
\cite{Fitzpatrick:2012ix}, which are related to the functions $W_{\alpha}^{\N\N'}$ in Eq.~(41) 
of \cite{Anand:2014kea} by
\begin{align}
F_{\alpha}^{\N\N'}(q)= \frac{4\pi}{2J+1} \, W_{\alpha}^{\N\N'}.
\end{align}
Moreover, the functions
$R^{[ij]}_{\alpha}(q,v^{\perp})$, with
\begin{align}
(v^\perp)^2 = v^2 - \frac{q^2}{4\mu_T^2},
\end{align}
are WIMP response functions related to the WIMP response functions $R^{\tau\tau'}_{\alpha}(v^{\perp2},q^2/m_N^2)$ in Eq.~(38) of \cite{Anand:2014kea} through the expressions
\begin{align}
R^{\tau\tau'}_{\alpha}\!\big(v^{\perp2},\frac{q^2}{m_N^2}\big) = \sum_{ij} c^{\tau*}_i c^{\tau'}_j R^{[ij]}_{\alpha}(q,v^{\perp}) .
\end{align}
%The WIMP response functions $R^{[ij]}_{\alpha}(q,v^{\perp})$ depend only on the WIMP properties provided they are written as functions of $q$ and $v^{\perp}$ (not as functions of $q$ and $v$). In other words, the value of $R^{[ij]}_{\alpha}(q,v^{\perp})$ depends on the target nucleus mass only through $v^{\perp}$. 
As examples of this formalism, for the spin-independent operator ${\cal O}_1$ one has
\begin{align}
P_{11}^{\N\N'} = F_M^{\N\N'}(q) ,
\end{align}
and for the spin-dependent operator ${\cal O}_4$  one has
\begin{align}
P_{44}^{\N\N'} = \frac{s_\chi(s_\chi+1)}{12} \left[ F_{\Sigma'}^{\N\N'}(q) + F_{\Sigma''}^{\N\N'}(q) \right] .
\end{align}
Comparison of these expressions with Eqs.~(\ref{eq:PabSI})--(\ref{eq:PabSD}) shows that both the usual SI/SD 
approach and the nonrelativistic effective operator approach of \cite{Fitzpatrick:2012ix} are included in the same 
formalism of Eq.~(\ref{eq:defScal2}) and that they differ only by the choice of nuclear structure functions appearing in 
$P_{ij}^{\N\N'}$ and the number of nonzero WIMP-nucleon coupling constants $G_i^\N$.

Specific choices of nuclear structure functions can be selected by calling \code{dsddset('sf',} \code{label)}, where the 
character variable \code{label} indicates the set of structure functions. For the default option ('\code{best}'),
e.g., the code automatically picks the best currently available structure function (depending on the nucleus). 
The function returning the value of $\tilde\sigma_{\chi T}$ is to be provided by an interface function 
\code{dsddsigma(v,Er,A,Z,sigij, ierr)} residing in the particle physics module, where on input \code{v=$v$}, 
\code{Er=$E_R$}, 
\code{A=$A$}, \code{Z=$Z$} and on output the 27$\times$27 array \code{sigij} contains the (partial) equivalent 
cross sections  $\tilde\sigma_{ij}$ in cm$^2$ and the integer \code{ierr} contains a possible error code. 
The order of the entries in \code{sigij} corresponds to that of the independent nonrelativistic operators 
$\mathcal{O}_i$; for the first 11 entries, we use the same operators and convention as in 
Ref.~\cite{Fitzpatrick:2012ix}, while for the last 16 entries  we add the additional operators discussed in 
Ref.~\cite{GondoloScopel}. In particular, \code{sigij(1,1)} is the 
usual spin-independent cross section and \code{sigij(4,4)} is the usual spin-dependent cross section.
In addition, the direct detection module in \ds\ provides utility functions that can be used in the computation of the 
cross section. For example, the subroutine \code{dsddgg2sigma(v,} \code{er,A,Z,gg,sigij,ierr)} computes the (partial) 
equivalent  scattering cross sections $\tilde\sigma_{ij}$ for nucleus $(A,Z)$ at relative velocity $v$ and recoil energy 
$E_R$ starting from values of the $G_i^\N$ constants in \code{gg}, with nuclear structure functions set by the 
previous call to \code{dsddset}. The actual nuclear recoil event rate as given in Eq.~(\ref{eq:dRdE}),
finally, is computed by the function \code{dsdddrde}.
The latter two functions are independent of the specific particle physics implementation and hence 
are contained in the core library.

%%%%%%%%%%%%%%%%%%%%%%%%%%%%%%%%%%%%%%%%%%%%%%%%%%%%%%%%%%%%%%%%%%%%%%
\section{Halo models}
\label{sec:halo}

The distribution of the DM density $\rho_\chi(\mathbf{r})$ in the Milky Way and external objects, like 
nearby dwarf spheroidal galaxies,
is a crucial information for calculating rates both in highly local direct detection 
experiments (as described in the previous Section) and for indirect DM searches that probe the DM 
distribution inside a much larger volume (as described in the following Sections). Observationally, 
however, the distribution of DM on scales relevant for DM searches is often only poorly constrained
(with the exception of classical dwarf spheroidal galaxies -- though even here uncertainties in halo 
properties relevant for DM searches may be somewhat larger \cite{Ullio:2016kvy} than what is typically 
quoted \cite{Bonnivard:2015xpq}).
The situation is somewhat improved when instead referring to the results of large $N$-body simulations 
of gravitational clustering, which consistently find that DM halos {\it on average} are well described
by Einasto \cite{1965TrAlm...5...87E} or Navarro-Frenk-White profiles \cite{Navarro:1995iw}, with
the halo mass being essentially the only free parameter (after taking into account that the halo 
concentration strongly correlates with the halo mass \cite{Maccio:2008pcd}). On the other hand, there is 
a considerable halo-to-halo scatter associated to these findings, so that it remains challenging
to make concrete predictions for individual objects -- in particular if they are located in 
cosmologically somewhat `special' environments like in the case of the Milky Way and its
embedding in the Local Group. Even worse,
baryonic physics can have a large impact on the DM profiles, especially on their inner parts 
most relevant for indirect detection, and even though hydrodynamic simulations taking into
account such effect have made tremendous progress in recent 
years \cite{Schaye:2014tpa, Schaller:2014uwa, Wang:2015jpa, Tollet:2015gqa}, there is 
still a significant uncertainty related to the modelling of the underlying processes.
In light of this situation, there is a considerable degree of freedom concerning halo models
and the DM density profiles, and a code computing observables related to DM should be
able to fully explore this freedom. One can also note that even in the Milky Way, the local halo density 
has some uncertainty. If we assume a given halo profile, the uncertainty goes down as we get constraints from the galaxy as a whole, but the fluxes of cosmic ray particles locally depend on the density not only 
here, but also further away which cause a dependence on the halo profile and not just the local density.

In \ds\ \dsver, the implementation of dark matter halo models and related quantities in the 
library \code{ds\_core} follows a new and highly flexible scheme, avoiding pre-defined 
hardcoded functions. For convenience a few pre-defined options are provided, however 
these can be either complemented by other profiles eventually needed, or the entire sample 
configuration can be simply replaced by linking to a user-defined setup 
-- in both cases without editing routines provided in the release version of the code. 
A further improvement compared to previous versions of \ds\ is that different dark matter density 
profiles, possibly referring to different dark matter detection targets, can be defined at the same 
time: E.g., one can easily switch back and forth from a computation of the local antiproton flux 
induced by dark matter annihilations or decays in the Milky Way halo to the computation of the 
gamma-ray flux from an external halo or a Milky Way satellite within the same particle physics 
scenario. A special label is reserved to indicate which profile corresponds to the Milky Way, 
such that rates that are Milky Way specific, such as the local contribution to antimatter fluxes, 
can be computed for this (and only this) profile.
Finally the present implementation simplifies the task of keeping track of consistent 
definitions for related quantities, such as, e.g., a proper connection between the dark matter profile 
and the source function for a given dark matter yield (see Section \ref{sec:yields}), or calling 
within an axisymmetric coordinate system a spherically symmetric function (and preventing the opposite). 

At the heart of the implementation, there is the subroutine \code{dsdmsdriver} which 
contains the complete set of instructions necessary to act as an interface to all quantities related 
to the DM density profiles.\footnote{
From a technical point of view, it is actually not  \code{dsdmsdriver} itself which acts as an interface
but the set of wrapper routines collected in \code{src/dmd\_astro} (which all call \code{dsdmsdriver}
in a specific way). Only those routines are called by 
cosmic-ray flux routines and other functions in \code{src}, never \code{dsdmsdriver} directly.
The routines in  \code{src/dmd\_astro} therefore provide examples of functions that {\it cannot be
replaced} individually in a consistent way, but only as a whole set (along with \code{dsdmsdriver},
in case the user wants to change the structure of the driver itself).  
}
 For example, it checks the scaling of the various DM source functions 
(Section \ref{sec:yields}) with the DM density $\rho_\chi$ -- namely $\rho$ for decaying dark matter 
and $\rho^2$ for dark matter pair annihilations %(in case the effect of substructures is neglected) 
-- and passes this information
to the routines for propagations of charged cosmic rays in the Galaxy (Section \ref{sec:cr});
the same goes for line of sight integration routines (needed, e.g. for the computation of 
gamma-ray fluxes, see Section \ref{sec:gamma}).
\code{dsdmsdriver} also contains information on whether the dark matter density profile is 
spherically symmetric or axisymmetric, and consistently handles requests of functions assuming
one or the other. Furthermore, it can be used for initialization calls, for instance to set parameters 
for a given parametric density profile, and test calls, for instance to print which dark matter profile 
is currently active within a set of available profiles. 

While a specific \code{dsdmsdriver} routine should match the user needs in the problem at hand, a sample 
version is provided with the \ds\ release, illustrating the flexibility of the setup. In particular, the simple version 
included in \ds\ \dsver\ assumes that the DM density profile is spherically symmetric 
and does not include dark matter substructure; it allows to choose as dark matter 
profile one among three parametric profiles, namely the 
Einasto \cite{1965TrAlm...5...87E}, the Navarro-Frenk-White \cite{Navarro:1995iw}
and the Burkert \cite{Burkert:1995yz} profiles, or a profile interpolated from a table of values of the 
dark matter density 
at a given radius. The present version also implements a halo {\it profile database} for later use, 
where the code allows to associate a given  {\it input label} to every full set of entries specifying a halo 
profile;
such a profile can be reloaded at any time when needed (but profiles can also be declared as `temporary',
in which case only the latest set temporary profile is available at any given time). For example, all indirect 
detection flux 
routines have this label among their input parameters, so that in case of several dark matter detection 
targets or several profiles for the same targets it becomes unambiguous which profile is being considered. 
For objects in this database, one can furthermore save or read tabulated quantities from the disc, 
such as the Green's function needed for the computation of the local positron flux.
In  Appendix \ref{sec:newhalo}, we provide more details on how to use this driver routine in praxis,
illustrated by example programs shipped with the \ds\ release,
as well as how to expand it such as to include different setups and profile parameterizations.

Let us finally mention that the local DM velocity profile that enters in the direct detection rate, 
Eq.~(\ref{eq:dRdE}), is in principle not independent of the chosen DM density profile. For 
a spherically symmetric and isotropic system, e.g., the two profiles are related by the 
Eddington equation \cite{1915MNRAS..76...37E,Catena:2011kv}. A fully self-consistent 
implementation of phase-space distributions will be available with a later \ds\ version; until
then, the user can freely choose a DM velocity distribution among those provided in 
\code{src/dmd\_vel} -- but should keep in mind this consistency requirement 
when comparing direct detection rates to, e.g., the gamma-ray flux from the galactic center 
(which requires choosing a density profile). Concretely, it is the function \code{dshmuDF}
that returns the 3D distribution function $f(\mathbf{v})$ needed by the direct detection routines. 
It allows to switch between various pre-implemented functional forms, including tabulated
velocity profiles, but can of course also be replaced by an arbitrary function supplied by the user
(c.f.~Section \ref{sec:replace}).

%%%%%%%%%%%%%%%%%%%%%%%%%%%%%%
\section{Dark matter self-interactions}
\label{sec:si}

The general expectations for DM profiles described in the previous section apply to cold, collisionless
DM. While remarkably successful at large cosmological scales, however, the CDM paradigm is less well tested
at smaller scales. In fact, DM self-interactions are phenomenologically still allowed to be
as strong as the interaction between nucleons, which is much stronger than current limits on
DM interacting with SM particles. Partially also triggered by  the absence of undisputed DM signals
in traditional searches, the possibility that DM could be self-interacting  \cite{Spergel:1999mh}
and thereby leave imprints on cosmological observables related to structure formation
has thus seen greatly increased interest in recent years, both from an astrophysical and a model-building
perspective. In fact, it has been pointed out \cite{Loeb:2010gj,Vogelsberger:2012ku,Peter:2012jh,
Zavala:2012us,Aarssen:2012fx,Elbert:2014bma,Kamada:2016euw,Robertson:2017mgj} that 
self-interacting DM (SIDM) could address the most pressing potential small-scale problems of $\Lambda$CDM cosmology 
\cite{Bullock:2017xww}, referred to as `core-vs.-cusp'' \cite{Flores:1994gz,Moore:1994yx}, `too-big-to-fail'
\cite{BoylanKolchin:2011de,BoylanKolchin:2011dk}, `diversity'  \cite{Oman:2015xda,Oman:2016zjn}
and (for late kinetic decoupling) `missing satellites' problems \cite{Moore:1999nt,Klypin:1999uc}.
While those specific discrepancies with the $\Lambda$CDM paradigm may well turn out to be either
due to  poorly modelled baryonic effects or observational uncertainties, it is clear that SIDM {\it can} leave observable 
imprints which, in case of an unambiguous detection, would significantly narrow down the range
of possible DM particle models.

In the literature, one typically uses the momentum transfer cross section,
\be
\label{eq:sigmat}
 \sigma_T \equiv \int d\Omega\left(1-\cos\theta\right)\frac{d\sigma}{d\Omega}\,,
\ee
as the relevant quantity to measure the impact of DM self-interactions on the halo 
structure, where $\sigma$ is the standard cross section for DM-DM scattering (which equals
$\sigma_T$ for isotropic scattering). 
This has the advantage of regulating large forward-scattering amplitudes, which should
not affect the DM distribution.\footnote{%
The disadvantage is that backward scattering is not treated on equal footings,
see Ref.~\cite{Tulin:2013teo,Kahlhoefer:2017umn} for a more detailed discussion.
}
In particular, 
\be
  \sigma_T/m_\chi \sim 1\,\mathrm{cm}^2/\mathrm{g}
\ee
is very roughly the typical scale of DM self-interactions of cosmological relevance:
cross sections in this ballpark may leave observable imprints and possibly address 
the afore-mentioned small-scale structure problems, while much smaller cross 
sections have no impact on the structure
of DM halos; much larger values of $\sigma_T/m_\chi$ are ruled out, on the other hand, in 
particular from observations of galaxy clusters. For a detailed 
review that summarizes both various constraints on DM self-interactions and
models that have been discussed in the literature, see Ref.~\cite{Tulin:2017ara}.

In \ds, $\sigma_T/m_\chi$ is computed by an interface function \code{dssigtm} provided
by the particle module. In general, this function can depend on the relative
velocity of the scattering DM particles -- which from a model-building point of view
has the advantage that one can evade the strong cluster constraints but
still have sizeable interaction rates at the scale of dwarf galaxies (where the 
supposed discrepancies with $\Lambda$CDM expectations have been reported).
In such a situation, one needs to average over the velocity distribution of DM particles
in the halo; this is done by  \code{dssigtmav} provided by the \code{core} library.
The \code{core} library furthermore provides several auxiliary routines, to be used by
any  particle module, for the commonly encountered specific transfer cross sections in the 
presence of (attractive or repulsive) Yukawa potentials. These result from the exchange 
of single mediator particles of mass $m_{\rm med}$ with coupling $\alpha_\chi\equiv g^2/4\pi$.
In particular, \code{dssisigmatborn} returns the resulting  $\sigma_T$ in the {\it Born limit},
$\alpha_\chi m_\chi \ll m_{\rm med}$, where the cross section can be calculated perturbatively~\cite{Feng:2009hw},
and \code{dssisigmatclassical} returns  $\sigma_T$ in the {\it classical limit},
$m_\chi v \gg m_{\rm med}$, for which we use  the parameterizations from Ref.~\cite{Cyr-Racine:2015ihg}.
For the {\it intermediate (resonant) regime},  \code{dssisigmatres} implements the analytical 
expressions from Ref.~\cite{Tulin:2013teo} that result when approximating the Yukawa potential by a 
Hulth\'en potential in the limit where scattering dominantly proceeds via an $s$-wave (which is 
only guaranteed for $m_\chi v \ll m_{\rm med}$). We stress that 
these specific {\it auxiliary} routines are only useful for models where the
self-interaction is mediated by a single scalar or vector particle. For other cases (see e.g.~Ref.~\cite{Bellazzini:2013foa} 
for a classification according to effective operators) the corresponding self-interaction cross sections are 
currently not pre-implemented and thus have to be provided along with the corresponding particle module.

%%%%%%%%%%%%%%%%%%%%%%%%%%%%%%
\section{Particle yields}
\label{sec:yields}

The annihilation or decay of DM particles in places with large astrophysical DM
densities (like in our Milky Way halo) produces SM particles which either propagate on straight lines like
gamma rays or neutrinos (Section \ref{sec:gamma}) or, in the case of charged particles,
are deflected by stochastically distributed inhomogeneities in the Galactic magnetic fields  
(Section \ref{sec:cr}). In either case, the DM signal ultimately only depends on the local 
injection rate of some stable (cosmic ray) particle $f$, per volume and energy,
\be
\label{psource}
\frac{d\mathcal{Q}(E_f,\mathbf{x})}{dE_f}=\sum_n \rho_\chi^n(\mathbf{x}) 
\left\langle\mathcal{S}_n(E_f)\right\rangle.
\ee
Here, $\rho_\chi$ is the DM density (of the respective component, in case of multi-component DM) 
and the ensemble average $\langle ...\rangle$ is taken 
over the DM velocities; in principle, it  depends on the spatial location $\mathbf{x}$, but in many 
applications of interest this can be neglected. 

The particle source terms $\mathcal{S}_n(E_f)$ encode the full information about 
the DM particle physics model. For a typical WIMP DM candidate, e.g., only the annihilation part ($n=2$)
contributes,
\be
 \mathcal{S}_2(E_f)=\frac{1}{N_\chi m_\chi^2}\sum_i \sigma_i v \frac{dN_i}{dE_f} \,,
 \label{eq:Sann}
\ee
where $\sigma_i$ is the annihilation cross section of two DM particles into final state $i$ and 
$dN_i/dE_f$ is the resulting number of stable particles of type $f$ per such 
annihilation and unit energy. $N_\chi$ is a symmetry factor that depends on the nature of DM; if 
DM is (not) its own antiparticle we have $N_\chi=2$ ($N_\chi=4$). For decaying DM, on the other 
hand, we have
\be
 \mathcal{S}_1(E_f)=\frac{1}{m_\chi}\sum_i \Gamma_i \frac{dN_i}{dE_f} \,,
 \label{eq:Sdec}
\ee
where $\Gamma_i$ denotes the partial decay widths. Let us stress, however, that 
Eq.~(\ref{psource}) is much more general in that it encapsulates also DM that is {\it both} 
annihilating and decaying, multi-component SM, as well as DM models with an internal $Z^n$ 
symmetry \cite{Belanger:2012zr,Belanger:2014bga,Ko:2014nha,Choi:2015bya}. 

In \ds\ \dsver, the particle source term $\mathcal{S}_n(E_f)$ therefore takes a central r\^ole as 
essentially the only interface between the particle physics modules and the indirect 
detection routines provided by the core library (apart from the routines providing neutrino yields 
from the Sun or Earth, see Section \ref{sec:se}, which rely on a slightly different setup). Concretely, 
there are two interface functions that
provide this quantity, \code{dscrsource} and \code{dscrsource\_line}. The former returns 
$\mathcal{S}_n(E_f)$ for any continuous energy distribution of the particle $f$, while
the latter returns location, strength and width of (almost) monochromatic contributions -- which
result if (at least) one of the two particles in a two-body channel $i$ is the stable particle $f$ itself.

In general, it is up to the particle module how to provide $\mathcal{S}_n(E_f)$. For convenience,
however, the core library contains a function \code{dsanyield\_sim} to provide $dN_i/dE_f$ for
any two-particle SM model final state. Those are based on \code{Pythia} 6.426 \cite{Sjostrand:2006za} runs. The Pythia 
simulations are run for 30 different masses from 3 GeV to 20 TeV and for annihilation into $d\bar{d}$, $u\bar{u}$, 
$s\bar{s}$, $c\bar{c}$, $b\bar{b}$, $t\bar{t}$, $gg$, $W^+ W^-$, $Z^0 Z^0$, $\mu^+ \mu^-$ and $\tau^+ \tau^-$.\footnote{ 
We have not included Higgs final states here as these are somewhat more model-dependent. In the \code{mssm}
module, e.g., the yields from both standard model and other Higgs bosons are obtained by letting those particles
decay in flight, and then boosting the tabulated yields provided by \code{ds\_core}. 
}
For each of these annihilation channels we store the resulting spectra of $\gamma$'s, positrons, antiprotons, anti-deuterons, $\nu_e$/$\bar{\nu}_e$, $\nu_\mu/\bar{\nu}_\mu$, $\nu_tau/\bar{\nu}_\tau$. For the muon neutrinos we also simulate neutrino-nucleon interactions at neutrino detector on Earth and calculate the muon yields at the neutrino-nucleon vertex (i.e.\ a yield/volume) and at a plane at the detector (i.e.\ a yield/area). These latter neutrino-nucleon interactions are simulated with \code{nusigma} \cite{Edsjo:2007ns}. All these yields are available in \code{dsanyield\_sim} which interpolates the supplied tables in both energy and mass.

For the anti-deuterons, we have performed a simulation with \code{Pythia} 6.426 where we from the event record connect anti-neutrons and anti-protons if they are within a coalescence momentum $p_0$. These yields can then in \ds\ be retrieved for different coalescence momenta $p_0$ from 0 to 300 MeV. For the same Pythia setup, we have also performed simulations of the yields of anti-deuterons on the $Z^0$ resonance that have been compared with ALEPH measurements \cite{Schael:2006fd}. The best fit coalescence momentum and its errors are available in \ds\ (and for the default simulation the best fit value is 202 MeV). The parameter \code{dbp0bar} determines which $p_0$ value to use. If zero, the best-fit $p_0$ value is used, if non-zero it determines how much above or below the best fit $p_0$ to use (in units of standard deviations from the fit, $\sigma$).

It is important to realize though that these yield functions in \ds\ are a prime example of a replaceable function. If the user wants to supply yields calculated in a specific scenario or adding a specific new process, or just using a different simulation/analytic calculation, it is fairly easy to replace either the yield function \code{dsanyield} in the chosen particle physics module or the simulation yields in \code{dsanyield\_sim} in \code{src/}. 

The setup described above is for WIMP annihilations in e.g.\ the galactic halo, a dwarf galaxy or any other similar environment where the background density is low. For annihilations in the Sun and the Earth, we have similar routines that will be described in Section~\ref{sec:se} below.

\bigskip
%%%%%%%%%%%%%%%%%%%%%%%%%%%%%%
\section{Gamma rays and Neutrinos from the halo}
\label{sec:gamma}

Gamma rays produced in the galactic halo propagate on straight lines and hence point directly
back to their sources. Along with the fact that they are expected to be copiously produced in 
many DM models, and can carry distinct spectral features that would allow to relatively easily 
distinguish signals from astrophysical backgrounds, this is the reason
they are sometimes referred to as the {\it golden channel} of indirect DM searches \cite{Bringmann:2012ez}.
Unless produced in celestial bodies like the Sun or the Earth (see Section \ref{sec:se}), the propagation
of neutrinos follows the same simple pattern. Indeed, while not as common as for gamma rays, and not as easily 
distinguished because of the poorer energy resolution of neutrino telescopes, neutrino spectra may also
carry characteristic features related to their DM origin \cite{Fukushima:2012sp, Ibarra:2013eba, Ibarra:2014vya, Bringmann:2017sko}.

For a telescope pointing in the direction $\psi$, the expected DM-induced differential flux in gamma 
rays or neutrinos -- i.e.~the expected number of particles per unit area, time and energy -- from a 
sky-region $\Delta \psi$  is thus given by a line-of-sight integral
\be
 \frac{d\Phi}{dE}= \frac1{4\pi}\int_{\Delta\psi} d\Omega \int_{\rm l.o.s.}\!\!\!\!d\ell\, \frac{d\mathcal{Q}}{dE}\,,
\ee
where the local injection rate $d\mathcal{Q}/dE$ was earlier introduced in Eq.~(\ref{psource}). 
For decaying DM, the above line-of-sight integral always factorizes into the particle source term
$S_1$ given in Eq.~(\ref{eq:Sdec}) and a term that only depends on the DM distribution,
\be
\label{eq:phiann}
\frac{d\Phi^{\rm dec}}{dEd\Omega}= \frac1{4\pi}J^{\rm dec} S_1\,, \qquad J^{\rm dec} \equiv \int_{\rm l.o.s.}\!\!\!\!d\ell\, \rho\,.
\ee
For annihilating DM, the corresponding factorization strictly speaking  {\it only} holds {\it if} the annihilation rate is independent of velocity:
\be
\label{eq:phidec}
\frac{d\Phi^{\rm ann}}{dEd\Omega}= \frac1{4\pi}J^{\rm ann} S_2\,, \qquad J^{\rm ann} \equiv \int_{\rm l.o.s.}\!\!\!\!d\ell\, \rho^2\,.
\ee
While notable exceptions exist (in particular for resonances \cite{Arina:2014fna}, 
$p$-wave annihilation \cite{Campbell:2010xc} and  Sommerfeld-enhanced annihilation \cite{Lattanzi:2008qa}), 
this is a commonly encountered situation and hence of general interest.

\ds\ therefore provides the functions \code{dscrgaflux\_dec} and \code{dscrgaflux\_v0ann} that take $J^{\rm dec}$ 
(or $J^{\rm ann}$)  as input and return the fluxes given in the above two equations for gamma rays 
(as well as corresponding routines \code{dscrnuflux\_dec} 
and \code{dscrnuflux\_v0ann} for neutrinos). Here, the subscript \code{\_v0ann} refers to the fact that, for the
purpose of those routines, $S_2$ is evaluated
in the limit of vanishing relative velocity of the annihilating DM pair. While this is the only situation of practical 
interest in many DM models, future \ds\ versions will offer support for a full velocity dependence of this quantity.
In analogy with \code{dscrsource\_line}, the \code{core} library furthermore provides
routines \code{dscrgaflux\_line\_dec} and \code{dscrgaflux\_line\_v0ann} (and correspondingly for neutrinos) 
that return strength, width and location of {\it monochromatic} (`line') signals.

The values for $J^{\rm dec}$ and $J^{\rm ann}$ can be supplied by the user, e.g.~as taken from the literature, 
or be calculated with the routines \code{dsjfactor} and \code{dsdfactor}, respectively. Those latter routines
take as input a halo label as described in Section \ref{sec:halo}, and only take into account the contribution
from the smooth halo profile (thus neglecting a possible enhancement of $J^{\rm ann}$ due to the existence
of DM substructure, which would have to be added by hand). If such a halo label is available, one can also 
more conveniently call \code{dsgafluxsph} instead. This function  directly returns the gamma-ray flux,
fully automatically calculating the required line-of-sight integrals and adding decaying and annihilating DM
components, depending on which particle model is initialized.

%%%%%%%%%%%%%%%%%%%%%%%%%%%%%%
\section{Cosmic ray propagation and antimatter signals}
\label{sec:cr}

There is a clean asymmetry between particles and antiparticles in the standard cosmic ray picture:
The bulk of cosmic rays -- protons, nuclei and electrons -- are mainly ``primary" species, i.e. particles
accelerated in astrophysical sources and then copiously injected in the interstellar medium;
``secondary" components, including antimatter, are instead produced in the interaction of primaries 
with the interstellar
medium during the propagation. It follows that there is a pronounced deficit of antimatter compared to matter 
in the locally measured cosmic ray flux (about 1 antiproton in $10^4$ protons). When considering instead a
source term due to DM annihilations or decays, a significant particle-antiparticle asymmetry is in general
not expected, and antiprotons, positrons and antideuterons turn out to be competitive indirect DM probes.

Charged particles propagate diffusively through the regular and turbulent components of Galactic magnetic 
fields. This makes it more involved for local measurements to track spectral and morphological imprints of 
DM sources than, e.g., for the gamma-ray and neutrino channels (though searches for spectral features
in CR positron fluxes still lead to very competitive limits \cite{Bergstrom:2013jra}).
In fact the transport of cosmic rays in the Galaxy is still a debated
subject: Most often one refers to the quasi-linear theory picture (with magnetic inhomogeneities as a perturbation 
compared to regular field lines) in which propagation can be described in terms of a (set of) equation(s) linear
in the density of a given species, containing terms describing diffusion in real space, diffusion in momentum
space (the so-called reacceleration), convective effects due to Galactic winds, energy or fragmentation
losses and primary and secondary sources (see, e.g., Ref.~\cite{Strong:2007nh} for a review).
Dedicated codes have been developed to solve numerically this transport equation, including GALPROP~\cite{Strong:1998pw}, DRAGON~\cite{Evoli:2016xgn} and PICARD~\cite{Kissmann:2014sia}. 

Here we follow instead a semi-analytical approach, analogous to that developed for the USINE 
code~\cite{Maurin:2001sj}. In particular, we model
the propagation of antiprotons and antideuterons by considering the steady state equation~\cite{Bergstrom:1999jc}
\begin{equation}
 \frac{\partial{N}}{\partial{t}} = 0 = \nabla \cdot
 \left(D\,\nabla N\right)
 - \nabla \cdot \left( \vec{u}\,N \right)
 - \frac{N}{\tau_N} + Q\,.
\label{eq:pbardiff}
\end{equation}
We solve it for situations where {\it i)} the diffusion coefficient $D$ can have an arbitrary dependence on the particle 
rigidity but can at most take two different values in the Galactic disc and in the diffusive halo, {\it ii)} the 
convective velocity $\vec{u}$ has a given fixed modulus and is oriented outwards and perpendicular to the disc,
{\it iii)} the loss term due to inelastic collisions has an interaction time $\tau_N$ which is energy dependent 
but spatially constant in the disc and going to infinity in the halo (corresponding to a constant target gas 
density in the disc and no gas in the halo), {\it iv)} the DM
source $Q$ is spatially axisymmetric and has a generic energy dendence. Under these
approximations and assuming, as is usually done, that the propagation volume is a cylinder centred at the disc 
and that particles can freely
escape at the boundaries of the diffusion region, Eq.~(\ref{eq:pbardiff}) can be solved analytically by expanding $N$
in a Fourier-Bessel series; the computation of the flux involves, at each energy, a sum over the %series of 
zeros of a Bessel function of first kind and order zero, and a volume integral of the spatially dependent part in the axisymmetric
source term $Q$, as introduced in Section \ref{sec:yields},
% (basically the DM density $\rho_\chi$ for decaying DM and its square for pair annihilating DM) 
times a
weight function depending on the given zero in the series (see~\cite{Bergstrom:1999jc} for further details).
We are neglecting in this equation energy changing operators, 
such as the reacceleration term and the adiabatic flow term connected to the 
convection operator; while these are know to have a relevant impact at 
low energy, 
their effect can be approximately mimicked by a proper reshaping of 
the scaling of the 
diffusion coefficient at low rigidities.

Since the path lengths for antiprotons and antideuterons are rather large, of the order of a few kpc, taking average
values for parameters in the transport equation rather than the more realistic modelling that can be implemented
in full numerical solutions, has no large impact in case of extended and rather 
smooth sources such as for DM. 
Eq.~(\ref{eq:pbardiff}) neglects reacceleration effects, which may in general be relevant at %low energies
rigidities below a few GV; however even this does not have a large impact in case of the species at hand,
see, e.g.,~\cite{Evoli:2011id} for a comparison of results with numerical and semi-analytical solutions for cosmic-ray
antiprotons. The power of our semi-analytic approach is that one can store %values of 
the solution of the transport equation for %obtained by assuming 
a given mono-energetic source -- provided by 
the functions \code{dspbtdaxi} and \code{dsdbtdaxi} for, respectively, antiprotons and antideuterons -- and 
then apply these as weights to any particle source term $\mathcal{S}_n(E_f)$ as introduced above.
For antiprotons and antideuterons, this latter step is done in the functions that compute the local galactic differential 
fluxes from DM annihilation and decay, \code{dspbdphidtaxi} and \code{dsdbdphidtaxi}, respectively.
%Note that while the outputs of \code{dspbtdaxi} and \code{dsdbtdaxi}  are labelled ``confinement time" in the code,
%since they do have a dimension of time and scale the dependence between source and flux, one cannot trade
%them for what is usually intended as confinement time for standard cosmic ray components, given that the
%morphology of the DM source is totally different from supernova remnant distributions usually implemented
%for describing ordinary primary components.

The structure we implemented gives a particularly clear advantage when the code is used to scan over many 
particle physics DM models, but only over a limited number of propagation parameters and DM density profiles.
For such an application,  it is useful to tabulate the 'confinement times' (returned by \code{dspbtdaxi} and 
\code{dsdbtdaxi}) over a predefined range of energy; this is
done in the functions \code{dspbtdaxitab} and \code{dsdbtdaxitab} when calling the flux routines with an appropriate
option. %(and only in case the DM halo profile currently active is within the halo profile database). 
Such tabulations are time-consuming at first run but can be saved and re-loaded for later use; 
here the proper table association is ensured by 
a propagation parameter label setting system in analogy to the one implemented for the halo profile database. 
%Computing such a table on the first call is rather CPU  consuming, especially for DM profiles that are 
%singular towards the Galactic center, so in case only a few flux
%computations are needed it may be better to switch off the tabulation option; this is true also in case the flux is
%needed at a small number of energy values, since the latest 100 (non-equivalent) calls to  
%\code{dspbtdaxi} and \code{dsdbtdaxi}
%are stored in memory (with the corresponding propagation parameters and halo model correctly linked).

The case for positrons is treated analogously, except that energy losses and spatial diffusion are
the most important effects for propagating cosmic ray leptons. The transport equation we solve 
semi-analytically therefore has the form~\cite{Baltz:1998xv}
\begin{equation}
 \frac{\partial{N}}{\partial{t}} = 0 = \nabla \cdot
 \left(D\,\nabla N\right)
 + \frac{\partial}{\partial{p}} \left( \frac{dp}{dt} N \right) + Q\,,
\label{eq:eplusdiff}
\end{equation}
where the functional form of $D$ and $Q$ can be chosen as for antiprotons and antideuterons, 
and the energy loss rate $dp/dt$ can have a generic
momentum dependence but needs again to be spatially constant. 
Assuming the same topology for the propagation volume and free escape 
conditions  at the vertical boundaries (the radial boundary, being much farther 
away, is actually irrelevant when computing local densities), the solution of the 
propagation equation is given in terms of a Green's function in energy (the function 
\code{dsepgreenaxi} in the code) 
to be convoluted over the source energy spectrum at emission for a 
given particle DM candidate. This last step is performed
by the function \code{dsepdndpaxi} returning the local positron number density,
while \code{dsepdphidpaxi} converts it to a flux and is the function which should be called from the main file.
The method to implement this solution is a slight generalization of the one described~\cite{Baltz:1998xv} and
generalizes the one introduced in~\cite{Colafrancesco:2005ji}
for a spherically symmetric configuration to an axisymmetric system.

The computation of the Green's function involves a volume integral over the spatially dependent part of
the DM source function $Q$ %(again basically the DM density $\rho_\chi$ for decaying DM and its square for pair
% annihilating DM) 
and implements  the so-called method of image charges.
%  the implementation via the so-called method of image charges (again a sum over a series)
% of the free escape boundary condition. 
It can again be CPU expensive for singular halo profiles, but its
tabulation is always needed since the Green function appears in a convolution. The main limiting factor with
respect to full numerical solutions is that one is forced to assume an (spatially) average value for $dp/dt$. 
However this has not a severe
impact on our results for the local DM-induced positron flux since, especially for energies above 10 GeV, 
the bulk of the DM contribution to the local flux stems from a rather
close-by emission volume; one thus just has to make sure to normalize $dp/dt$ to the mean value for 
{\it local} energy losses, as opposed to the mean value in the Galaxy, which are mainly due to the 
synchrotron and inverse Compton processes.

While the transport equations (\ref{eq:pbardiff}) and (\ref{eq:eplusdiff}) are essentially the same as considered
in previous releases of the code, their implementation in the present release is completely new and appears to be
numerically more stable. In particular cases with very singular DM profiles still give numerically accurate
results and converge faster.
% (in case of antiprotons and antideuterons implementing a procedure which applies to point sources); 
Note however that the case of very singular DM profiles is also the one in which
% our models or 
{\it any} propagation model is subject to a significant uncertainty related to the underlying
physics,
%can treat less carefully the physical problem, 
since propagation in the
Galactic center region is difficult to model and probably rather different from what can be tested in the
local neighbourhood by measurements of primary and secondary cosmic rays. Finally, the 
implementation in \ds\ \dsver\ is more flexible regarding parameter choices, such as for the 
rigidity scaling of the diffusion
coefficient and energy scaling of energy losses, in a framework which is now fully consistent for
antiprotons, antideuterons and positrons.

%%%%%%%%%%%%%%%%%%%%%%%%%%%%%%
\section{Neutrinos from the Sun and Earth}
\label{sec:se}

%%%%%%%%%%%%%%%%%%%%%%%%%%
\begin{figure}[t!]
\centering
\includegraphics[width=0.49\textwidth]{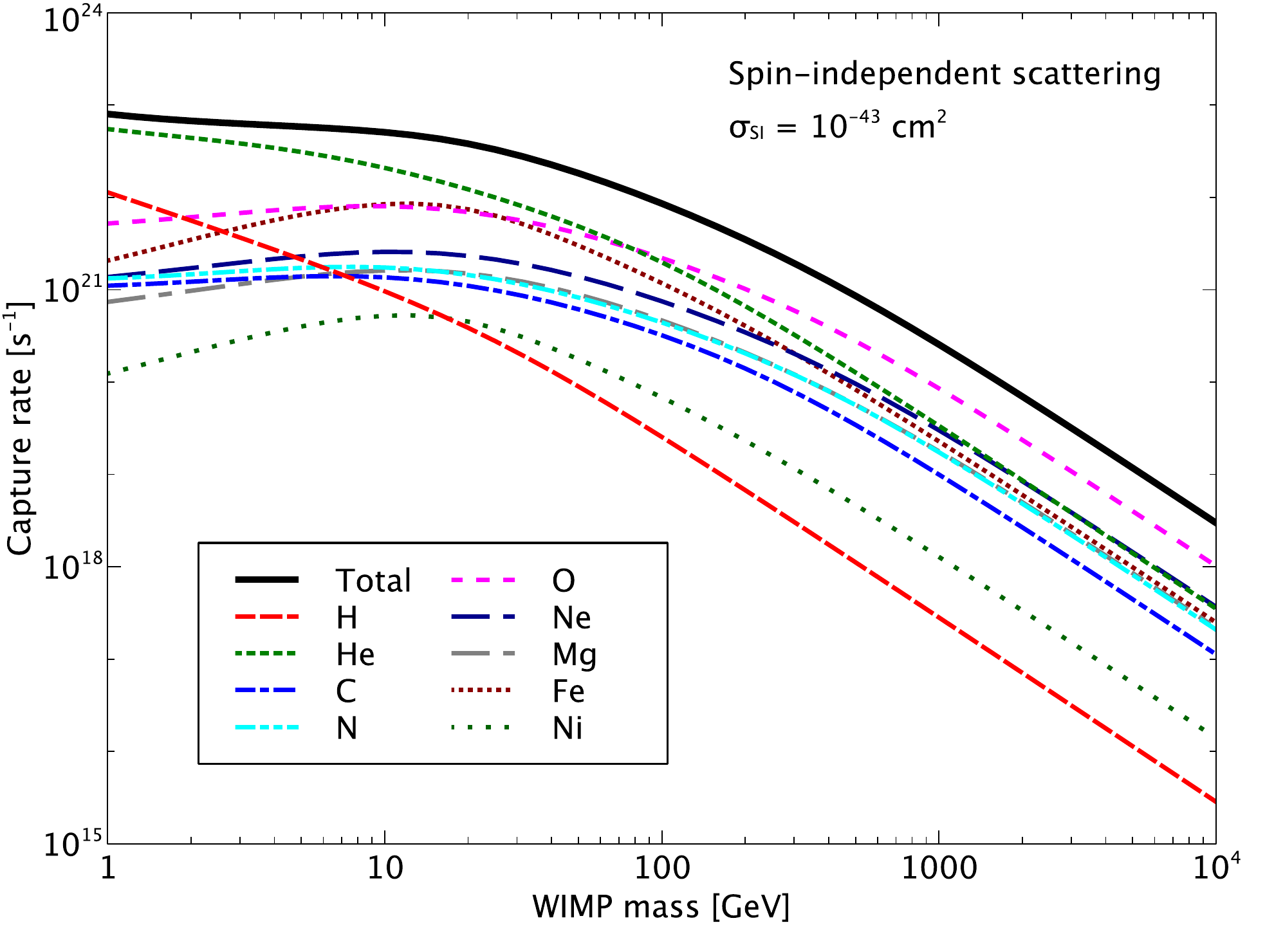}
\includegraphics[width=0.49\textwidth]{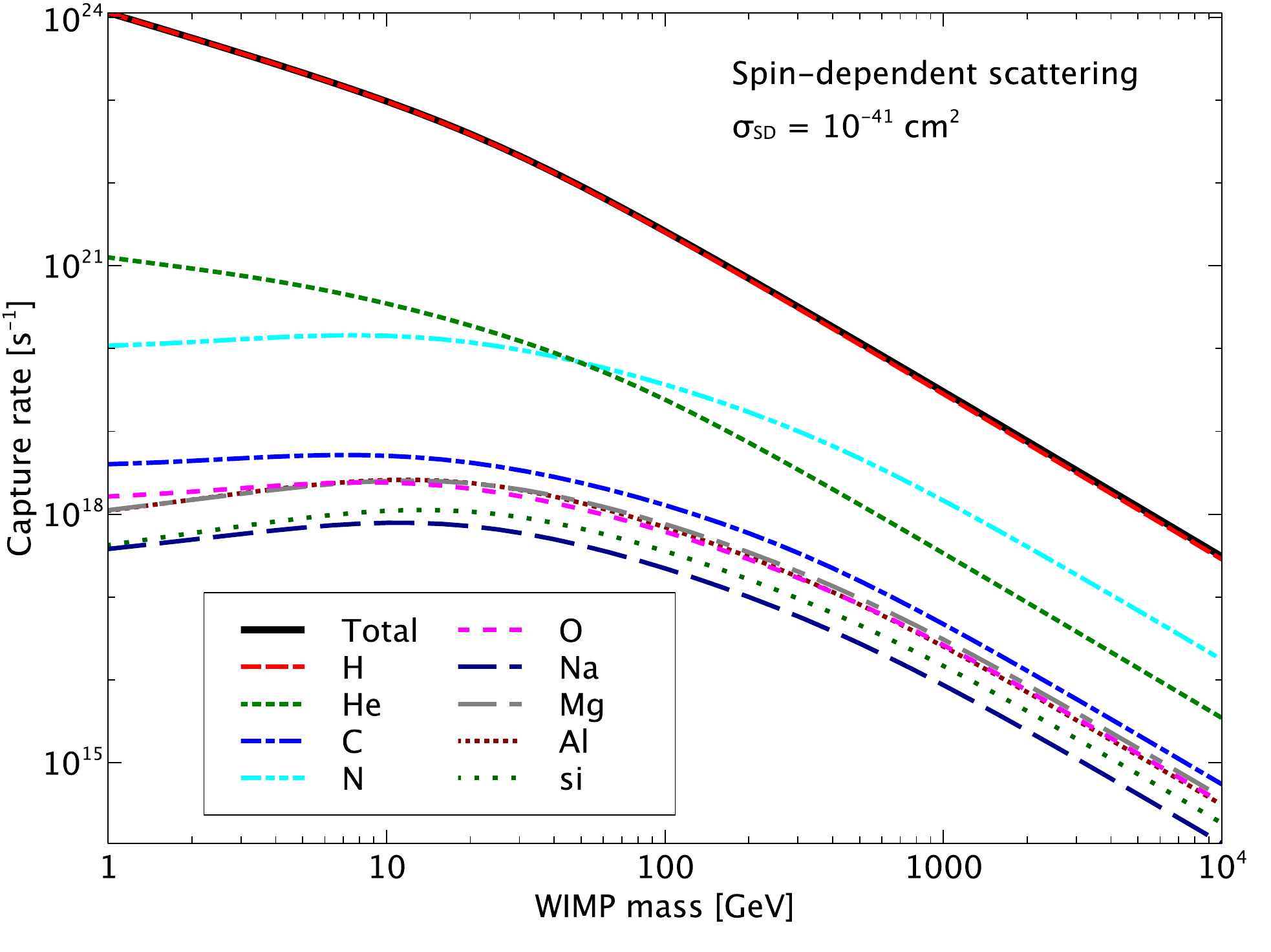}
\caption{WIMP capture rates in the Sun for spin-independent scattering (left) and spin-dependent scattering (right). The contributions from different elements are shown. Notice the different cross sections used in the left and right plots.}
\label{fig:suncapture}
\end{figure}
%%%%%%%%%%%%%%%%%%%%%%%%%%		

We can write the capture rate in the Sun/Earth as (see e.g.\ \cite{Gould:1987ir})
\begin{equation}
C =  \sum_i^N \int_{V} dV \frac{dC_{i}}{dV}= \sum_i^N \int_{V} \mathrm{d}V \int_{0}^{u_{max}}\mathrm{d}u\frac{f(u)}{u}w\Omega_{v,i}(w)\,,
\label{eq:dCdV}
\end{equation}
where $\Omega_{v,i}(w)$ is the capture probability on element $i$, $f(u)$ is the velocity distribution of the WIMPs, 
$u$ is their velocity (far away from the Sun/Earth), $w=\sqrt{u^2+v^2}$ is the velocity at the interaction radius, $v$ is the escape velocity at that point and the sum is over all the elements in the Sun/Earth. The capture probability $\Omega_{v,i}(w)$ depends on the scattering cross sections, including form factors of the different nuclei, 
the kinematics of the scattering process and the element abundance (as a function of position) in the Sun/Earth.

In \ds\ we have different ways to calculate the capture rate and different solar models available. For the Sun, the most general calculation available contains a full numerical integration over the solar radius, the velocity distribution and the momentum exchange and sums over all the elements as given by the solar model by Serenelli et al \cite{Serenelli:2009yc} (up to 289 isotopes for spin-independent scattering and 112 for spin-dependent scattering). In Fig.~\ref{fig:suncapture} we show the capture rate in the Sun for spin-independent (left) and spin-dependent (right) scattering and the most important elements that contribute to the total capture rate. The example program to calculate the curves in Fig.~\ref{fig:suncapture} is available in \code{examples/aux}. We include a simple setup where the user can easily choose how many elements to include. To speed up the calculation when scanning over particle physics models, we write the total capture rate as
\begin{equation}
C = C_{1} (G_1^p)^2 + C_{2} (G_1^n)^2 + C_{3} (G_1^p G_1^n) + C_4 (G_4^p)^2 + C_5 (G_4^n)^2 + C_6 (G_4^p G_4^n)\,,
\end{equation}
so that we can tabulate the $C_i$:s as a function of mass once and for all for a given halo and solar composition setup. The default is to use these tabuled capture rates, and if not available for the chosen halo model, they are recreated. \ds\ also contains simpler (and faster) routines where e.g.\ the momentum transfer is integrated analytically if the form factor is Gaussian. The main routine to calculate the capture rate in the Sun including the full form factor integration is \code{dssenu\_capsunnumff}, whereas the tabulated version is \code{dssenu\_capsuntabff} (which will call \code{dssenu\_capsunnumff} to set up the tables if they do not exist). \ds\ obtains the couplings $G_i^{p/n}$ above from a call to the interface routine \code{dsddgpgn}. In principle, these routines could use the more general routine \code{dsddsigma} as the direct detection routines do, but this would make tabulation very difficult, hence we currently require the particle physics module to provide \code{dsddgpgn} for the best capture rate calculation.

For the Earth, we use essentially the same set of routines, but currently assume that the form factor is Gaussian. This is motivated by the escape velocity being much lower in the Earth and hence the momentum transfer is also smaller, meaning that the dependence on the form factors is smaller as well.  Also for the Earth, \ds\ tabulates the capture rates to speed up the calculation when the halo model does not change. The main routine to calculate the capture rate in the Earth is \code{dssenu\_capearthnum}.

To get the number of WIMPs in the Sun/Earth, we have to solve the evolution equation
\begin{equation}
 \frac{dN}{dt} = C-C_{A}N^{2}-C_{E}N\,.
 \label{eq:Nevol}
\end{equation}
The annihilation term $C_A N^2$ depends on the potential and temperature in the Sun/Earth and is proportional to the annihilation cross section. The evaporation term $C_E N$ can be neglected for masses above a few GeV.
Eq.~(\ref{eq:Nevol}) can then be solved and the WIMP annihilation rate $\Gamma_{A}$ is then
\begin{align}
 \Gamma_{A}& =\frac{1}{2}C_{A}N^{2} = \frac{1}{2}C \mathrm{tanh}^{2}(t/\tau)\,, \label{eq:gammaa} \\
 \tau& = 1/\sqrt{C C_{A}}, \label{eq:tau}
\end{align}
The annihilation rate today is then obtained for $t=t_{\odot}\simeq4.5\cdot10^{9}$ years. The annihilation rates \ds\ return are the ones given from the solution above. It is instructive though to consider what happens when $t_{\odot}/\tau \gg1$. We then have equilibrium between capture and annihilation, i.e.\ $\frac{dN}{dt}=0$, and  we then have
\begin{equation}
 \Gamma_{A}=\frac{1}{2}C.
\end{equation}
In e.g.\ MSSM models we typically have equilibrium in the Sun, but not in the Earth.

Once we have the annihilation rate, we can calculate the flux of neutrinos from WIMP annihilations in the Sun/Earth. This is simulated with \code{WimpSim} \cite{Edsjo:2007ws,Blennow:2007tw} which uses \code{Pythia} \cite{Sjostrand:2006za} to simulate the yields from WIMP annihilations in the Sun/Earth, and then takes care of interactions and oscillations on the way out of the Sun/Earth and propagates the neutrinos to the detector. Once at the detector, interactions are again simulated and we obtains both neutrino-induced lepton volumetric fluxes (particles created per volume element) at the neutrino-nucleon vertex and lepton fluxes (particles passing per area element) at the detector. We also obtain hadronic showers both from charged and neutral current interactions. The neutrino-nucleon interactions are simulated with \code{nusigma} \cite{Edsjo:2007ns}. For neutrino oscillations we use the normal ordering best fit points of \cite{Esteban:2016qun,nufitonline}. In the same way as for halo yields, these yields (i.e.\ how many particles we get at the detector per annihilation in the Sun/Earth) are tabulated for a range of masses and annihilation channels. These simulation results can be accessed via the function \code{dsseyield\_sim}. The particle physics module is expected to provide a function \code{dsseyield} that
returns the yield for the current particle physics model. This routine can of course use the simulation results from \code{dsseyield\_sim} summing over different branching fractions and include cascade decays. For the MSSM model this is for example done including yields from Higgs bosons decaying in flight.

The routine \code{dssenu\_rates} performs all the steps of the calculation including the calculation of the capture rates, solution of the time evolution equation (\ref{eq:Nevol}) and inclusion of the yields from \code{dsseyield}. The end result is then a differential or integrated rate of events.  The default in \ds\ is to use the full numerically integrated capture rates in \code{dssenu\_rates}, but tabulated to speed up the calculation. For a more sophisticated likelihood analysis for IceCube data, the external package \code{nulike} \cite{IC22Methods,IC79_SUSY} can be used in conjunction with \ds.

%%%%%%%%%%%%%%%%%%%%%%%%%%%%%%%%%%%%%%%%%%%%%%%%%%%%%%%%%%%%%%%%%%%%%%
\section{Conclusions}
\label{sec:conc}

In this article, we have presented a fully revised new version \dsver\ of the numerical package
\ds\ to compute dark matter (DM) properties and observables. Compared to earlier versions of the code, 
the main new feature is a manifestly modular and 
flexible structure. In particular, \ds\ is no longer restricted to supersymmetry and neutralino
DM, but allows to handle a large variety of DM candidates from particle physics. Each such particle model is 
contained in a
separate \ds\ library, or module, and communicates with the (particle-physics independent) core library via a 
small set of interface functions.

With \ds \dsver, one can compute a large variety of very accurate predictions for DM observables, 
using state-of-the-art techniques. This includes the relic density of thermally produced DM and the 
associated cutoff scale in the spectrum of matter density perturbations, 
predictions for DM self-interaction rates, as well as comprehensive 
direct detection routines that can handle not only the traditional spin-dependent and spin-independent
cross sections, but arbitrary effective non-relativistic operators describing DM scattering on nuclei.
A large focus of \ds\ is furthermore on indirect DM detection, where the code can in principle simultaneously 
handle a large number of halo objects without re-computing time-consuming quantities, in particular 
cosmic ray fluxes in positrons, antiprotons or anti-deuterium after propagation. For the same
general halo setup, efficient line-of-sight integration routines are set up to ensure accurate predictions
for the DM-induced flux in gamma rays and cosmic neutrinos. For some particle models, in particular 
the MSSM, the spectra from DM annihilation fully include radiative corrections beyond the simplifying,
but often adopted, `model-independent' approach. Sophisticated DM capture rate routines, furthermore,
take into account the individual scatterings of DM on the different elements in the Sun and Earth,
resulting in highly reliable predictions for the flux of neutrinos from the center of these objects.

In this article we have briefly described the main ingredients and structure of \ds. In the Appendix,
we have provided details on the currently implemented particle modules -- whose number will
increase with upcoming releases -- and provided various examples of what can be calculated 
with \ds. For further details, we encourage the reader to download \ds, take a look at the manual,
and start using the code. We believe that this comprehensive package can be of great use to the 
physics community, complementary to codes that are similar in scope 
(like {\sf micrOMEGAs} \cite{Belanger:2013oya}). We also note that precision observables computed 
with \ds\ are particularly useful to feed into accurate experimental likelihoods, as provided 
e.g.~by  {\sf DarkBit} \cite{Workgroup:2017lvb}, to 
reconstruct or constrain individual signals, or to be used in global scans like in the
{\sf GAMBIT} \cite{Athron:2017ard} framework.

\bigskip
\vfill
%%%%%%%%%%%%%%%%%%%%%%%%%%%%%%%%%%%%%%%%%%%%%%%%%%
\section*{Acknowledgements}

Besides the authors of this article, several people have 
contributed smaller and larger pieces to the code in its present form. Special thanks goes 
therefore to Francesca Calore (electroweak corrections in the MSSM), Mia Schelke 
(sfermion co-annihilations), Pat Scott (ultracompact 
minihalos) as well as Gintaras Duda, Edward Baltz, Mathias Garny, Michael Gustafsson,
 Erik Lundstr\"om and Parampreet Walia. 
We finally thank all our users of previous releases of \ds. Their feedback has been very 
valuable when developing \dst\ \dsver. TB wishes to thank McGill university for hospitality,
where part of this manuscript was completed.

\newpage
%%%%%%%%%%%%%%%%%%%%%%%%%%%%%%%%%%%%%%%%%%%%%%%%%%%%%%%%%%%%%%%%%%%%%%
\appendix
%%%%%%%%%%%%%%%%%%%%%%%%%%%%%%%%%%%%%%%%%%%%%%%%%%%%%%%%%%%%%%%%%%%%%%

%\bigskip
%%%%%%%%%%%%%%%%%%%%%%%%%%%%%%%%%%%%%%%%%%%%
%%%%%%%%%%%%%%%%%%%%%%%%%%%%%%%%%%%%%%%%%%%%
\section{The generic WIMP module}
\label{app:genwimp}
%%%%%%%%%%%%%%%%%%%%%%%%%%%%%%%%%%%%%%%%%%%%%%%%
\subsection{Model parameters}
The module \code{generic\_wimp} provides the simplest example of a particle physics
library that the \ds\ core library can link to. Rather than being based on an actual
particle physics model, it mostly serves to provide an illustration of how the functionalities
of \ds\ can be used in phenomenological studies of `vanilla' WIMP DM, when only providing 
the absolute minimum of input parameters.

A generic WIMP model in \ds\ is set up by a call to  
\code{dsgivemodel\_generic\_wimp}, and hence fully defined by the input parameters of that 
routine: the mass $m_\chi$ of the DM particle and a flag stating whether the DM particle is 
its own anti-particle or not; a constant annihilation rate $\sigma v$,\footnote{
We note that a minimal way of implementing $s$-wave annihilation would rather enforce
$\sigma v\propto E_\mathrm{CMS}^2$, which is only to leading order independent of the relative
velocity $v$ of the annihilating particles. The implementation used here follows instead the
common practice of parameterizing the annihilation rate as $\sigma v=a + bv^2+\mathcal{O}(v^4)$. 
}
 along with the dominant 
annihilation channel into SM particles; the spin-independent scattering cross section 
$\sigma_\mathrm{SI}$ of DM with nucleons. 

While the purpose of this example module
is to provide the most simple realization of a generic WIMP, we note that extensions in
various directions are straightforward to implement (see Appendix \ref{sec:newmodel}
for technical instructions). How to provide a list of final states, rather than only one dominant
annihilation channel, is for example illustrated in the   \code{generic\_decayingDM} module
described in Appendix \ref{app:gendecay}.

%%%%%%%%%%%%%%%%%%%%%%%%%%%%%%%%%%%%%%%%%%%%%%%%
\subsection{DM annihilation}

The interface function \code{dssigmav0tot} simply returns the input model parameter for 
$\sigma v$ -- but only if the stated dominant annihilation channel is kinematically accessible, 
otherwise it returns zero. In a similar fashion, the module provides the interface function
\code{dsanwx} which calculates $W_\mathrm{eff}$ according to Eq.~(\ref{Weffsimp}), 
if kinematically accessible, from the input value for $\sigma v$. 
The module also provides the interface functions  \code{dscrsource}, \code{dscrsource\_line} 
and  \code{dsseyield}, which link to the tabulated yields provided by the core library.
As explained in Section \ref{sec:yields}, these interface functions are sufficient to allow
full access to all indirect detection routines of \ds.

%%%%%%%%%%%%%%%%%%%%%%%%%%
\begin{figure}[t!]
\centering
\includegraphics[width=0.7\textwidth]{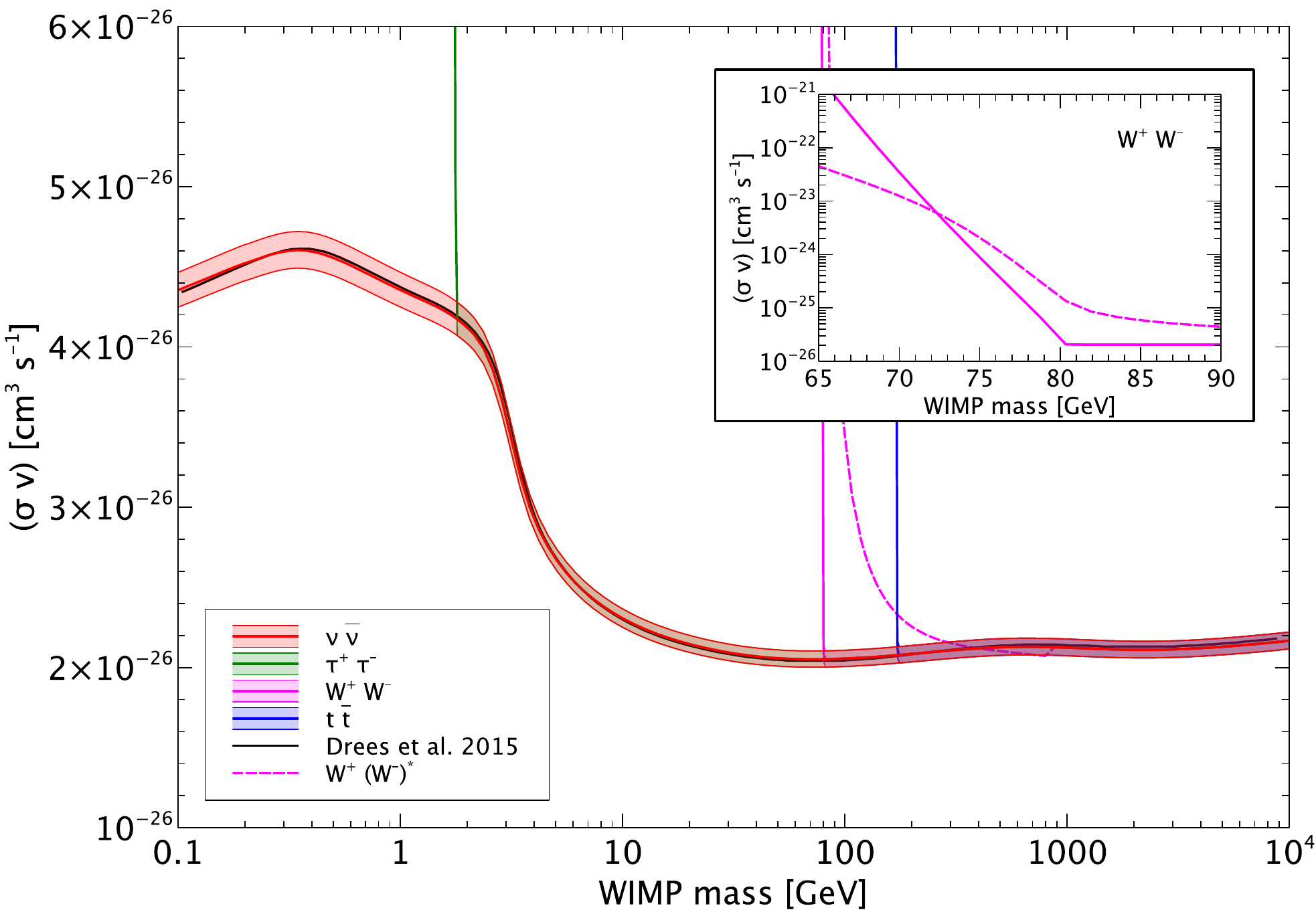}
\caption{`Thermal' annihilation rate $\sigma v$ of a generic WIMP, as a function of its mass 
$m_\chi$, assuming that 
it only couples to the indicated standard model particles. The width of the band corresponds to 
the accuracy, at the $2\sigma$ 
level, with which the relic density is determined by Planck \cite{Ade:2015xua}. The dashed line shows the results for a model with virtual final state particles (see text for details).
For comparison, we also show the results by Drees et al \cite{Drees:2015exa}.}
\label{fig:RDwimp}
\end{figure}
%%%%%%%%%%%%%%%%%%%%%%%%%%		

As an example of a simple result obtained with this particle module, we show in 
Fig.~\ref{fig:RDwimp} the annihilation rate $\sigma v$ that is required to obtain a DM relic 
abundance in accordance with current observations, where we use the results from 
Planck \cite{Ade:2015xua}, $\Omega_\chi h^2 = 0.1193 \pm 0.0028\, (2\sigma)$. We show 
separately the cases where DM annihilates only to $\bar{\nu} \nu$, $\mu^+ \mu^-$, $W^+ W^-$ 
and $\bar{t} t$. As introduced  above, the \code{generic\_wimp} particle module does not allow
off-shell final state particles. For DM masses below the kinematic threshold, the cross section
thus sharply drops to zero, and the thermal average in Eq.~(\ref{eq:sigmaveffdef}) only picks 
up contributions from the tail of the thermal distribution. This leads to a rapid increase in the 
resulting relic density, explaining the almost vertical lines in the figure (for better visibility, 
we zoom in on the threshold region for $W^+W^-$ final states). 
Apart from this effect, the non-constant nature of the
`thermal' annihilation cross section is entirely due to changes in the effective number of 
relativistic degrees of freedom after decoupling. As a default, \ds\ adopts tabulated values
from Drees et al \cite{Drees:2015exa}, but it is straight-forward 
for the user to choose different options or provide own tables. The program to produce the curves 
in Fig.~\ref{fig:RDwimp} is available in \code{examples/aux}.

As an illustration of the concept of replaceable functions, we also used the identical program in
 \code{examples/aux}, but at compile time replaced \code{dsanwx} with a version of $W_{\rm eff}$ 
 that takes into account the effect of allowing for virtual particles in the final states. The result
for $W^+W^-$ final states is indicated in Fig.~\ref{fig:RDwimp}  as dashed lines. Concretely,
we assumed that for $m_\chi\gg m_W$ the constant annihilation cross section of \code{generic\_wimp} 
is recovered, $(\sigma v)_{\chi\chi\to W^+W^-}=(\sigma v)_{\rm gen.WIMP}$, but that close to 
threshold we still have the phase-space suppression of 
$(\sigma v)_{\chi\chi\to W^+W^-}\propto \lambda^\frac12(s,m_W^2,m_W^2)=s\sqrt{1-4m_W^2/s}$
that is unavoidable in any realistic model. Around the threshold of unstable final states, 
general expressions  for the corresponding $s$-wave annihilation cross section can be found in 
Ref.~\cite{Bringmann:2013oja}. Assuming $W$ to decay to massless particles, with 
$\Gamma_W\propto m_W$ and the above scaling of the two-body cross section, in particular,
we have
\be
\frac{(\sigma v)_{\chi\chi\to W^+(W^-)^*}}{(\sigma v)_{\rm gen.WIMP}}
=\int_0^{(\mu_W^{-1/2}-1)^2}
\frac{d\mu}{\pi}\frac{\gamma \mu \sqrt{1-4\mu_W}}{(\mu-1)^2+\gamma^2}\,,
\ee
where $\mu_W\equiv m_W^2/s$ and $\gamma\equiv \Gamma_W/m_W$. For $m_\chi>m_W$, this
ratio is smaller than unity due to the phase-space suppression and the required annihilation
rate to obtain the observed relic density is thus larger than in the \code{generic\_wimp} case.
Even for $m_\chi<m_W$, on the other hand, the ratio does not vanish completely, such that the
`thermal' cross section becomes smaller than in the \code{generic\_wimp} case 
for $m_\chi\lesssim72$\,GeV.

%%%%%%%%%%%%%%%%%%%%%%%%%%%%%%%%%%%%%%%%%%%%%%%%
\subsection{DM scattering}
The interface function required for direct DM detection, \code{dsddsigma}, returns
the usual 27$\times$27 array \code{sigij} containing the equivalent cross sections 
$\tilde \sigma_{ij}$. Here, \code{sigij(1,1)} contains the parameter value of $\sigma_{SI}$ used 
to initialize the model and the other elements of $\tilde\sigma_{ij}$ equal zero (note in particular
that the spin-dependent nucleon coupling is assumed to vanish in this simple model.). 
The module also provides a routine \code{dsddgpgn}, which translates the input parameter 
$\sigma_\mathrm{SI}$ to the corresponding couplings to nucleons, $g^{a/s}$, assuming
a four-point fermion interaction. 

For the kinetic decoupling routines, the
interface function \code{dskdm2simp} provides the squared amplitude for DM-SM particle
scattering. Here, the implementation assumes that DM couples only to one type of
SM particles, as in the case of annihilation, and that the same constant amplitude describes
both annihilation and scattering.

\bigskip
%%%%%%%%%%%%%%%%%%%%%%%%%%%%%%%%%%%%%%%%%%%%
%%%%%%%%%%%%%%%%%%%%%%%%%%%%%%%%%%%%%%%%%%%%
\section{The generic decaying dark matter module}
\label{app:gendecay}
%%%%%%%%%%%%%%%%%%%%%%%%%%%%%%%%%%%%%%%%%%%%%%%%
\subsection{Model parameters}

Similar in spirit to the generic WIMP case, the module \code{generic\_decayingDM} provides
an illustration of how to use the \ds\ core library with a minimal phenomenological framework
where DM is not stable but assumed to decay on cosmological time scales. In this
case, there is no generic way of thermally producing DM, and we choose to be agnostic about 
the genesis of DM in the early universe.

Phenomenological studies for such a minimal decaying DM candidate are thus restricted to 
indirect searches, see 
below, and fully characterized by the decay rate into standard model particles. More concretely,
a generic decaying DM candidate in this module is initialized by a call to 
\code{dsgivemodel\_decayingDM}, which takes as input the DM mass $m_\chi$, the total decay 
rate $\Gamma$, as well as a list of decay channels (specified by branching ratios and 
PDG \cite{Groom:2000in} codes of the final state standard model particles).

%%%%%%%%%%%%%%%%%%%%%%%%%%%%%%%%%%%%%%%%%%%%%%%%
\subsection{Indirect detection routines}
The only interface routines provided by the module \code{generic\_decayingDM} are the 
source functions needed by the various indirect detection routines, namely \code{dscrsource}
and \code{dscrsource\_line}. The former loops over the list of decay
channels provided in the model setup and then adds the tabulated yields $dN_f/dE$ from the 
corresponding SM final states $f$ by a call to \code{dsanyield\_sim\_ls}, which is 
provided as an auxiliary routine in the \code{core} library. The quantity returned by
\code{dscrsource} is then simply  $\mathcal{S}_1$ given in Eq.~(\ref{eq:Sdec}), i.e.
\be
 \mathcal{S}_1(E)=\frac{\Gamma}{m_\chi}\sum_fBR_f \frac{dN_f}{dE}\,,
\ee
where $E$ is the energy of the messenger requested by the corresponding indirect detection
routine (e.g.~photons for gamma rays, or antiprotons for the charged cosmic ray propagation
routines).
If a final state particle in a two-particle decay channel $f$ is stable, this results in a 
monochromatic contribution to the corresponding cosmic-ray yield. These cases are caught by a 
call to \code{dscrsource\_line}, which returns $\Gamma BR_f/m_\chi$ (times 2 if there are two 
identical particles in the final state), along with the energy and width of the `line' signal (where the
 width equals the total decay rate of the other final-state particle).

Note that for decaying DM no interface function is provided for neutrino yields from the interior of 
the Sun or the Earth, because the capture rate routines require a non-vanishing DM-SM 
scattering rate.

\bigskip
%%%%%%%%%%%%%%%%%%%%%%%%%%%%%%%%%%%%%%%%%%%%
%%%%%%%%%%%%%%%%%%%%%%%%%%%%%%%%%%%%%%%%%%%%
\section{The Minimal Supersymmetric Standard Model (MSSM) module}
\label{app:mssm}

%%%%%%%%%%%%%%%%%%%%%%%%%%%%%%%%%%%%%%%%%%%%%%%%
\subsection{Model framework}
\label{sec:MSSMdef}

Linking the core library to the particle module \code{mssm} allows \ds\ \dsver\ to access
the same particle-physics specific functionality as earlier versions of the code. In 
particular, the conventions 
for the superpotential and soft supersymmetry-breaking potential are the same as 
implemented in \cite{ds4} (following \cite{Bergstrom:1995cz} and similar to \cite{Haber:1984rc,Gunion:1984yn}).
The full set of input parameters to be provided at the weak scale thus consists of the 
pseudoscalar mass ($m_A$), 
the ratio of Higgs vacuum expectation values ($\tan\beta$), the Higgsino ($\mu$) and 
gaugino ($M_1$, $M_2$, $M_3$) mass parameters, trilinear couplings ($A_{Eaa}$,
$A_{Uaa}$, $A_{Daa}$, with $a=1,2,3$) as well as  soft sfermion masses ($M^2_{Qaa}$, 
$M^2_{Laa}$, $M^2_{Uaa}$, $M^2_{Daa}$, $M^2_{Eaa}$, with $a=1,2,3$).\footnote{
Note that currently only diagonal matrices are allowed.  While not being the most general
ansatz possible, this implies the absence of flavour changing neutral currents at tree-level  
in all sectors of the model.
}
Internally, those values are stored in \code{mssm} common blocks. The user may either 
provide them directly or by setting up pre-defined phenomenological MSSM models
with a reduced number of parameters
through a call to a routine like \code{dsgive\_model} or  \code{dsgive\_model25}
(followed by a call to \code{dsmodelsetup}). The 
former sets up the simplest of those models, defined by the input parameters  
$\mu$, $M_2$, $m_A$, $\tan\beta$, a common scalar mass $m_0$, and trilinear 
parameters $A_t$ and $A_b$; $M_1$ and $M_3$ are then calculated by assuming the 
GUT condition, and the remaining MSSM parameters are 
 given by ${\bf M}_Q = {\bf M}_U = {\bf M}_D = {\bf M}_E = {\bf M}_L = m_0{\bf 1}$, 
 ${\bf A}_U = {\rm diag}(0,0,A_t)$, ${\bf A}_D = {\rm diag}(0,0,A_b)$, ${\bf A}_E = {\bf 0}$. 
 Similarly, \code{dsgive\_model25} sets up a pMSSM model with 25 free parameters
 (see the header of that file for details). Alternatively, all those values can be set by 
 reading in an SLHA file, or providing GUT scale parameters in the case of cMSSM
 models (via an interface to the \code{ISASUGRA} code, as included in ISAJET \cite{Paige:2003mg,isajet_www}).
 
Compared to previous versions of the code, \ds\ \dsver\ has a new interface to read and write SUSY 
Les Houches Accord (SLHA) files \cite{Skands:2003cj,Allanach:2008qq}. To use the general 
structure of \ds\ to its fullest, SLHA2 files are preferred, but \ds\ will also work with SLHA1 files, 
but then of course within the limitations of those (no inter-generation sfermion mixing e.g.). \ds\ 
can both read and write SLHA files from/to other codes and uses the SLHA I/O library of 
Ref.~\cite{Hahn:2006nq} as supplied with 
\code{FeynHiggs} \cite{Heinemeyer:1998yj,feynhiggs_www,Heinemeyer:1998jw,
 Heinemeyer:1998np,Heinemeyer:1999be} to perform this.

 All particle and sparticle masses are stored in a common block array \code{mass()}. For 
 neutralino masses, we include the leading loop corrections \cite{Drees:1996pk,
 Pierce:1993gj,Lahanas:1993ib}
 but neglect the relatively small corrections for charginos \cite{Drees:1996pk} 
 (in both cases, masses cannot be negative in our convention).\footnote{
 Unless, of course, those values are provided by an SLHA file. This comment also applies to the following 
 simplifications concerning both sparticle masses and widths.
 }
 Likewise, all mixing matrices and decay widths are available as common block arrays. 
 The latter are currently only computed for the Higgs particles (via an interface to the
 \code{FeynHiggs} \cite{Heinemeyer:1998yj,feynhiggs_www,Heinemeyer:1998jw,
 Heinemeyer:1998np,Heinemeyer:1999be} package),
 while the other sparticles have fictitious widths of 0.5\% of the sparticle mass (for the 
 sole purpose of regularizing annihilation amplitudes close to poles). Again, the 
 conventions for masses and mixings follow exactly those of Ref.~\cite{ds4}, to which we 
 refer for further details.

%%%%%%%%%%%%%%%%%%%%%%%%%%%%%%%%%%%%%%%%%%%%%%%%
\subsection{Experimental constraints on supersymmetry}
\label{sec:constraints}

There are various experimental constraints on supersymmetric models that are not directly 
connected to the DM observables that can be computed with \ds. Furthermore, \ds\ generally 
focusses on theoretical predictions for such observables, given a DM model realization,
rather than on the implementation of experimental likelihoods and the possibility to derive
statistically well-defined limits from those. For the latter, we instead refer to packages like 
{\sf DarkBit} \cite{Workgroup:2017lvb} (or  {\sf ColliderBit} \cite{Balazs:2017moi} for 
accelerator-based constraints), which are particularly useful in conjunction with 
global scans to determine the viable parameter space of a given particle physics model, 
for example with advanced tools like  {\sf GAMBIT} \cite{Athron:2017ard}, 
Still, it is useful to have at least a rough idea of whether a particle model is already clearly excluded before
performing advanced calculations to determine DM-related observables. For this purpose,
the \code{mssm} module allows to perform simple checks on {\it i)} the theoretical viability of
a given combination of model parameters, {\it ii)} a very approximate implementation of current accelerator 
bounds, as well as to calculate two traditional observables that have turned out to be particularly 
useful in constraining supersymmetry, namely {\it iii)} the anomalous decay $b\to s\gamma$ and {\it iv)} the 
anomalous magnetic moment of the muon $(g-2)_\mu$.

The theoretical viability of a given model is immediately indicated after the initializing call 
to the subroutine \code{dsmodelsetup(unphys,warning)}, where a non-zero flag \code{unphys} on 
return indicates a problem of type {\it i)}, e.g.~no electroweak symmetry breaking or the appearance of 
tachyonic particles (while  a non-zero flag \code{warning} in the case of the \code{mssm} module
indicates the breakdown of approximations used in radiative corrections in the Higgs sector).
The subroutine \code{dsacbnd} allows to check whether any (and which) of the above-mentioned 
observables {\it ii) -- iv)} most likely violates current bounds. 
Compared to previous \ds\ versions, we use in particular updated limits from 
{\sf HiggsBounds} \cite{Bechtle:2008jh} on  the mass of the MSSM Higgs bosons, as well as
approximate bounds on squark and gluino masses from LHC 8 TeV data \cite{Aad:2014wea}.
For $b \rightarrow s \gamma$, we keep our genuine routines for this rare decay (see Ref.~\cite{ds4} for
a more detailed description) but now use as a default the result from {\sf SuperIso} \cite{Mahmoudi:2007vz};
we compare this to the current limit of $\mathcal{B}(B \to X_s\gamma) = (3.27 \pm 0.14) \times 10^{-4}$ 
as adopted in {\sf FlavBit} \cite{Workgroup:2017myk}, based on data from BarBar and Belle 
\cite{Lees:2012wg,Lees:2012ym,Belle:2016ufb}. {\sf SuperIso} also computes the rate for the
rare leptonic decay  $B_s^0\to\mu^+\mu^-$, which we compare to the LHCb measurement of 
$\mathcal{B}(B_s^0 \to \mu^+\mu^-) = (3.0 \pm 0.6^{+0.3}_{-0.2}) \times 10^{-9}$  \cite{Aaij:2017vad}.
Finally, $a_\mu\equiv (g-2)_\mu/2$ is calculated by \code{dsgm2muon}, based on \cite{Moroi:1995yh};
in  \code{dsacbnd}, this is compared to the observed
value of $a_{\mu,\,{\rm obs}} = (11659208.9 \pm 6.3) \times 10^{-10}$ \cite{Bennett:2006fi} 
after subtracting the SM expectation as specified in {\sf PrecisionBit} \cite{Workgroup:2017bkh}.

%%%%%%%%%%%%%%%%%%%%%%%%%%%%%%%%%%%%%%%%%%%%%%%%
\subsection{Annihilation rates}

%%%%%%%%%%%%
\subsubsection{The invariant rate at tree level}
\label{sec:mssm_Weff}

In order to allow an efficient numerical integration, all tree-level diagrams contributing
to the invariant rate $W_\mathrm{eff}$ introduced in Eq.~(\ref{eq:weff}) have been 
classified according to their topology ($s$-, $t$- or $u$-channel) and to the spin of the
involved particles.  For initial states with two fermions, all corresponding helicity amplitudes are calculated analytically and 
then summed to give
\begin{equation} \label{eq:helsum}
   \frac{d W_{\rm eff}}{ d \cos\theta } =
\sum_{ijkl}
\frac{p_{ij} p_{kl}}{32 \pi S_{kl} \sqrt{s} }
\sum_{\rm helicities}
    \left| \sum_{\rm diagrams}  {\cal M}(ij \to kl) \right|^2,
\end{equation}
where $\theta$ is the angle between particles $k$ and $i$. The strength of 
this method lies in obtaining relatively compact analytic expressions that are
summed {\it at the amplitude level} and then squared numerically. 
For initial states containing bosons, the squared amplitude is directly calculated
in the standard way.
Finally, a numerical integration over $\cos\theta$ is performed
by means of an adaptive method \cite{1983qspa.book.....P}. An important new feature of 
this \ds\ version is that we have parallelized this integration, leading to a spead-up
by a factor of up to about six in particular for models with many coannihilations. We have also generalized the annihilation rate routines to that they now allow for a fully general flavour changing structure (i.e.\ we always include the full $6\times6$ sfermion mass matrices and do not treat them generation by generation.)

In \ds, all coannihilations between neutralinos, charginos and sfermions as calculated in 
\cite{Edsjo:2003us} are included. For advanced users, there exists the possibility to manually 
decide exactly which coannihilation processes to include.

%%%%%%%%%%%%
\subsubsection{Internal bremsstrahlung}
\label{sec:chi_radcorr}

The emission of an additional boson in the final state can greatly enhance neutralino 
annihilation rates for small DM velocities \cite{Bergstrom:1989jr,Bergstrom:2005ss}.
For photons, this is fully implemented in the \code{mssm} module, using analytic expressions 
for $|\mathcal{M}|^2$ for all processes $\tilde\chi^1_0\,\tilde\chi^1_0\to X^+ X^- \gamma$ in the 
limit of vanishing relative velocity of the annihilating neutralinos, 
 where $X={q,\ell,H^\pm,W^\pm}$  \cite{Bringmann:2007nk}. While this may lead to striking 
spectral features in the spectrum of gamma rays \cite{Bringmann:2012vr} or positrons 
\cite{Bergstrom:2008gr} relevant for indirect  detection, the effect 
on relic density calculations is very small. In \ds\, these contributions are thus not added to 
the interface function \code{dsanwx}, but only to \code{dscrsource}. More specifically, only the 
primary contributions from final state $\gamma$ and $e^+$ are added, because the secondary 
contributions from the hadronization or decay of the other final states $X^\pm$ are subdominant.
A gauge-invariant procedure subtracts the soft/collinear (final state radiation) part that is already 
included in the tabulated yields from two-body final states  \cite{Bringmann:2007nk}.
For performance reasons, 
these processes are  per default only calculated for the most relevant models -- 
e.g.~in the case of sfermions almost degenerate in mass with the lightest neutralino -- but this 
behaviour can be steered with a call to a dedicated subroutine \code{dsibset}.

For the internal bremsstrahlung of gluons, $\tilde\chi^1_0\,\tilde\chi^1_0\to \bar q q g$, \ds\
uses the above mentioned full analytic expressions for photon IB and rescales the  
results for quark final states by $Q^2\alpha_\mathrm{em}\to(4/3)\alpha_\mathrm{s}$. 
In this case, the much larger coupling strength implies that even the integrated annihilation 
rate can be significantly affected; both gluon IB and QCD loop corrections (which contribute 
at the same order in $\alpha_\mathrm{s}$, see below) are therefore added to \code{dsanwx}.
While the energy distribution of the final state gluons can be sharply peaked just as in the case of 
photons, the resulting spectrum of gamma-rays or antiprotons (after gluon and quark 
fragmentation) is relatively feature-less -- though noticeably different to that resulting from 
two-body quark final states \cite{Bringmann:2015cpa}. Rather than simulating these spectra for 
each MSSM model point with {\sf Pythia}, \ds\ interpolates between pre-tabulated versions of the 
most extreme 3-body 
spectra by following the method developed in Ref.~\cite{Bringmann:2015cpa}; 
for typical MSSM models, this procedure approximates the true gamma-ray and antiproton 
spectra to an accuracy of a few percent. These contributions are per default included in
the particle source term provided by \code{dscrsource}.

The implementation of electroweak internal bremsstrahlung is considerably more involved, not
the least due to the shear number of contributing diagrams. For all processes with a fermion pair
and either a Higgs or an electroweak gauge boson in the final state, the helicity amplitudes have 
been  determined fully analytically and implemented in \ds\
\cite{Bringmann:2013oja,Bringmann:2017sko}, extending Eq.~(\ref{eq:helsum}) to three-body final states.
Another complication is that for these processes -- unlike in the case of the $U(1)$ and $SU(3)$
corrections discussed above -- intermediate particles can go on-shell, so a sophisticated 
subtraction scheme has been implemented both at the cross section and at the yield level to 
avoid double-counting when including contributions from $\bar t t$ or bosonic two-body final states 
\cite{Bringmann:2017sko}. In \ds\ \dsver\  electroweak internal bremsstrahlung is now fully implemented for {\it all} 
cosmic ray yields; while disabled per default due to performance reasons (it takes 
$\mathcal{O}(10\,\mathrm{s})$ to compute a full cosmic ray spectrum), these contributions are
automatically added to \code{dscrsource} after a corresponding call to \code{dsib2set}.

%%%%%%%%%%%%%%%%%%%%%%%%%%%%%%%%%%%%%%%%%%%%
\begin{figure}[t!]
	\centering
		\includegraphics[width=0.49\columnwidth]{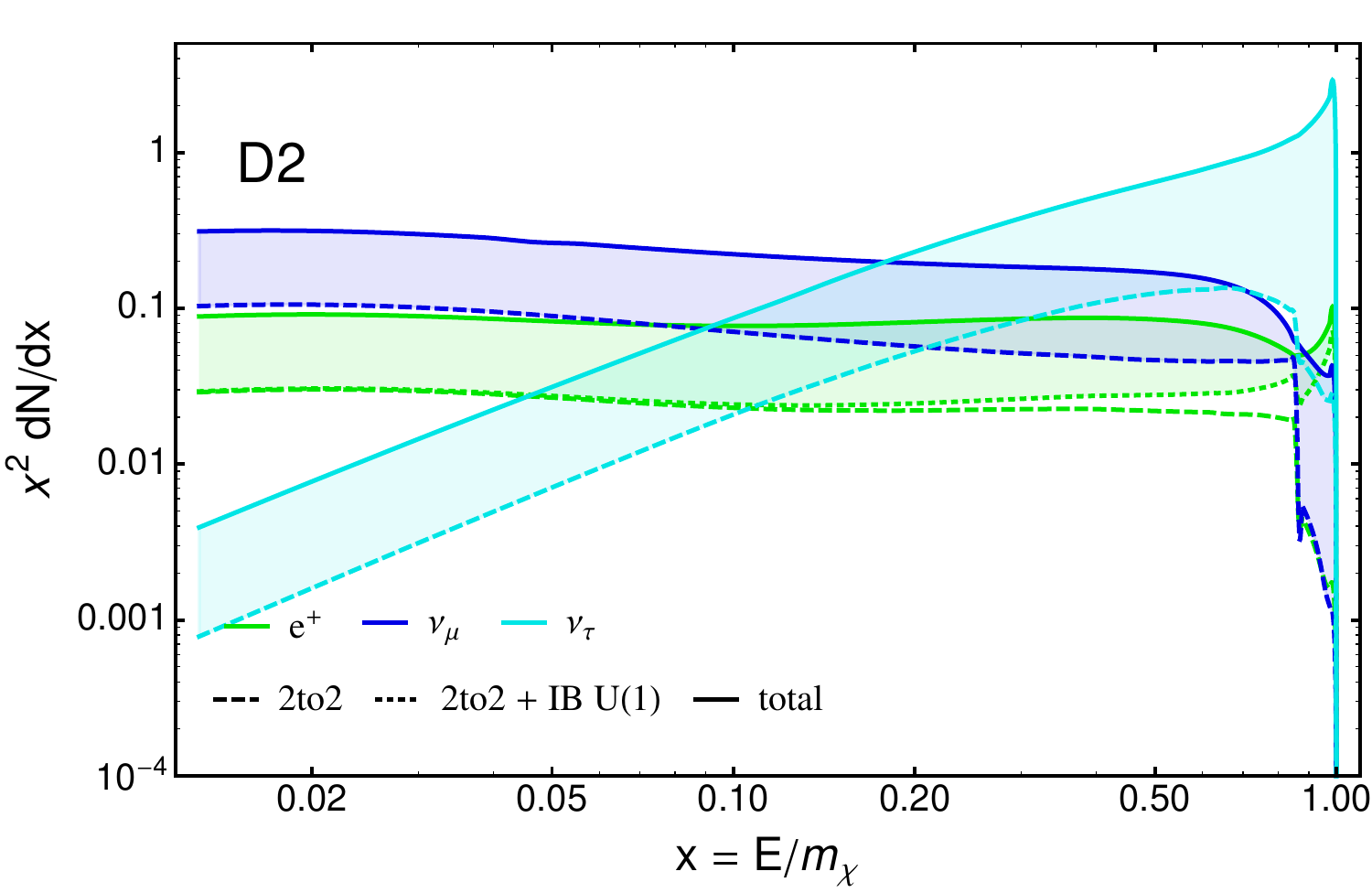}
		\includegraphics[width=0.49\columnwidth]{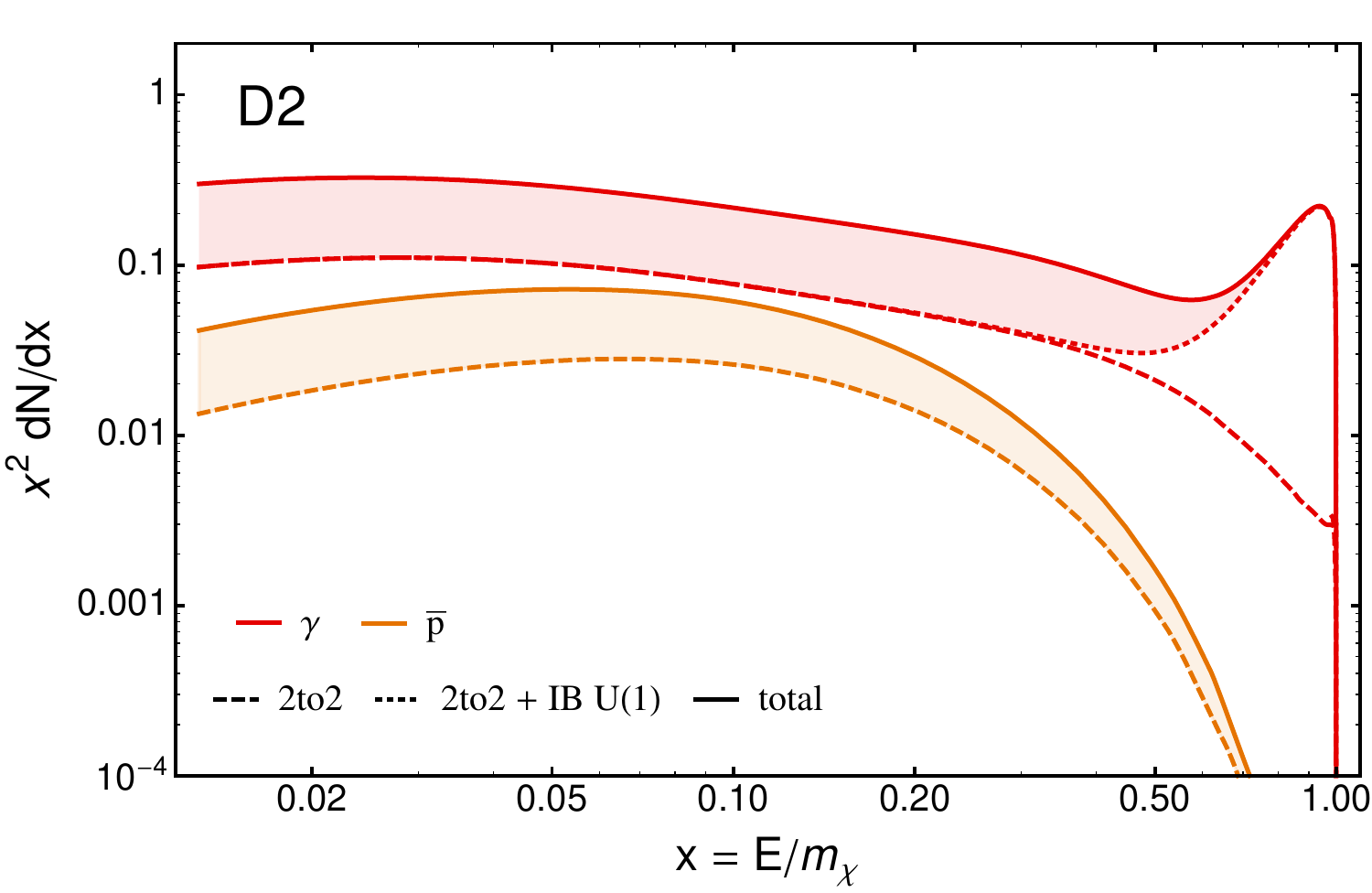} 
	\caption{Cosmic ray spectra for an MSSM benchmark model with a degenerate sfermion spectrum,
	              featuring a lightest neutralino with mass $m_\chi\sim3.4$\,TeV and the correct thermal 
	              relic abundance
	              (model D2 in \cite{Bringmann:2017sko}). The \textit{left panel} shows
		     leptonic cosmic ray particles $e^\pm$, $\nu_\mu$, and $\nu_\tau$ 
		     (green, blue and cyan line respectively), and the \textit{right panel} the case of photons
		      (red) and antiprotons (orange). Solid lines indicate the total spectrum, while 
		      dashed lines represent the spectrum expected when only taking into account 2-body final
		      states. Dotted lines show the effect of adding {\it photon} bremsstrahlung to the
		      2-body result, so shaded areas highlight the difference due to the inclusion of 
		      {\it electroweak} corrections. 
		}
	\label{fig:MSSMspectra}
\end{figure}
%%%%%%%%%%%%%%%%%%%%%%%%%%%%%%%%%%%%%%%%%%%%

We stress that the resulting continuum spectra can differ significantly from two-body final states,
or the spectra that result from including `model-independent' electroweak corrections as implemented
in \cite{Ciafaloni:2010ti,Cirelli:2010xx}, and even may lead to striking
spectral signatures in positron or neutrino yields. For any detailed phenomenological study
we thus strongly recommend to switch on these corrections, despite the additional computation 
time that this requires.
For illustration, we show in Fig.~\ref{fig:MSSMspectra} the example of cosmic-ray spectra
computed with \ds\ \dsver\,, showing separately the impact of electromagnetic and 
electroweak IB (for a pMSSM-7 benchmark model taken from \cite{Bringmann:2017sko}).

%%%%%%%%%%%%
\subsubsection{Loop and higher-order corrections}
\label{sec:chi_loop}

In \ds, the full analytic one-loop expressions for the processes 
$\tilde\chi^1_0\,\tilde\chi^1_0\to \gamma\gamma$ and 
$\tilde\chi^1_0\,\tilde\chi^1_0\to g g$ \cite{Bergstrom:1997fh,Bern:1997ng}, as well as 
$\tilde\chi^1_0\,\tilde\chi^1_0\to \gamma Z$ \cite{Ullio:1997ke},  are implemented in the limit 
of vanishing relative velocity of the annihilating neutralino pair, both as contributions to the 
particle source function \code{dscrsource} (and \code{dscrsource\_line}) and the invariant 
annihilation rate \code{dsanwx}. The annihilation to a gluon pair, e.g., can be the dominant 
process in setting the relic density for light squarks \cite{Bringmann:2015cpa}. For 
implementation details, we refer to Ref.~\cite{ds4}.
Let us mention that these one-loop expressions formally violate unitarity for diagrams
with electroweak gauge-boson exchange because they do not take into account the 
nonperturbative effects related to multiple electroweak gauge boson exchange (`Sommerfeld
effect') as described in \cite{Hisano:2003ec,Hisano:2004ds}. In practice, one can still trust 
the result for sub-TeV neutralinos -- but should keep in mind that the implemented cross 
sections become unphysical in the limit $m_W/m_\chi \to 0$.

For neutralino annihilation to quark final states, we implement the prescription of 
Ref.~\cite{Bringmann:2015cpa} to capture the leading effects of QCD corrections in the limit
of vanishing relative velocity of the incoming DM particles. This essentially amounts to modelling
the incoming neutralino pair as a decaying pseudo-scalar, for which we compute QCD 
corrections by re-summing leading logarithms as in Refs.~\cite{Braaten:1980yq, Drees:1990dq}, 
and to add separately the potentially large correction from gluon IB discussed above.

%%%%%%%%%%%%
\subsection{Neutralino scattering rates}
\label{sec:chi_scatter}

%%%%%%%%%%%%
\subsubsection{Direct detection}
\label{sec:chi_direct}

Neutralino-nucleus elastic scattering is provided as in the previous version of \ds\, apart from small corrections 
reflecting the fact that the neutralino scattering matrix does {\it not} have a pole at $m_{\tilde b}  = m_\chi + m_b$
\cite{Gondolo:2013wwa}. We note that a very similar reasoning also applies to all other quarks.\footnote{%
For {\it heavy} quarks the argument is indeed identical: given that
the probability to find a heavy quark in the nucleon is negligible, it has to be co-produced with the squark
-- which leads to a pole in the unphysical region, at $m_\chi = m_q + m_{\tilde q}$. The same reasoning applies to
{\it light} quarks that are not valence quarks. For valence quarks, on the other hand, the parton distribution function at the
momentum transfer required to produce a squark in the $s$-channel, $Q^2\sim m_\chi^2$, is negligible, and so is the probability
of forming a resonance.
}
The default option in the  \code{mssm} module is therefore not to include poles in direct detection amplitudes,
but otherwise adopt the prescription by Drees and Nojiri \cite{Drees:1993bu}.
This can be changed with a call to \code{dsddset\_mssm}, allowing both to switch to tree-level
amplitudes -- with or with our without poles -- and to adopt the original prescription by Drees and Nojiri 
(which includes poles except for $b$ and $t$ quark scattering) \cite{Drees:1993bu}.

Presently, only the traditional  spin-independent and spin-dependent cross sections are implemented, 
i.e.~the interface function \code{dsddsigma} returns the (partial) equivalent 
cross sections  $\tilde\sigma_{ij}$ off a target with all elements but \code{sigij(1,1)}
and  \code{sigij(4,4)} set to zero. While it is typically preferable to directly call \code{dsddsigma} from 
a main program, the module also provides a function \code{dsddgpgn} that returns the individual 
couplings to nucleons as given in Eqs.~(\ref{eq:gpn1}, \ref{eq:gpn2}).

%%%%%%%%%%%%
\subsubsection{Kinetic decoupling}
\label{sec:chi_tkd}
The interface function \code{dskdm2} provides the full amplitude for neutralinos
scattering with fermions in the limit of vanishing momentum transfer, as determined
in Ref.~\cite{Bringmann:2009vf};  \code{dskdm2} implements analytical expressions for the
same quantity in the simplified limit where the scattering partners are relativistic, and their
energies are well below any kinematically accessible threshold. The subroutine 
\code{dskdparticles} which is usually called directly from \code{dskdtkd}, finally, determines all 
relevant resonances for the scattering processes
(stemming from $s$-channel sleptons).

\bigskip
%%%%%%%%%%%%%%%%%%%%%%%%%%%%%%%%%%%%%%%%%%%%
%%%%%%%%%%%%%%%%%%%%%%%%%%%%%%%%%%%%%%%%%%%%
\section{The Silveira-Zee module (scalar singlet)}
\label{app:singlet}
%%%%%%%%%%%%%%%%%%%%%%%%%%%%%%%%%%%%%%%%%%%%%%%%

As an example of implementing a nonsupersymmetric model in \ds\ 6.0 we present the scalar singlet 
model of Silveira and Zee \cite{Silveira:1985rk}. Dark matter particles in this model go under many 
names, among them: scalar phantoms (which is the original name in \cite{Silveira:1985rk}) and singlet 
Higgs dark matter.  To avoid confusion in the naming of \ds\ modules, we have adopted the unambiguous 
name \code{silveira\_zee} for the particle-physics module containing it.

\subsection{Model parameters}

The Silveira-Zee model \cite{Silveira:1985rk} adds a gauge-singlet real scalar field $S$ to the standard 
model onto which it imposes a $Z_2$ symmetry $S\to -S$. Its Lagrangian is
\begin{align}
{\cal L}_{\rm SZ} = {\cal L}_{\rm SM} + \frac{1}{2} \partial_\mu S \partial^\mu S - \frac{1}{2} \mu^2 S^2 - \frac{1}{2} \lambda S^2 H^\dagger H,
\end{align}
where $H$ is the Standard Model Higgs doublet. After electroweak symmetry breaking, the $S$ boson 
acquires a tree-level mass
\begin{align}
m_S = \sqrt{\mu^2 + \frac{1}{2} \lambda v_0^2}\,,
\end{align}
where $v_0=(\sqrt{2} G_F)^{-1/2}=246.2$~GeV is the Higgs vacuum expectation value.
The module uses the $S$ mass $m_S$ and the $S$-Higgs coupling constant 
$\lambda$ as model parameters. These parameters are set with a call to
\code{dsgivemodel\_silveira\_zee}, followed as usual by a call to \code{dsmodelsetup}
to initialize a given model.

%%%%%%%%%%%%%%%%%%%%%%%%%%%%%%%%%%%%%%%%%%%%%%%%
\subsection{DM annihilation}

We have included the following annihilation channels in the annihilation of a pair of $S$ bosons: 
$SS\to \ell^+ \ell^-$, $\rm q\bar{q}$, $\gamma\gamma$, $\rm W^+W^-$, $\rm ZZ$, 
$\rm Z\gamma$, $\rm gg$, $HH$.
All these annihilation channels but $SS\to HH$ are mediated exclusively by Higgs exchange in the 
$s$-channel. We have computed the invariant annihilation rate and obtained
\begin{align}
W_{SS\to XY} = 2 \lambda^2 v_0^2 \sqrt{s} \, |D_H(s)|^2 \, \Gamma_{H\to XY}(\sqrt{s}) \,.
\label{eq:silvzeeW}
\end{align}
Here $\Gamma_{H\to XY}(\sqrt{s})$ is the partial decay width of a Standard-Model Higgs boson of 
mass $\sqrt{s}$, and
\begin{align}
\label{DHdef}
|D_H(s)|^2 = \frac{1}{(s-m_H^2)^2+m_H^2 \Gamma_H^2 }\,,
\end{align}
i.e., the square of the Higgs boson propagator. In it, the Higgs width $\Gamma_H$ must include all 
Standard Model channels and the $H\to SS$ channel if open. Eq.~(\ref{eq:silvzeeW}) coincides with the 
analogous result in \cite{Cline:2013gha} once it is recognized that their quantity 
$\sigma v_{\rm rel} = W/s$ (notice that their 
$\sigma v_{\rm rel} = 2 \sigma v/(1+\sqrt{1+v^2}) \ne \sigma v$).
For $\Gamma_{H\to XY}(\sqrt{s})$ we follow the procedure in \cite{Cline:2013gha}, i.e.~we use tabulated 
values for $\sqrt{s} < 300$~GeV and analytic expressions at higher $\sqrt{s}$, but instead of the tables 
in \cite{Dittmaier:2011ti} we have produced our own tables using HDecay 6.51, and we use the analytic 
expressions in \cite{Cline:2013gha,Ilisie:2011}.

The annihilation channel $SS\to HH$ is a sum of Higgs-mediated $s$-channel, $S$-mediated $t$-
channel, $S$-mediated $u$-channel, and contact $SSHH$ diagrams. We have computed the invariant 
annihilation rate and find
\begin{align}
%W_{SS\to HH} = \frac{\lambda^2 v_H}{8\pi} 
%TB: Note that I changed an overall factor of 1/2 here!
W_{SS\to HH} = \frac{\lambda^2 v_H}{16\pi} 
\left[ a_R^2 + a_I^2 + \frac{8y^2}{1-x^2} - \frac{2 y (a_R-y)}{x} \log\!\left( \frac{1+x}{1-x} \right)^2 \right] ,
\label{eq:silvzeeW2}
\end{align}
where
\begin{align}
a_R & = 1 + 3 m_H^2 (s-m_H^2) \, | D_H(s) |^2 ,
&
a_I & = 3 m_H^2 \, \sqrt{s} \, \Gamma_H  \, | D_H(s) |^2 ,
\\
y & = \frac{\lambda v_0^2}{s-2m_H^2},
&
x & = \frac{2 v_S v_H}{1+v_H^2} ,
\\
v_S & = \sqrt{ 1 - \frac{4m_S^2}{s}},
&
v_H & = \sqrt{ 1 - \frac{4m_H^2}{s}} .
\end{align}
When comparing Eq.~(\ref{eq:silvzeeW2}) with the analogous expression in \cite{Cline:2013gha}, there 
appears to be a typo in their Eq.~(A4): the sign in front of their log term is incorrect.

Besides \code{dsanwx}, which returns the invariant rate $W$, the module also provides
the annihilation cross section times relative velocity (as it appears e.g.~in indirect detection 
calculations),
\begin{align}
\label{Weffsvconv}
\sigma v = \frac{W}{2(s-2m_S^2)}\,,
\end{align}
in terms of interface functions \code{dssigmav} and  \code{dssigmavpartial}.
This allows in the usual way access to all indirect detection routines of the \code{core} library,
via interface functions  \code{dscrsource}, \code{dscrsource\_line} 
and  \code{dsseyield}.

%%%%%%%%%%%%%%%%%%%%%%%%%%%%%%%%%%%%%%%%%%%%%%%%
\subsection{DM scattering}

For the scattering cross section of $S$ bosons off nuclei, we follow \cite{Cline:2013gha} and use the 
following nonzero $S$-nucleon coupling constants from Higgs-boson exchange with quarks and gluons,
\begin{align}
G_{1}^{\rm p} = G_{1}^{\rm n} = \frac{\lambda f_N m_N}{2m_S m_H^2}\,,
\end{align}
where we take $f_N=0.30$ as in the Erratum of \cite{Cline:2013gha}. These $G$'s are computed in a 
function \code{dsddgpgn}. Based on this, the interface function \code{dsddsigma} returns the (partial) equivalent 
cross sections  $\tilde\sigma_{ij}$ off a target, as explained in Section \ref{sec:direct}. Presently, only the 
traditional  spin-independent and spin-dependent cross sections are implemented, i.e.~\code{sigij(1,1)}
and  \code{sigij(4,4)} are the only non-vanishing components of  \code{sigij} as returned by \code{dsddsigma}
in this module.

For the kinetic decoupling routines, we need the full scattering amplitude with fermions, averaged 
over the transferred momentum  as stated in Eq.~(\ref{maverage}). This was recently calculated 
in \cite{Binder:2017rgn},
\be
\label{eq:mav_singlet}
\left\langle \left| \mathcal{M}\right|^2 \right\rangle_t=
\frac{N_f\lambda^2m_f^2}{8k^4}\left [
\frac{2k_{CM}^2-2m^2+m_H^2}{1+m_H^2/(4k_{CM}^2)}
-\left( m_H^2-2m_f^2\right) \log\left(1+4k_{CM}^2/m_H^2 \right)
\right]\,,
\ee
and we provide it in terms of the interface function \code{dskdm2}. Here, $m_f$ is the mass
of the fermionic scattering partner and the colour factor is $N_f= 3$ for quarks and $N_f= 1$ 
for leptons. We note that this is an example where the $t=0$ prescription for the scattering
amplitude (see the description in Section \ref{sec:minihalo}) would fail rather badly, 
given that {\it only} a $t$-channel process contributes. In particular, we find
\be
\left\langle \left| \mathcal{M}\right|^2 \right\rangle_t \quad \xlongrightarrow[m_f\ll k\ll m_h]{} \quad
\frac43N_f\lambda^2\left( \frac{m_f}{m_H}\right)^4 \left( \frac{k}{m_f}\right)^2\,,
\ee
while $ \left| \mathcal{M}\right|^2_{t=0} =2 N_f\lambda^2(m_f/m_H)^4 $ in the same limit, 
which features a different scaling with energy ($k\simeq\omega$ for relativistic scattering partners).
Given that there are no $s$-channel resonances in the scattering amplitude with fermions, 
the interface function \code{dskdparticles} is trivial for the \code{silveira\_zee} module.

%%%%%%%%%%%%%%%%%%%%%%%%%%
\begin{figure}[t!]
\centering
\includegraphics[trim=0 20 0 50,clip,width=0.8\columnwidth]{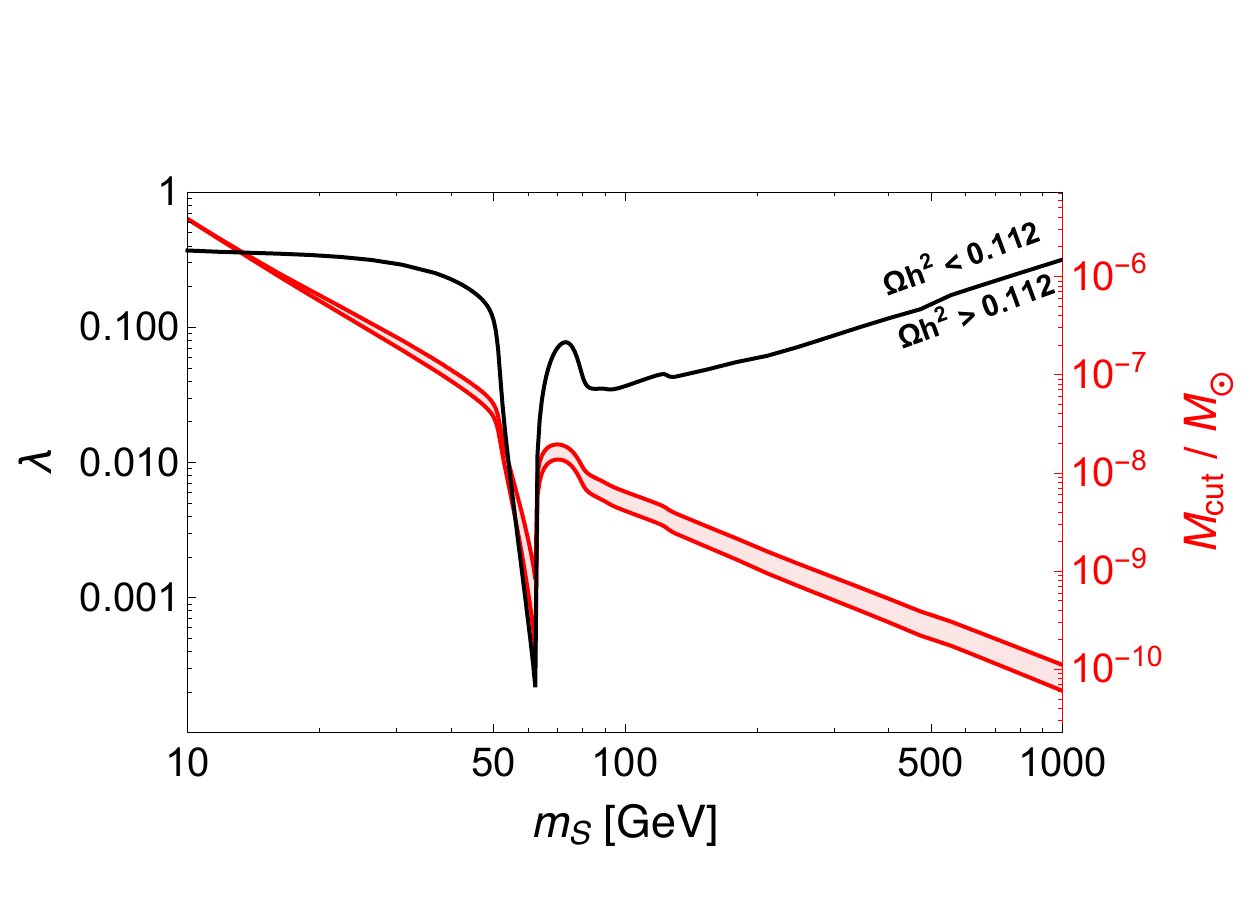}
\caption{Value of $\lambda$ that results in $\Omega h^2=0.112$, along with the corresponding
mass scale of the smallest protohalos, as a function of the Scalar Singlet mass $m_S$. 
The shaded band reflects the uncertainty in $M_{\rm cut}$ due to quark scattering:
the lower curve follows \cite{Gondolo:2012vh} and assumes that {\it all} quarks contribute
as free particles for $T>T_{\rm QCD}\equiv154$\,MeV, with no DM-quark scattering 
for $T<T_{\rm QCD}$, while the upper curve follows the conservative treatment of 
\cite{Bringmann:2009vf} and assumes that only {\it light} quarks contribute at $T>4T_{\rm QCD}$.
}
\label{fig:ScalarSinglet}
\end{figure}
%%%%%%%%%%%%%%%%%%%%%%%%%%		

\medskip
As an example of a result obtained with \ds\ using this specific particle module, we plot in 
Fig.~\ref{fig:ScalarSinglet} the value of the coupling $\lambda$ that is required to 
obtain the correct relic density, as a function of the DM mass $m_S$. 
% consistent with \cite{Cline:2013gha} at 5-10\% level
For these values of $\lambda$, we also show the mass of the smallest protohalos in this model,
given by the cutoff $M_{\rm cut}$ in the power spectrum of matter density perturbations.
We note that close to the resonance, $m_S\sim m_H/2$, the assumption of kinetic equilibrium
during chemical freeze-out (which the standard treatment \cite{Gondolo:1990dk,Edsjo:1997bg} 
discussed in Section \ref{sec:RelDens} relies on) is not satisfied; as a result, the actual value
of $\lambda$ resulting in the correct relic density differs by a factor of up to about 2 from what 
is shown in this figure \cite{Binder:2017rgn}. Extended relic density routines to capture this
effect will be available with a future \ds\ release.

\bigskip
%%%%%%%%%%%%%%%%%%%%%%%%%%%%%%%%%%%%%%%%%%%%
%%%%%%%%%%%%%%%%%%%%%%%%%%%%%%%%%%%%%%%%%%%%
\section{A dark sector module with velocity-dependent self-interactions}
\label{app:vdsidm}
%%%%%%%%%%%%%%%%%%%%%%%%%%%%%%%%%%%%%%%%%%%%%%%%

With the \code{vdSIDM} module, we provide an example for a class of models confined to a dark 
sector with no or only little interaction with the SM. These simplified models contain a CDM particle,
a relativistic dark radiation (DR) particle that is thermally distributed, and one additional 
-- typically light -- particle that 
mediates a renormalizable interaction between them. Such models have been extensively discussed 
in the literature as they affect structure formation in a way that leads to interesting cosmological 
observables for DM particles that are complementary to the traditional direct, indirect or collider
searches for DM. In fact, it has been argued that such simple setups may be sufficient to {\it simultaneously} 
mitigate (combinations of) the main known observational discrepancies with the $\Lambda$CDM 
concordance model 
\cite{Loeb:2010gj,Vogelsberger:2012ku,Aarssen:2012fx,Dasgupta:2013zpn,Bringmann:2013vra,Huo:2017vef,Binder:2017lkj}.

\subsection{Model parameters}

Similar in spirit to the simplified model analysis presented in Ref.~\cite{Bringmann:2016ilk}, the module 
provides a structure that allows any of the three particles just mentioned to be either a scalar,
a fermion or a vector, with arbitrary masses and couplings. In the present version 6.1 of the code, the 
module has two concrete example models fully implemented, where DM is a massive Dirac fermion
$\psi_\chi$, DR is a massless fermion $\psi_{\tilde \gamma}$, and the mediator is either a vector $V$
or a scalar $\phi$. The interaction parts of the respective
Lagrangians are thus given by 
\bea
\label{sidm_vector}
\Delta \mathcal{L}_\mathrm{vector}& =&  g_\chi \bar\psi_{\chi} \slashed{V} \psi_{\chi}+g_{\tilde \gamma} \bar\psi_{\tilde \gamma} \slashed{V} \psi_{\tilde \gamma}
\eea
and
\bea
\label{sidm_scalar}
\Delta \mathcal{L}_\mathrm{scalar}& =&   g_\chi \bar\psi_\chi \psi_\chi \phi +g_{\tilde \gamma} \bar\psi_{\tilde \gamma} \psi_{\tilde \gamma} \phi \,.
\eea
DM mass, mediator mass  and couplings are set by calls 
to \code{dsgivemodel\_vdSIDM\_vector} or \code{dsgivemodel\_vdSIDM\_scalar}, followed as usual by 
a call to \code{dsmodelsetup}.

In these models, the mediator thus only decays into invisible particles. Additional decay channels into SM particles are in principle 
straightforward to add, in a similar fashion as done in the \code{generic\_decayingDM} module, and will be included in a later
version of the code. We caution, however, that such decay channels are strongly constrained from direct detection experiments
(for the case of scalar mediators \cite{Kaplinghat:2013yxa}) or cosmology (for the case of vector mediators \cite{Bringmann:2016din}).

A dark radiation component with temperature equal to that of the photons would not be cosmologically viable, as the new DS particles 
would result in an unacceptably large additional contribution to the radiation density. This is conventionally stated in terms of
\be
 \Delta N_\mathrm{eff}\equiv \frac{\rho_\mathrm{DS}}{\rho_\mathrm{1\nu}}=  
 \frac47g_*^\mathrm{DS}\left(\frac{T_{\tilde\gamma}}{T_\nu}\right)^4\,,
\ee
where $\rho_\mathrm{1\nu}$ is the energy density contributed by one massless neutrino species, and $T_\nu$ is the neutrino
temperature (which differs from the photon temperature after $e^\pm$ annihilation). 
In the \code{vdSIDM} module, we provide a 
function \code{dsrdxi} which assumes that the dark sector has been
in thermal contact with the SM heat bath at very early times, but then decoupled at a temperature when all SM particles 
were still relativistic.  This function thus automatically replaces the trivial version of \code{dsrdxi}  provided by the \code{core}
library. In the expression in footnote \ref{foot:xi}, we thus have $g_*^\mathrm{SM}(T_\mathrm{dc})=106.75$
and $g_*^\mathrm{DS}(T_\mathrm{dc})= 10\,(8)$ for a vector (scalar) mediator.
Aggressively assuming that the mediators are still relativistic during BBN, this results in
\be
 \left.\Delta N_\mathrm{eff}\right|_{T\sim1\,\mathrm{MeV}} = 0.30\, (0.25)
\ee
for the effective number of relativistic degrees of freedom, which in both cases respects current 
limits at 2-3$\sigma$ \cite{Cyburt:2015mya,Hufnagel:2017dgo}. At late times, when only the DR particles are present,
we find $\Delta N_\mathrm{eff} = 0.38\, (0.28)$, largely compatible with the current CMB bound of $\Delta N_\mathrm{eff}<0.35$ (95\%\,C.L.) 
from the Planck collaboration \cite{Ade:2015xua}. For convenience, the module provides a function \code{dsrddeltaneff} to compute 
$\Delta N_\mathrm{eff}$ as a function of (photon) temperature.

%%%%%%%%%%%%%%%%%%%%%%%%%%%%%%%%%%%%%%%%%%%%%%%%
\subsection{DM annihilation}

At tree-level, DM annihilates via the $t$- and $u$-channel to a pair of mediators, which in the early universe are 
kept in thermal equilibrium with the DR particles. We have calculated and implemented the full expressions for the
invariant rate, and only quote here the result in the limit of small CMS energies $\sqrt{s}\to2m_\chi$:
\bea
 W_{\chi\chi\to VV} & \xlongrightarrow[p_\mathrm{CM}\ll m_\chi]{} & 4\pi \alpha_\chi^2\left(1-\frac{m_V^2}{m_\chi^2} \right)^{3/2} \left(1-\frac{m_V^2}{2m_\chi^2} \right)^{-2}\\
 W_{\chi\chi\to \phi\phi} & \xlongrightarrow[p_\mathrm{CM}\ll m_\chi]{} & 6\pi \alpha_\chi^2p_\mathrm{CM}^2\left(1-\frac{m_\phi^2}{m_\chi^2} \right)^{1/2} \left(1-\frac{m_\phi^2}{2m_\chi^2} \right)^{-4}\left(1-\frac89\frac{m_\phi^2}{m_\chi^2}+\frac29\frac{m_\phi^4}{m_\chi^4}   \right),
\eea
where $\alpha_\chi\equiv g_\chi/(4\pi)$ and $p_\mathrm{CM}$ is the initial CMS momentum of (each of) the DM particles.
Just as in Eq.~(\ref{Weffsvconv}), the invariant rate can be translated to the annihilation cross section times relative velocity
(which is returned by \code{dssigmav}). We note that the resulting expressions, in the above limit, differ from the corresponding 
lowest-order results stated in Eq.~(28) of Ref.~\cite{Tulin:2013teo}.
%mediator mass dependence is very different
% for m_med->0, the s-wave agrees, but our sv for the p-wave is a factor of 1/2 smaller.
Furthermore, DM can annihilate to a pair of DR particles, by a mediator exchange in the $s$-channel.
For this we find the following contributions to the invariant rate:
\bea
 W_{\chi\chi\to V^*\to\tilde\gamma\tilde\gamma} & = & 8\pi \alpha_\chi s^2\left(1+\frac{2m_\chi^2}{s} \right)
   \frac{\Gamma_{V\to\tilde\gamma\tilde\gamma}}{m_V}\left|D_{V}\right|^2\\
  W_{\chi\chi\to \phi^* \to\tilde\gamma\tilde\gamma} & = & 8\pi \alpha_\chi s^2\left(1-\frac{4m_\chi^2}{s} \right)
   \frac{\Gamma_{\phi\to\tilde\gamma\tilde\gamma}}{m_\phi}\left|D_{\phi}\right|^2\,,
\eea 
where $D_{V,\phi}$ are defined in analogy to Eq.~(\ref{DHdef}) as the inverse of the denominator of the 
respective propagator.
For the total widths that enter in $D_{V,\phi}$ we add to the decay rate of the mediator into dark radiation the partial width
of the mediator decaying into DM particles (if this is kinematically allowed).

If the mediator is much lighter than the DM particle, the annihilation rates for small relative velocities are strongly enhanced by the 
Sommerfeld effect \cite{Sommerfeld,ArkaniHamed:2008qn}. The interface function \code{dsanwx} therefore adds to the full tree-level
expressions an enhancement factor \code{dsansommerfeld} to multiply the leading-order expressions in the $v\to0$ limit.
This factor depends on whether the process is $s$-wave dominated (as in the vector mediator case) or $p$-wave dominated
(as in the scalar mediator case), and implements in the present version the analytic expressions from Ref.~\cite{Cassel:2009wt,Tulin:2013teo}, 
which result from approximating the Yukawa potential with a Hulthén potential.

Since the mediator in this simplest realization is assumed to exclusively decay into invisible DR particles, all source functions
for indirect detection routines in the \code{vdSIDM} module currently return zero.

%%%%%%%%%%%%%%%%%%%%%%%%%%%%%%%%%%%%%%%%%%%%%%%%
\subsection{DM scattering}

Since the mediators do not couple to SM particles, the DM scattering cross section off nuclei
is zero in this module. There is, however, significant scattering of the DM particles with the 
DR particles, through $t$-channel exchange of mediator particles, which can lead to kinetic 
decoupling much later than in the case of ordinary WIMPs. For the momentum-averaged 
scattering amplitude returned by the interface function \code{dskdm2}, we implement the 
expressions from Ref.~\cite{Bringmann:2016ilk}:
\bea
\left\langle \left| \mathcal{M}\right|^2 \right\rangle_t^\mathrm{vector} &=&
\frac{256}{3}g_\chi^2g_{\tilde \gamma}^2 \left(\frac{m_\chi}{m_V}\right)^4\left(\frac{\omega}{m_\chi}\right)^2\,,\\
\left\langle \left| \mathcal{M}\right|^2 \right\rangle_t^\mathrm{scalar} &=&
\frac{512}{3}g_\chi^2g_{\tilde \gamma}^2 \left(\frac{m_\chi}{m_\phi}\right)^4\left(\frac{\omega}{m_\chi}\right)^2\,,
\eea
where $\omega$ is the energy of the DR particle. We also add the amplitudes for DM scattering directly 
off heat bath vector mediators, $\left\langle \left| \mathcal{M}\right|^2 \right\rangle_t=64g_\chi^4/3$, and scalar mediators,   
$\left\langle \left| \mathcal{M}\right|^2 \right\rangle_t=16g_\chi^4/3$ \cite{Bringmann:2016ilk}, though these are 
typically subdominant for the parameter ranges of interest.

The fact that the mediators generate a Yukawa potential does not only lead to the already mentioned
Sommerfeld effect enhancing the annihilation of DM particles, but also implies strong and velocity-dependent 
DM self-interactions if the mediators are light (which is the reason for the `\code{vd}' in the module name).
The interface function \code{dssisigtm} returns the resulting momentum transfer cross section per unit mass,
$\sigma_T/m_\chi$, as introduced in Section \ref{sec:si}, by determining whether scattering takes place in 
the classical, Born or resonant regime (following Ref.~\cite{Tulin:2013teo}). For a vector mediator, it averages 
between the corresponding scattering rates for a repulsive and an attractive potential to properly take into 
account both components of the Dirac DM particles $\chi$ and $\bar \chi$ (for asymmetric DM, only the 
repulsive part would contribute). For a scalar mediator, an attractive potential is implemented.

%%%%%%%%%%%%%%%%%%%%%%%%%%
\begin{figure}[t!]
\centering
\includegraphics[trim=0 0 0 0,clip,width=0.48\columnwidth]{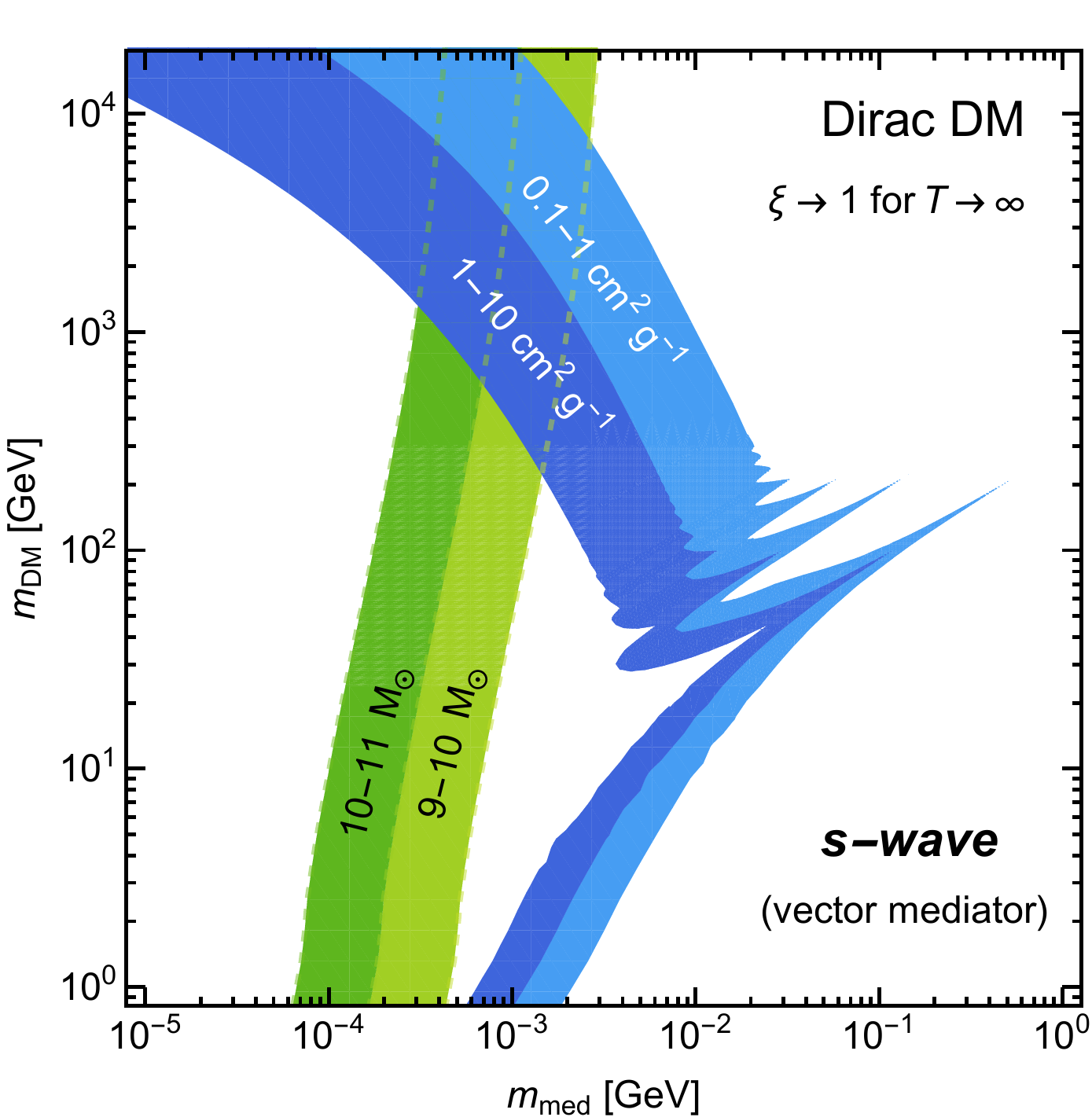}~~~
\includegraphics[trim=0 0 0 0,clip,width=0.48\columnwidth]{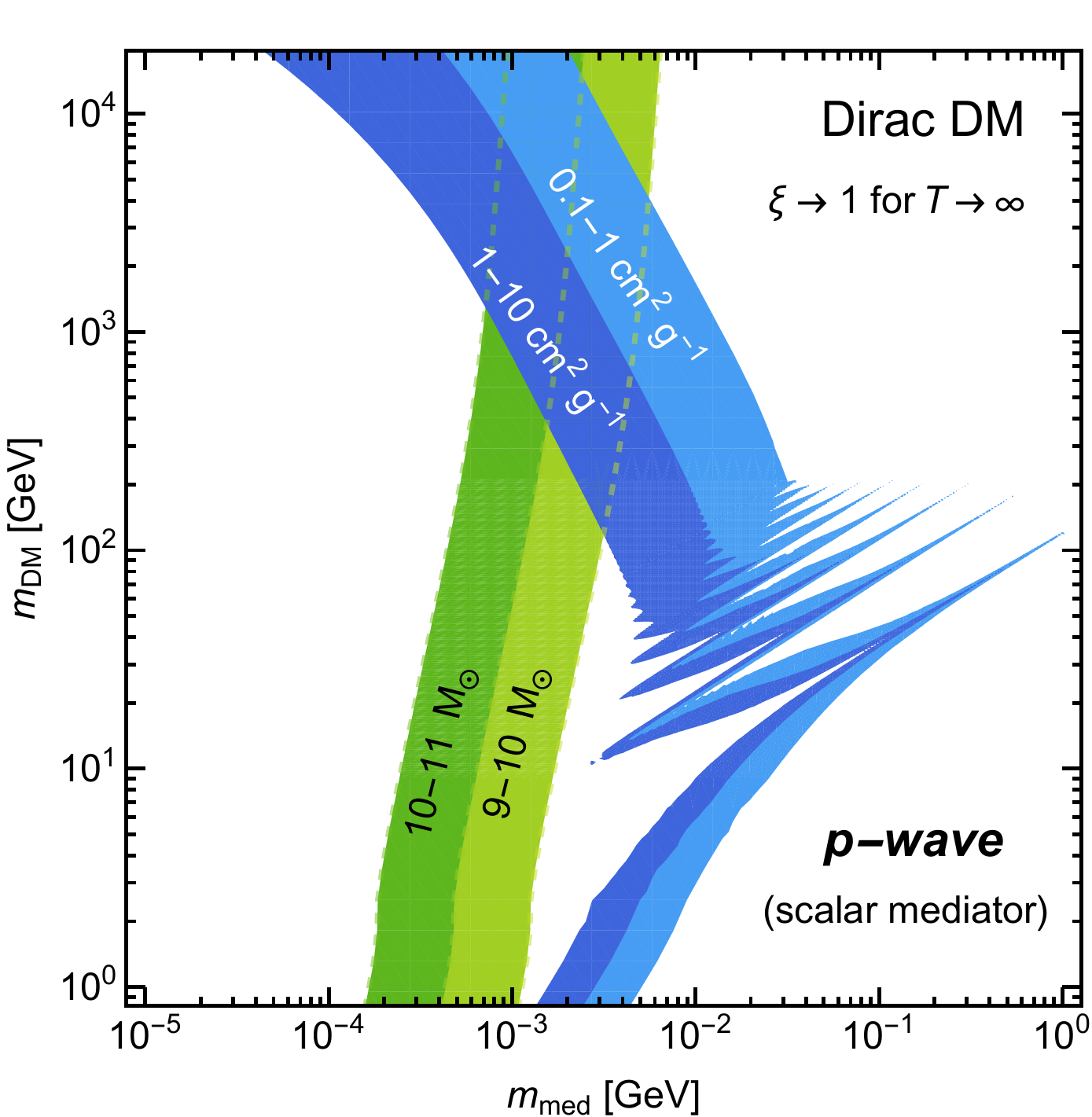}
\caption{Thermally produced Dirac DM interacting via a vector (left) or scalar (right) mediator both
with itself and with a dark radiation component.  
The blue band shows  $\langle\sigma_T\rangle/m_\chi$
for typical velocities in dwarf spheroidal galaxies, indicating the range of values relevant to 
addressing various small-scale issues. 
%The red curve shows where $\langle\sigma_T\rangle/m_\chi=1$\,cm$^2$/g
%at cluster scales; everything to the left and below this line is excluded. 
% These constraints are marginal...
The green band indicates
the cutoff mass $M_\mathrm{cut}$ resulting from late kinetic decoupling; everything to the 
left is firmly excluded by Lyman-$\alpha$ data. These figures illustrate how even DM models 
interacting only invisibly can be observationally constrained. See text for further details.
}
\label{fig:vdSIDM}
\end{figure}
%%%%%%%%%%%%%%%%%%%%%%%%%%		

\bigskip
In Fig.~\ref{fig:vdSIDM} we provide a simple example of a \ds\ result obtained using the \code{vdSIDM} module.
For each point in the DM mass vs.~mediator mass plane, the coupling constant $g_\chi$ is determined by
requiring that thermal production leads to a  relic density of $(\Omega_{\chi}+\Omega_{\bar\chi})h^2=0.112$, 
thus matching the observed DM density (fully including the Sommerfeld enhancement). The blue band shows
$\langle\sigma_T\rangle/m_\chi$, averaged over a Maxwellian distribution with a most probable velocity of
$v_0=30$\,km/s. This corresponds to typical velocities in dwarf spheroidal galaxies, where it has been argued
that a value of $\langle\sigma_T\rangle/m_\chi\sim1$\,cm$^2$/g can mitigate both the cusp-core problem, the 
too-big-to-fail problem and the diversity problem (see section \ref{sec:si}). 
%The red line indicates where $\langle\sigma_T\rangle/m_\chi=1$\,cm$^2$/g for $v_0=1000$\,km/s.
%This corresponds to typical velocities in 
%galaxy clusters, where larger cross sections (to the left and below this line) are firmly excluded as no large 
%cores are observed \cite{Tulin:2017ara}. 
%[These constraints are marginal in this plane and hence anyway not shown] 
One can clearly see the resonant regime of the self-interaction cross section -- the main reason for the different
structure being that in the vector mediator case we average between attractive and repulsive potentials
(the latter of which does not result in resonances).
The green band, finally, indicates the cutoff mass $M_\mathrm{cut}$ resulting from late kinetic decoupling for
a range that may be relevant in mitigating the missing satellite problem (while too large values are excluded by 
Lyman-$\alpha$ data, see Refs.~\cite{Bringmann:2016ilk, Huo:2017vef} for a recent discussion). For the purpose of this figure, we set 
$g_{\tilde\gamma}=g_\chi$, noting that $M_\mathrm{cut}\propto g_{\tilde\gamma}^{3/2} m_{\tilde\gamma}^{.3}$ 
for a given DM mass $m_\chi \gtrsim m_{\phi,V}$ \cite{Bringmann:2016ilk}.
As explained above, the temperature of the dark sector is not a constant in this model; during DM freeze-out, we have 
$\xi\approx 0.54\,(0.57)$ for vector (scalar) mediators, but  $\xi\approx 0.47\,(0.44)$ during late kinetic decoupling.

We note that the relic density calculation that enters in fixing the coupling constant $g_\chi$ assumes that 
the combinations of the model parameters $(g_\chi,m_\chi,m_{V,\phi})$ do not result in a Sommerfeld enhancement very close
to a resonance. In that case, DM would re-enter chemical equilibrium after kinetic decoupling, with
a subsequent second freeze-out period \cite{Dent:2009bv,Zavala:2009mi,Feng:2010zp,vandenAarssen:2012ag}. 
The necessary coupled set of Boltzmann equations to accurately 
describe such a scenario \cite{vandenAarssen:2012ag} will be available with a future release of \ds. While requiring some fine-tuning
in the model parameters, such a scenario would provide additional phenomenological interest as it may not only
address the common $\Lambda$CDM small-scale issues, but additionally reconcile
the observed tension in $\sigma_8$ and $H_0$ measurements between high- and low-redshift cosmological 
observables \cite{Binder:2017lkj, Bringmann:2018jpr}.

\bigskip
%%%%%%%%%%%%%%%%%%%%%%%%%%%%%%%%%%%%%%%%%%%%%%%%%%%%%%%%%%%%
\section{Technical details}
\label{app:technical}

\subsection{Getting started}

To get started, first download \ds\ from \url{www.darksusy.org} and unpack the tar file. To compile, run the
following in the folder where you unpacked it

\begin{verbatim}
./configure
make
\end{verbatim}

You will then have compiled the main \ds\  library, as well as all supplied particle physics 
modules. To test whether the installation was successful, type

\begin{verbatim}
cd examples/test
./dstest
\end{verbatim}

The program will take up to about a minute to run and reports if there are any problems.\footnote{
Strictly speaking, this is only a test of the default particle physics module, \code{mssm}.
To see what happens behind the scenes, you can run in verbose mode by 
replacing `\code{testlevel/2/}' with  `\code{testlevel/1/}' at the beginning of dstest.f. Then type \code{make} and run \code{dstest} again.
}
Even if you now have \ds\ running, it is more fun to start doing some calculations on your own, so the next steps might be to

\begin{itemize}
\item Look at the code \code{examples/dsmain\_wimp.F} which contains an example main 
program to calculate various observables for WIMP DM candidates. It can be the starting point 
if you want to write  your own programs %(make a copy of it, add the build rules to 
%\code{examples/makefile.in} and then configure again).
(simply copy both the program and the makefile in \code{examples} to a user-defined directory, 
and run \code{make} again in that new directory)

\item Use another particle physics module, e.g.~the generic WIMP or the scalar singlet 
(Silveira-Zee) model. To do this, 
just change your makefile to select the particle physics module you wish to use.
Again, a good starting point to test this is the example program 
\code{examples/dsmain\_wimp.F}:  In this case, you can run the \emph{same} program 
for different particle modules simply by changing the first line in  
\code{examples/dsmain\_wimp.driver}. Alternatively, you can call
\code{make -B DS\_MODULE=<your\_module\_choice>} to override the entry 
in the driver file.

\item If you want to modify some existing \ds\ function or subroutine, please \emph{don't do it!}
Instead, add your own routine as a replaceable function, by running the script \code{scr/make\_replaceable.pl} 
on the routine you wish to have a user-replaceable version of (just run the script without arguments for more details).
You will then find a dummy version of the routine in the corresponding \code{user\_replaceables} folder, 
where you can edit it to your liking.\footnote{
Following these steps, it is guaranteed that the newly created user-replaceable function
is properly included in the library where the original DS function used to be, with all 
makefiles being automagically updated. An alternative 
way of using user-replaceable functions is to leave the DS libraries untouched, and to 
instead link to the user-supplied function only when making the main program; this option
is indicated in the top left part of Fig.~\ref{fig:concept} and explained in more detail in the manual.
}
\end{itemize}

\centerline{\bfseries Happy running!}
\bigskip

%%%%%%%%%%%%%%%%%%%%%%%%%%%%%%%%%%%%%%%%%%%%%%%%%%%%%%%%%%%%%%%%%%%%%%
\subsection{How to add a new particle physics module}
\label{sec:newmodel}

To create a new particle physics module, the easiest way is to start from an existing one as a template 
and create a new one from that one. To help you in this process we provide a script 
\code{scr/make\_module.pl} that takes two arguments, the module you want to start from and 
the new one you 
want to create (for further instructions, just call the script without arguments). 
It will then copy the module to a new one, change its name 
throughout the module and make sure that it is compiled by the makefiles and included properly when 
requested by the main programs. If you specify the option \code{-i} only interface functions will be copied 
(which creates a cleaner starting point, but also will most likely not compile without modifications). When 
creating a new module this way, the best is to copy from a module that is as similar as possible to your 
new model. If you want a clean setup, you can always copy from the \code{empty} module. A general
advice is to view the modules we provide as a starting point as inspiration for your new modules.

Even though a particle physics module does not need to include all interface functions (which ones 
are needed only depends on the observables you try to calculate in your main programs), it needs to provide 
an initialization routine 
\code{dsinit\_module.f}. This routine should set a global variable \code{moduletag} to the name of 
the module so that routines that need to check if the correct module is loaded can do so. When using the 
script \code{scr/make\_module.pl} this routine is always created and \code{moduletag} set as it should.

%%%%%%%%%%%%%%%%%%%%%%%%%%%%%%%%%%%%%%%%%%%%%%%%%%%%%%%%%%%%%%%%%%%%%%
\subsection{How to choose pre-defined halo profiles and add new ones}
\label{sec:newhalo}

As described in more detail in Section \ref{sec:halo}, the halo profiles are handled by the \code{dsdmsdriver} 
function in \code{src/dmd\_mod}. This driver function is like a container, or an interface, for all presently
active halo profiles. It ensures that all parts of the code use the halo parameterization in a consistent
and optimized way (e.g.~routines using spherical vs.~axial symmetry for improved performance, storage
of computationally intensive quantities, etc.). Given the complexity of this function, we provide various
example main program in \code{examples/aux} to explicitly demonstrate possible use cases. 
The example \code{DMhalo\_predef.f} illustrates how to use the default version of \code{dsdmsdriver} 
(provided with the \ds\ \dsver\ release) to load additional profiles into the currently active halo database
by using its pre-defined halo parameterizations, while the example \code{DMhalo\_table.f} shows
how to load a profile from a table. In \code{DMhalo\_new.f}, we demonstrate instead how to correctly extend 
\code{dsdmsdriver} when adding a new profile parameterization in order to consistently make it available to all
\ds\ routines that rely on the DM density (in this concrete example, we add the spherical Zhao 
profile~\cite{Zhao:1995cp},
aka $\alpha\beta\gamma$ profile). Given the complexity of the previous example, finally, the program 
\code{DMhalo\_bypass.f}  demonstrates a work-around of completely bypassing the default 
\code{dsdmsdriver} setup when switching to a user-provided new DM density profile; in this approach,  the advanced 
\ds\ system of automatic tabulation of quantities related to DM rates cannot be easily exploited.

%%%%%%%%%%%%%%%%%%%%%%%%%%%%%%%%%%%%%%%%%%%%%%%%%%%%%%%%%%%%%%%%%%%%%%
\subsection{Main technical changes  compared to previous \dst\ releases}
\label{app:changes}
For those familiar with \ds\ 4 and 5, we list here some of the most important 
\emph{technical} changes introduced in \dst\ \dsver. Most physics differences/
improvements are described earlier in this document, with a summary of highlights 
provided in Section \ref{sec:phys}, and for the general changes and the overall structure
see section \ref{sec:Philo}. 

\begin{itemize}
\item The way DM halos are initialized and used has fundamentally changed. See Sections \ref{sec:halo}
and Appendix \ref{sec:newhalo}.

\item The separation between the particle physics model independent \code{ds\_core} and the 
particle physics modules. This has the effect that e.g.\ the routines that calculate yields of 
particles from WIMP annihilation, the former \code{dshayield} and \code{dswayield}, are now 
split into two parts. The most important routines are the interface functions that reside 
in the particle physics module (\code{dscrsource[\_line]} and \code{dsseyield}). 
Those {\it can} call auxiliary routines 
such as \code{dsanyield\_sim} in \code{ds\_core} to get results from the simulated yield tables. 
Cascade decays (e.g.\ decaying Higgs bosons) are handled in the particle physics module as 
this is particle physics dependent. 

\item The yield functions have changed name from \code{dshayield} to \code{dsan\_yield} for annihilation 
in the halo and from \code{dswayield} to \code{dsse\_yield} for annihilation in the Sun/Earth.

\item In general, many routines have changed names to clarify what they do and which belong 
together.

\item The standard model resides in the particle physics library. The reason for this is that one most 
often wants to implement a particle physics model and the standard model at the same time, and 
separating it out to \code{ds\_core} would not be very practical. At the same time, the standard model
is obviously not a BSM theory in itself, with a viable DM candidate. Particle definitions and basic
properties have therefore, for convenience, simply been collected in the directory 
\code{src\_models/common/}  and can be included by any (BSM) particle physics module if the 
user so wishes.

\item The particle codes (\code{k}-variables) are now treated as internal codes. They can be used by the 
particle physics module if the module so wishes, but the interface functions and routines in 
\code{ds\_core} instead use PDG \cite{Groom:2000in} codes when referring to particles.

\item A more stream-lined make system where most makefiles and the configuration script can be 
updated via a simple script. This makes it easier if the user wants to add particle physics modules or 
other contributions to the code.

\item A new test routine \code{dstest} that tests the code and compares with expected results for a set of 
given observables.

\item \code{dsmain.F} is now particle physics module \emph{independent} in the sense that 
it can be linked to different particle physics modules. This assumes that the module has implemented the 
observables one is asking for -- which is why \code{dsmain.F} presently
comes in two versions (\code{dsmain\_wimp.F} and \code{dsmain\_decay.F}).
\item In \code{examples/aux}, we provide various examples of main programs that are more specific
than the general-purpose demonstration of the scope of \ds\ given in (\code{dsmain\_wimp.F} 
and \code{dsmain\_decay.F}). This includes demonstrations of how to use and set up halo profiles
(already explained in Appendix \ref{sec:newhalo}), as well as an example of how to replace 
\ds-supplied functions with user-defined functions (note the two options in the makefile
for how to compile \code{generic\_wimp\_oh2}). Physics-wise, we supply examples of how to
calculate the relic density in both simple and more complex models -- which were in fact used to 
generate Figs.~\ref{fig:RDwimp}, 
\ref{fig:ScalarSinglet} and {fig:vdSIDM} in this article --  as well as the simple program
that was used to produce Fig.~\ref{fig:suncapture}, \code{caprates\_ff.f}. Further examples include programs to calculate fluxes
(\code{wimpyields.f}) and convert between them (\code{flxconv.f}), as well as to compute various
quantities related to UCMHs (\code{ucmh\_test.f}). With future releases, this list of explicit examples
will be continuously extended.
\end{itemize}

%%%%%%%%%%%%%%%%%%%%%%%%%%%%%%%%%%%%%%%%%%%%%%%%%%%%%%%%%%%%%%%%%%%%%%
\subsection{Interfaces to other codes}
\label{sec:interface}

For some of its calculations, \ds\ uses  external codes. We here briefly mention which codes 
we use and how they are interfaced:
\begin{itemize}
\item \code{FeynHiggs} \cite{Heinemeyer:1998yj} is used for Higgs boson mass calculations in the MSSM model. We interface with \code{FeynHiggs} by calling their setup routines directly.
\item \code{HiggsBounds} \cite{Bechtle:2008jh}  and  \code{HiggsSignals} \cite{Bechtle:2013xfa} are used for LEP and LHC limits and constraints on the observed Higgs boson. We interface with both these codes by piggy-backing on the \code{FeynHiggs} interface following the example programs in \code{HiggsBounds} and \code{HiggsSignals}.
\item \code{SuperIso} \cite{Mahmoudi:2007vz} is used for rare decays and is interfaced via our SLHA interface (i.e.\ by writing an SLHA file to disk and read this in from \code{SuperIso}).
\item \code{IsaSugra}  is used for RGE running in CMSSM models and is interfaced by manually extracting all the low-energy parameters from the \cite{Paige:2003mg,isajet_www} common blocks. 
\end{itemize}

\bigskip

\bibliography{DS}

\end{document}